\documentclass[twocolumn,prx,showpacs,amsmath,superscriptaddress]{revtex4-1}

\usepackage[dvips]{graphicx}
\usepackage{overpic}
\usepackage[utf8]{inputenc}
\usepackage{txfonts}
\usepackage{mathrsfs}

\newcommand{\rd}[3][]{\frac{\partial^{#1} #2}{{\partial #3}^{#1}}}
\makeatletter
\renewcommand{\p@subsection}{\thesection-\thesubsection\expandafter\@gobble}
\makeatother

\begin{document}

\title{Unattainability of Carnot efficiency in thermal motors: \\ Coarse-graining and entropy production of Feynman-Smoluchowski ratchet}
\author{Yohei Nakayama}
 \affiliation{Department of Physics, Chuo University, Tokyo 112-8551, Japan}
 \email[]{nakayama@phys.chuo-u.ac.jp}
\author{Kyogo Kawaguchi}
 \affiliation{Department of Systems Biology, Harvard Medical School, Boston, MA02115, USA}
 \email[]{Kyogo\_Kawaguchi@hms.harvard.edu}
\author{Naoko Nakagawa}
 \affiliation{College of Science, Ibaraki University, Ibaraki 310-8512, Japan}
 \email[]{naoko.nakagawa.phys@vc.ibaraki.ac.jp}
\date{\today}

\begin{abstract}
We revisit and analyze the thermodynamic efficiency of the Feynman-Smoluchowski (FS) ratchet, a classical thought experiment describing an autonomous heat-work converter.
Starting from the full kinetics of the FS ratchet and
deriving the exact forms of the hidden dissipations resulting from coarse-graining, we restate the historical controversy over its thermodynamic efficiency.
The existence of hidden entropy productions implies that the standard framework of stochastic thermodynamics applied to the coarse-grained descriptions fails in capturing the dissipative feature of the system.
In response to this problem, we explore an extended framework of stochastic thermodynamics to reconstruct the hidden entropy production from the coarse-grained dynamics.
The approach serves as a key example of how we can systematically address the problem of thermodynamic efficiency in a multi-variable fluctuating non-equilibrium system.
\end{abstract}

\maketitle

\section{Introduction}
The framework of stochastic thermodynamics has not only allowed experimental characterization of small thermodynamic systems \cite{toyabe_experimental_2010}, but has also established a unified scheme to address fundamental questions in thermodynamics.
Identities and inequalities formulated for general stochastic dynamics have been given thermodynamical interpretations such as the second law \cite{evans_probability_1993,jarzynski_nonequilibrium_1997,wang_experimental_2002}, role of information feedback \cite{sagawa_second_2008,toyabe_experimental_2010}, bound on efficiencies of engines at finite time operations \cite{esposito_universality_2009,shiraishi_universal_2016}, and laws extended to nonequilibrium setups \cite{hatano_steady-state_2001,komatsu_steady-state_2008,maes_nonequilibrium_2014}.

The crucial concept behind the developments in stochastic thermodynamics is the entropy production, which is typically introduced through local detailed balance using the logarithmic ratio of transition probabilities \cite{seifert_stochastic_2012}.
This quantity is equivalent, at least in several models, to the energy exchanged with the heat bath divided by the temperature of the bath \cite{crooks_entropy_1999}, and satisfies the second law-like inequality.
Recent works, however, have clarified that fluctuating nonequilibrium systems can carry hidden entropy productions \cite{gomez-marin_lower_2008,puglisi_entropy_2010,roldan_estimating_2010,celani_anomalous_2012,crisanti_nonequilibrium_2012,esposito_stochastic_2012,mehl_role_2012,kawaguchi_fluctuation_2013,strasberg_thermodynamics_2013,munakata_entropy_2014,zimmermann_effective_2015,wang_entropy_2016},
and even under the properly-controlled limit of coarse-graining the coarse-grained model  may not preserve the thermodynamic properties of the original system \cite{puglisi_entropy_2010,celani_anomalous_2012,kawaguchi_fluctuation_2013,bo_multiple-scale_2017}.

\begin{figure}[b]
 \begin{overpic}[width=.925\hsize, trim= 60 85 30 0, clip]{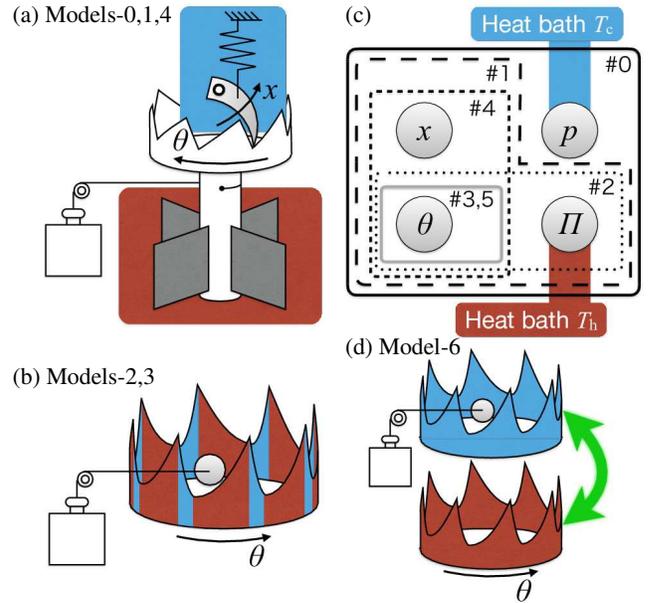}
  \put(-5, 96){(a) Models-0,1,4}
  \put(-5, 36){(b) Models-2,3}
  \put(50, 96){(c)}
  \put(50, 41){(d) \ref{e:decomp}}
 \end{overpic}
 \caption{Schematics of the FS ratchet and its coarse-grained descriptions.
 (a) In Models-0, 1 and 4, the FS ratchet consists of a vane, a gear and a pawl. A spring presses the pawl against the gear, and an external load applies
torque to the axle. The vane and pawl are attached to different heat baths.
 (b) In Models-2 and 3, Langevin equations with the effective mechanical potential, \(U_\mathrm{eff}(\theta)\), inhomogeneous friction, \(\mathcal{G}(\theta)\), and temperature, \(T_\mathrm{eff}(\theta)\), describe the dynamics of the FS ratchet. 
 (c) The  scheme of coarse-graining and the degrees of freedom in Models-0, 1, 2, 3, 4 and 5.
 (d) Langevin equation with stochastic switching between the two heat baths (\ref{e:decomp}).}
 \label{f:schematic}
\end{figure}

In this paper, we focus on the analysis of the Feynman-Smoluchowski (FS) ratchet (\figurename~\ref{f:schematic}a) \cite{feynman_feynman_2010} as a model case to understand how the thermodynamic efficiency can seemlingly change according to the different coarse-grained descriptions of the dynamics.
The FS ratchet is 
one of the most celebrated thought experiments in thermodynamics, where there has been a controversy over its thermodynamic efficiency.
The FS ratchet, due to its asymmetric design, may appear as if it can convert the thermal fluctuation of a single heat bath into work by unidirectional rotation, violating the second law of thermodynamics.
After Smoluchowski showed that there is no rotation and extracted work if the FS ratchet is placed in an isothermal environment \cite{smoluchowski_experimentell_1912}, in \textit{The Feynman Lectures on Physics} \cite{feynman_feynman_2010}, Feynman considered whether it is possible for the ratchet to operate as a Carnot efficient engine.
It was claimed,  based on the analysis of a simplified discrete-stepping model, that the ratchet may attain Carnot efficiency at the stalled state between two heat baths with different temperatures.

Parrondo and Espa\~nol, however, critisized Feynman's argument and pointed out that the momentum variable is non-negligible for the dissipation in the FS ratchet by using a different method of simplifying the model \cite{parrondo_criticism_1996}.
In addition, the unattainability of Carnot efficiency was established in another autonomous Brownian heat engine, B\"uttiker-Landauer (BL) motor  \cite{derenyi_efficiency_1999,hondou_unattainability_2000,benjamin_inertial_2008}, which is a model that has been thought to be closely related to the FS ratchet.
These studies have thus formed a consensus that the FS ratchet cannot attain Carnot efficiency \cite{shiraishi_attainability_2015,acikkalp_analysis_2016,a.martinez_colloidal_2017}.

 Thermodynamic efficiencies in coarse-grained models are not only a theoretical concern but also important in the interpretation of experimental data, since the measurements are typically restricted to a small set of slow variables.
The potential existence of hidden entropy productions (i.e., dissipation owing to the unobserved variables) will make it virtually impossible to draw any conclusions about thermodynamic efficiency in a nonequilibrium small system experiment. 
Therefore, it is of interest to extend the framework of stochastic thermodynamics to be able to re-interpret the coarse-grained data in order to obtain the original thermodynamic properties.

Here we aim to provide a unified understanding and a workaround to the FS ratchet problem, through a systematic procedure of coarse-graining which does not involving any empirical simplifications.
We derive the coarse-grained descriptions of the original FS ratchet including the previously known models \cite{buttiker_transport_1987,millonas_self-consistent_1995,sekimoto_kinetic_1997,hondou_irreversible_1998,jack_intrinsic_2016} together with new models.
 We then ask how the entropy productions may differ in the series of models by obtaining the explicit expressions for the hidden entropy productions, and discuss their relations to the previous arguments on the controversial thermodynamic efficiencies.
 Finding that most of the coarse-grained dynamics do not preserve the thermodynamic property of the original FS ratchet, we further explore and find a way to quantify the hidden dissipation based on a limited number of coarse-grained observables. 




This paper is organized as follows.
In Sec.~\ref{s:setup} we introduce the original FS ratchet model (\ref{m:origin}).
In Sec.~\ref{s:coarse-graining} the coarse-grained descriptions (Models-2,3 and 5) are derived by taking the time-scale separation limits.
In Sec.~\ref{s:hidden entropy production} we calculate the behavior of the dissipation through the framework of stochastic thermodynamics, in the limits where the coarse-grained descriptions are obtained. We derive the explicit forms of hidden entropy productions as the first main result of the paper. In addition, we clarify that what Feynman did can be regarded as applying stochastic thermodynamics to the coarse-grained description.
In Sec.~\ref{s:numerical} we present the results of numerical simulations which clarify the impact of hidden entropy production on the thermodynamic efficiencies (\figurename~\ref{f:efficiency}).
These results confirm that although the kinetics of the FS ratchet can be coarse-grained systematically, most of the coarse-grained models do not reproduce the entropy production of the original system.
In 
Sec.~\ref{s:decomposition} we describe our proposal of a workaround to the problem of hidden entropy production by demonstrating that even when using the coarse-grained variables, the fine-grained entropy production can be reconstructed by the decomposition of the Langevin dynamics (\ref{e:decomp}).
In Sec.~\ref{s:conclusion} we give concluding remarks.
Some technical details are described in Appendices.


\section{Setup} \label{s:setup}
As shown in \figurename~\ref{f:schematic}a, the FS ratchet consists of a vane and a gear connected by a rigid axle, and a pawl meshing with the gear.
A spring pushes the pawl against the gear.
The vane and the pawl are in contact with different heat baths with temperatures \(T_h\) and \(T_c\).
An external load couples with the axle, and applies a constant torque, \(f\).
By assuming the interaction between the pawl and the gear to be mechanical,
the equations of motion for the angle \(\theta\) of the coaxial vane and gear and the height \(x\) of the pawl reads
\begin{align}
 \begin{split}
  \dot \theta &= \frac{\Pi}{m},
  \\
  \dot \Pi &= - \frac{\Gamma}{m} \Pi + f - \rd{U(\theta, x)}{\theta} + \sqrt{2\Gamma T_h} \xi,
  \\
  \dot x &= \frac{p}{m_x},
  \\
  \dot p &= - \frac{\gamma}{m_x} p - \rd{U(\theta, x)}{x} + \sqrt{2\gamma T_c}\zeta,
 \end{split}
 \tag{Model-0}
 \label{m:origin}
\end{align}
where \(\Pi\) and \(p\) are the momentum conjugated to \(\theta\) and \(x\), respectively.
 Here, \(m\) is the corresponding moment of inertia, and \(m_x\) is the mass of the pawl.
We take Langevin heat baths where \(\Gamma\) and \(\gamma\) are the viscous frictional coefficients.
\(\xi\) and \(\zeta\) are independent white Gaussian noises with zero means and unit variances.
The Boltzmann constant is set to unity.

In a straightforward manner, we may obtain a coarse-grained description,
\begin{align}
 \begin{split}
 \dot \theta &= \frac{\Pi}{m}
 ,
 \\
 \dot \Pi &= - \frac{\Gamma}{m} \Pi + f - \frac{\partial U(\theta, x)}{\partial \theta} + \sqrt{2\Gamma T_h} \xi
 ,
 \\
 \gamma \dot x &= - \frac{\partial U(\theta, x)}{\partial x} + \sqrt{2\gamma T_c}\tilde\zeta
 ,
 \end{split} \tag{Model-1} \label{m:1}
\end{align}
where the momentum degree of freedom, \(p\), is eliminated by considering the overdamped limit for the pawl.
The symbol \(\tilde\zeta\) is an independent white Gaussian noise with zero mean and unit variance.

We assume the mechanical potential
\begin{align}
U(\theta, x) = U_0(x) + U_ {I}(x - \phi(\theta)),
\end{align}
where \(U_0(x)\) is the elastic potential of the spring attached to the pawl and \(U_ {I}(x - \phi(\theta))\) is the trapping potential between the tip of pawl and the surface of gear. \(\phi(\theta)\) is a periodic function which represents the shape of gear, with the period \(L\).

\section{Coarse-graining} \label{s:coarse-graining}
\begin{figure}[btp]
 \includegraphics[width=\hsize, trim= 0 400 560 0, clip]{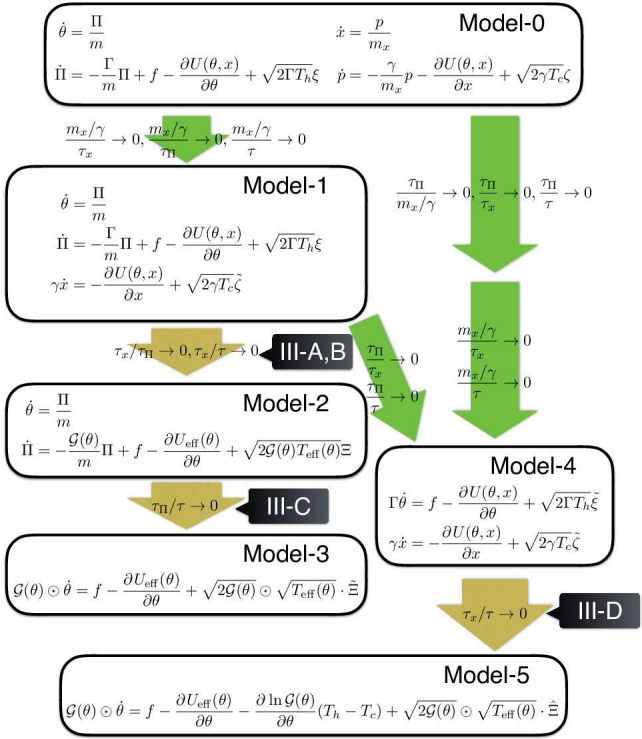}
 \caption{A complete picture of the routes of coarse-graining. 
 The arrows represent the processes of coarse-graining where the necessity limit conditions to carry out the coarse-graining are specified.
 The yellow ones represent the coarse-graining processes derived in this paper, and the green ones correspond to the trivial coarse-graining eliminating the momentum variable attached to the isothermal environment.
 \(\tau_x = L_x^2 \gamma / T_c\) and \(\tau_\Pi = m / \Gamma\) are the time scales of the pawl and momentum degree of freedom, respectively, and \(\tau\) is a time scale representing the other time scales of the system.
 }
 \label{f:complete}
\end{figure}

We here explicitly derive the coarse-grained descriptions of \ref{m:1} by taking the limits where the time-scales of the variables are separated.
This is in contrast to the approaches taken for example in \cite{feynman_feynman_2010} where the discrete stepping model and the BL motor were introduced on the basis of phenomenological arguments.
%
We start from \ref{m:1} and consider two limits, a ``tightly confined limit'' and a ``overdamped limit'', where \(x\) and \(\Pi\) are eliminated, respectively.
Through this two-step coarse-graining, we arrive at a closed equation of motion for \(\theta\).
We note that the order of elimination of \(x\) and \(\Pi\) matters.
As summarized in \figurename~\ref{f:complete}, we find that taking the tightly-confined limit first will result in a different model to when the overdamped limit is taken first (\ref{e:od} vs \ref{e:another}).

Hereafter, we denote the time scale of the relaxation in the trapping potential as \(\tau_x := L_x^2 \gamma / T_c\) with the length scale \(L_x\) of the trapping potential.
The time scale for the relaxation of the momentum of the vane and the gear is \(\tau_\Pi = m / \Gamma\).
We assume the other time scales to be of the same order, represented by \(\tau\).
In order to satisfy this assumption, we fix the functional forms of \(\phi(\theta)\) and \(U_0(x)\) and the ratios \(\Gamma / \gamma\), \(T_c/T_h\) and \(fL / T_h\).
We are interested in the cases where $\tau_x$ and $\tau_\Pi$ are separated from \(\tau\).

Here we note on why we chose the interaction between the tip of the pawl and the surface of the gear \(U_I (x - \phi (\theta))\) as a trapping potential. 
In the original FS ratchet, this interaction was a hard-core repulsion, so the tight confinement of the pawl could only be realized by increasing the force exerted by the spring [$U_0(x)$], since the length scale of the confinement is proportional to \(T_h /[ \partial U(\theta, x) / \partial x]\).
However, the energy required to lift the pawl becomes larger than the thermal energy when increasing the spring force, which means that the FS ratchet will stop rotating in this limit.
By introducing \(U_I (x - \phi (\theta))\) as a trapping potential, we may take the the tightly confined limit by keeping the height of the potential barrier constant.
This modification to the original dynamics is the key in conducting the following coarse-graining procedures.

We mainly conduct coarse-graining in terms of the master equation,
i.e., partial differential equation satisfied by probability densities.
The master equation for \ref{m:1} reads
\begin{align}
 &\rd{P(\theta, \Pi, x)}{t} = -\rd{}{\theta}\left[ \frac{\Pi}{m} P(\theta, \Pi, x) \right]
 \nonumber \\
 &-\rd{}{\Pi} \left[\left(-\frac{\Gamma}{m} \Pi + f - \rd{U(\theta, x)}{\theta}\right) P(\theta, \Pi, x)- \Gamma T_h \rd{P(\theta, \Pi, x)}{\Pi}\right]
 \nonumber \\
 &-\rd{}{x} \left(-\frac{1}{\gamma} \rd{U(\theta, x)}{x} P(\theta, \Pi, x) - \frac{T_c}{\gamma} \rd{P(\theta, \Pi, x)}{x}\right)
 .
 \label{e:KFP}
\end{align}
Here, \(P(\theta, \Pi, x)\) represents the joint probability density of \(\theta, \Pi\) and \(x\).

\subsection{Coarse-grained Description at Tightly Confined Limit} \label{ss:TC}
We first consider the limit where the tip of the pawl is tightly confined to the surface of the gear.
In this limit, \(\tau_x\) is  assumed to be separated from \(\tau_\Pi\) and \(\tau\).
Assuming that the ratio \(\tau_\Pi / \tau\) is fixed in this section, we introduce a small parameter representing the separation of time scales, \(\varepsilon := \tau_x / \tau_\Pi \sim \tau_x / \tau\).
 Here, we summarize the derivation of the coarse-grained description, and give the details in Appendix \ref{ss:SPT-1}.

In this tightly confined limit, the height of the pawl $x$ is eliminated as the fast variable.
Therefore, the coarse-grained dynamics is described by the master equation for the joint probability density, 
\(
 P(\theta, \Pi) = \int   dx P(\theta, \Pi, x)
\).
Throughout this paper, the integrals with respect to \(\theta, \Pi\) and \(x\) 
are taken over the domain of integrand.
The time derivative of \(P(\theta, \Pi)\) is obtained by integrating Eq.~(\ref{e:KFP}) with respect to \(x\):
\begin{align}
 \rd{P(\theta, \Pi)}{t} =& -\rd{}{\theta}\left[ \frac{\Pi}{m} P(\theta, \Pi) \right]
 \nonumber \\
 &-\rd{}{\Pi} \left[\left(-\frac{\Gamma}{m} \Pi + f\right) P(\theta, \Pi)- \Gamma T_h \rd{P(\theta, \Pi)}{\Pi}\right]
 \nonumber \\
 &-\rd{}{\Pi} \left[ - \int   dx \rd{U(\theta, x)}{\theta} P(\theta, \Pi, x) \right]
 .
 \label{e:formal}
\end{align}
The closed equation for \(P(\theta, \Pi)\) is obtained by evaluating the last line of Eq.~(\ref{e:formal}) in the limit of \(\varepsilon\to0\).
Employing the singular perturbation theory to avoid the divergence caused by the secular terms, we obtain
\begin{align}
 &\rd{P(\theta, \Pi)}{t} = - \rd{}{\theta} \left[ \frac{\Pi}{m} P(\theta, \Pi)\right]
 \nonumber \\
 &- \rd{}{\Pi}\left[ -\frac{\mathcal{G}(\theta)}{m} \Pi - \rd{U_\mathrm{eff}(\theta)}{\theta} + f
 - \mathcal{G}(\theta) T_\mathrm{eff}(\theta) \rd{}{\Pi}\right]P(\theta, \Pi),
 \label{e:Kramers}
\end{align}
which is the Kramers equation.
Here, \(U_\mathrm{eff}(\theta)\), \(\mathcal{G}(\theta)\) and \(T_\mathrm{eff}(\theta)\) are the effective potential, the effective frictional coefficient and the effective temperature:
\begin{align}
 U_\mathrm{eff}(\theta) &= U_0(\phi(\theta))
 ,
 \label{e:U_eff}
 \\
 \mathcal{G}(\theta) &= \Gamma + \gamma \phi'(\theta)^2
 ,
 \label{e:G_eff}
 \\
 T_\mathrm{eff}(\theta) &= \frac{\Gamma T_h + \gamma \phi'(\theta)^2 T_c}{\Gamma + \gamma \phi'(\theta)^2}
 \label{e:T_eff}
 .
\end{align}

The master equation (\ref{e:Kramers}) is equivalent to the Langevin equation
\begin{align}
 \begin{split}
 \dot \theta &= \frac{\Pi}{m}
 ,
 \\
 \dot \Pi &= -\frac{\mathcal{G}(\theta)}{m} \Pi + f - \frac{\partial U_\mathrm{eff}(\theta)}{\partial \theta} + \sqrt{2 \mathcal{G}(\theta) T_\mathrm{eff}(\theta)} \Xi \notag
 ,
 \end{split}
 \tag{Model-2}
 \label{e:ud}
\end{align}
where \(\Xi\) is a white Gaussian noise with zero mean and unit variance.
\ref{e:ud} describes the Brownian motion of a single degree of freedom under the effective potential, frictional coefficient, and temperature (\figurename~\ref{f:schematic}b), which is known as the B\"uttiker-Landauer motor \cite{buttiker_transport_1987}. 

\subsection{Quick Derivation of \ref{e:ud}} \label{ss:quick}
We here give a quick derivation of \ref{e:ud} in the special case where we set \(U(\theta, x) = k x^2 / 2 + \lambda [x - \phi(\theta)]^2 / 2\), based on a temporal coarse-graining method \cite{zwanzig_nonlinear_1973,sekimoto_temporal_1999}.
The details will be given in Appendix \ref{ss:TCG}.
Since the equation of motion
\begin{align}
 \gamma \dot x &= - (k + \lambda) x + \lambda \phi(\theta) + \sqrt{2\gamma T_c} \tilde\zeta
 ,
 \label{em:x}
\end{align}
is linear in $x$, we may formally solve \(x\) as the functional of \(\phi(\theta)\) and \(\tilde\zeta\), and eliminate \(x\) by substituting the formal solution of \(x\) into the equation of motion of \(\Pi\):
\begin{align}
 \dot \Pi_t =& - \frac{\Gamma}{m} \Pi_t + f + \lambda \phi'(\theta_t) \left[
 -\frac{k}{k + \lambda} \phi(\theta_t) - \frac{\gamma \lambda}{(k + \lambda)^2} \phi'(\theta_t) \frac{\Pi_t}{m} \right]
 \nonumber \\
 &+ \frac{ \lambda \phi'(\theta_t) \sqrt{2\gamma T_c} }{\gamma}\int_{-\infty}^t   dt' e^{-\frac{k + \lambda}{\gamma}(t - t')} \tilde\zeta_{t'}
 + \sqrt{2\Gamma T_h} \xi_t
 ,
 \label{em:p formal}
\end{align}
where we explicitly show the time-dependence of the variables as the subscript.
Note that Eq.~(\ref{em:p formal}) describes a non-Markovian dynamics with colored noise.
The Markovian property is recovered in the limit of \(\varepsilon = \tau_x / \tau_\Pi \to0\).
In this limit, we may introduce a time interval \(\Delta t\), which is shorter than \(\tau_\Pi\) and \(\tau\), 
but longer than \(\tau_x = \gamma / (k + \lambda)\). 
By integrating Eq.~(\ref{em:p formal}) over the time interval \(\Delta t\) and taking the limit of \(\varepsilon\to0\), 
we obtain \ref{e:ud}.

\subsection{Coarse-grained Description at Overdamped Limit}
Next, we discuss the overdamped limit.
In this limit \(\tau_\Pi = m/\Gamma\) is separated from the other time scales represented by \(\tau\).
The underdamped Brownian motion corresponding to \ref{e:ud} can be coarse-grained to an overdamped Langevin equation \cite{jayannavar_macroscopic_1995,sekimoto_stochastic_2010}:
\begin{align}
 \mathcal{G}(\theta) \odot \dot \theta =& f - \frac{\partial U_\mathrm{eff}(\theta)}{\partial \theta}
 + \sqrt{2\mathcal{G}(\theta)} \odot \sqrt{T_\mathrm{eff}(\theta)} \cdot \tilde\Xi
 ,
 \tag{Model-3}
 \label{e:od}
\end{align}
where \(\tilde\Xi\) is a white Gaussian noise with zero mean and unit variance.
The symbols \(\cdot\) and \(\odot\) indicate the product in the sense of It\^o and anti-It\^o, respectively, which specify the evaluation of the quantity on the left:
\begin{align}
 \sqrt{2\mathcal{G}(\theta)} \odot \sqrt{T_\mathrm{eff}(\theta)} \cdot \tilde\Xi
 = \lim_{\delta t\to0}\sqrt{2\mathcal{G}(\theta_{t+\delta t}) T_\mathrm{eff}(\theta_t)} \frac{1}{\delta t}\int_t^{t+\delta t} \!\!\!\!\! \tilde\Xi_s   ds
 .
\end{align}

\ref{e:od} is a generalized version of the B\"uttiker-Landauer motor to the case of \(\theta\)-dependent friction.
The net velocity of rotation of this motor is obtained as \cite{buttiker_transport_1987}
\begin{align}
 \langle \dot{\theta} \rangle = \frac{L [1-\exp(-L \Delta)]}{ \int_0 ^L dy \exp[-\psi (y)] \int_y ^{y+L} dy' \exp[\psi (y')] \mathcal{G}(y') / T_\mathrm{eff} (y')}
 , \label{e:vel}
\end{align}
where \(\langle \cdot \rangle\) represents the steady-state ensemble average and
\begin{align}
\psi (y) &:= \int^ y dy' \frac{f - U'_\mathrm{eff} (y') -T'_\mathrm{eff} (y')}{T_\mathrm{eff} (y')},
 \label{e:psi}
\\
\Delta &:=  \psi (y) - \psi (y+L).
\end{align}
Equation (\ref{e:vel}) indicates that there is unidirectional motion if $\Delta \neq 0$.
The system works as a Brownian heat engine when the rotation is opposite to the direction of the constant torque: $\langle \dot{\theta} \rangle < 0$ $(\Delta < 0)$ when $f>0$.
In Appendix \ref{ss:SPT-OD}, we present the derivation of \ref{e:od} based on the singular perturbation theory \cite{celani_anomalous_2012}.

\subsection{Other Routes of Coarse-graining}
As shown in \figurename~\ref{f:complete}, there is another path to obtain the closed equation of motion for $\theta$; we can take the overdamped limit before the tightly confined limit.
In a straightforward manner, we obtain a coarse-grained description of \ref{m:1}
\begin{align}
 \begin{split}
  \Gamma \dot \theta =& f - \rd{U(\theta, x)}{\theta} + \sqrt{2\Gamma T_h} \tilde\xi
  ,
  \\
  \gamma \dot x&= - \rd{U(\theta, x)}{x} + \sqrt{2\gamma T_c} \tilde\zeta
  ,
 \end{split}
 \tag{Model-4}
 \label{e:od-od}
\end{align}
which is equivalent to the master equation:
\begin{align}
 \rd{P(\theta, x)}{t} =& -\rd{}{\theta}\left[\frac{1}{\Gamma}\left(f - \rd{U(\theta, x)}{\theta}\right) P(\theta, x) - \frac{T_h}{\Gamma}\rd{P(\theta, x)}{\theta}\right]
 \nonumber \\
 & -\rd{}{x}\left[-\frac{1}{\gamma} \rd{U(\theta, x)}{\theta} P(\theta, x) - \frac{T_c}{\gamma}\rd{P(\theta, x)}{x}\right]
 .
 \label{e:od-od FP}
\end{align}
Here, \(\tilde\xi\) is an independent white Gaussian noise with zero mean and unit variance.
\ref{e:od-od} may also be obtained from Model-0 by taking the overdamped limit for the vane and the gear first (\figurename~\ref{f:complete}).
This type of model has also been analyzed in the context of the FS ratchet \cite{sekimoto_kinetic_1997,hondou_irreversible_1998,jack_intrinsic_2016}.

The time evolution of \(P(\theta) = \int   dx P(\theta, x)\) is obtained by taking the tightly-confined limit in Eq.~(\ref{e:od-od FP}).
In the form of the Langevin equation, the coarse-grained description is obtained as
\begin{align}
 \begin{split}
 \mathcal{G}(\theta) \odot \dot \theta =& f - \rd{U_\mathrm{eff}(\theta)}{\theta} - \rd{\ln \mathcal{G}(\theta)}{\theta} (T_h - T_c)
 \nonumber \\
 &+ \sqrt{2\mathcal{G}(\theta)} \odot \sqrt{T_\mathrm{eff}(\theta)} \cdot \hat\Xi
 .
 \end{split}
 \tag{Model-5}
 \label{e:another}
\end{align}
where \(\hat\Xi\) is a white Gaussian noise with zero mean and unit variance.
Unidirectional motion is driven by the same mechanism as \ref{e:od}.
The details of the derivation are given in Appendix \ref{ss:SPT-another}.

\ref{e:od} and \ref{e:another} are similar but slightly different; there is an extra term in \ref{e:another} that  vanishes when \(T_h = T_c\) but affects the average velocity when \(T_h \neq T_c\).
Formally, this means that the two limits, \(\tau_x \ll \tau_\Pi \ll \tau\) (\ref{e:od}) and \(\tau_\Pi \ll \tau_x \ll \tau\) (\ref{e:another}) are different under a non-equilibrium setup.

We here note on some of the previous works related to the coarse-graining of FS ratchet-like dynamics.
In \cite{magnasco_feynmans_1998}, the authors coarse-grained \ref{e:od-od} based on phenomenlogical arguments, and correctly obtained the effective potential and temperature [Eqs.~(\ref{e:U_eff},\ref{e:T_eff})].
However, they did not arrive at the inhomogeneous friction [Eqs.~(\ref{e:G_eff})] and the force proportional to temperature-difference (the third term of the right hand side of \ref{e:another}).
In \cite{gomez-marin_ratchet_2006}, a similar attempt was made to obtain a phenomenological model by neglecting the temporal correlation of the fast variable, which resulted in an unphysical model that does not satisfy the fluctuation-dissipation relation.
In \cite{millonas_self-consistent_1995}, Millonas considered a non-equilibrium bath variable ($x$ in our model) that couples to a motor, and essentially obtained all of Eqs.~(\ref{e:U_eff}-\ref{e:T_eff}).
The aspects of dissipation and thermodynamic efficiency, however, were not discussed.


\section{Stochastic Thermodynamics of Feynman-Smoluchowski Ratchet} \label{s:hidden entropy production}
We here discuss the thermodynamic properties of the FS ratchet by applying the framework of stochastic thermodynamics.
We first describe the entropy production for each model, and then take the coarse-graining limits in each case to see if there is descrepancy (i.e., hidden entropy production) between the different scales of descriptions.


\subsection{Entropy Production Rates}
The standard prescription of stochastic thermodynamics \cite{seifert_stochastic_2012} connects the entropy production rate in the heat baths with the transition probabilities of the model.
The entropy production rates for the models are written as
\begin{align}
 \sigma_{1}(\theta_{t'}, \Pi_{t'}, x_{t'}|\theta_{t}, \Pi_{t}, x_{t}) :=& \frac{1}{t' - t} \ln \frac{W_1(\theta_{t'}, \Pi_{t'}, x_{t'}|\theta_{t}, \Pi_{t}, x_{t})}{W_1(\theta_{t}, -\Pi_{t}, x_{t}|\theta_{t'}, -\Pi_{t'}, x_{t'})}
 \nonumber \\
 =& \frac{Q_1^h}{T_h} + \frac{Q^c}{T_c}
 ,
 \label{e:W-first}
 \\
 \sigma_{2}(\theta_{t'}, \Pi_{t'}|\theta_{t}, \Pi_{t}) :=& \frac{1}{t' - t}\ln \frac{W_2(\theta_{t'}, \Pi_{t'}|\theta_{t}, \Pi_{t})}{W_2(\theta_{t}, -\Pi_{t}|\theta_{t'}, -\Pi_{t'})}
 \nonumber \\
 =& \frac{1}{T_\mathrm{eff}(\theta)} \circ Q_2
 ,
 \\
 \sigma_{3}(\theta_{t'}|\theta_{t}) :=& \frac{1}{t' - t} \ln \frac{W_3(\theta_{t'}|\theta_{t})}{W_3(\theta_{t}|\theta_{t'})}
 \nonumber \\
 =& \frac{1}{T_\mathrm{eff}(\theta)} \circ \left(Q_3 - \rd{T_\mathrm{eff}(\theta)}{\theta} \circ \dot \theta\right)
 ,
 \label{e:W-OD1}
 \\
 \sigma_{4}(\theta_{t'}, x_{t'}|\theta_{t}, x_t) :=& \frac{1}{t' - t} \ln \frac{W_4(\theta_{t'}, x_{t'}|\theta_{t}, x_t)}{W_4(\theta_{t}, x_{t}|\theta_{t'}, x_{t'})}
 \nonumber \\
 =& \frac{Q_4^h}{T_h} + \frac{Q^c}{T_c}
 ,
\end{align}
\begin{align}
 \sigma_{5}(\theta_{t'}|\theta_{t}) :=& \frac{1}{t' - t} \ln \frac{W_5(\theta_{t'}|\theta_{t})}{W_5(\theta_{t}|\theta_{t'})}
 \nonumber \\
 =& \frac{1}{T_\mathrm{eff}(\theta)} \circ \left[Q_5
 - \left(\rd{T_\mathrm{eff}(\theta)}{\theta} + \rd{\mathcal{G}(\theta)}{\theta} (T_h - T_c)\right) \circ \dot \theta\right]
 ,
 \label{e:W-last}
\end{align}
where \(W_i\) are the transition probabilities of  each model (\(i = 1, \ldots, 5\)) whose explicit expressions are given in Appendix \ref{ss:Ws}.
The symbol \(\circ\) represents the product in the sense of Stratonovich, and the time increment \(t' - t\) is taken to be smaller than the time scales of each model.
The heat flux \(Q^h_i\), \(Q^c\) and \(Q_i\) from the system to each heat bath are defined as
\begin{align}
 Q_1^h =& -\left(\dot \Pi + \rd{U(\theta, x)}{\theta} - f\right) \circ \frac{\Pi}{m},
 \\
 Q^c =& - \rd{U(\theta, x)}{x} \circ \dot x,
 \\
 Q_2 =& -\left(\dot \Pi + \rd{U_\mathrm{eff}(\theta)}{\theta} - f\right) \circ \frac{\Pi}{m},
 \\
 Q_3 = Q_5 =& \left(-\rd{U_\mathrm{eff}(\theta)}{\theta}  + f\right) \circ \dot \theta,
 \label{e:OD1 heat}
 \\
 Q_4^h =& \left(-\rd{U(\theta, x)}{\theta} + f\right) \circ \dot \theta.
\end{align}

The entropy production rates obtained from the transition probabilities are not equal to the heat flux divided by the effective temperature \(T_\mathrm{eff}(\theta)\) in some cases [Eqs.~(\ref{e:W-OD1},\ref{e:W-last})].
This is because the heat flux are defined based on the consistency with the energy balance \cite{sekimoto_stochastic_2010}.
An alternative definition of heat flux and its effect on the thermodynamic efficiency will be discussed in Appendix \ref{a:heat}.

We note that the coarse-graining from \ref{m:origin} to \ref{m:1} or \ref{e:od-od} will not involve hidden entropy productions, since the elimination of the momentum degree of freedom attached to an isothermal heat bath does not involve any hidden entropy production \cite{celani_anomalous_2012}.
Therefore, we here focus on the analysis of the entropy productions in Models 1-5.



\subsection{Hidden Entropy Production in Coarse-graining to \ref{e:od}} \label{ss:1-2-3}
In this subsection, we focus on the entropy production rates in the limit where \ref{e:od} is derived. Since we have the asymptotic behavior of the probability density function \(P(\theta, \Pi, x)\) in the limit of tight confinement (cf. Appendix \ref{ss:SPT-1}), the ensemble average of \(\sigma_1\) can be written as
\begin{align}
 \langle\sigma_{1}\rangle =& \langle\sigma_{2}\rangle 
 + \left\langle \frac{\Gamma (\mathcal{G}(\theta) - \Gamma)}{\mathcal{G}(\theta)T_\mathrm{eff}(\theta)}
   \left(\frac{1}{T_c} - \frac{1}{T_h}\right) (T_h - T_c) \frac{\Pi^2}{m^2} \right\rangle
   ,
   \label{e:1-2}
\end{align}
The derivation of Eq.~(\ref{e:1-2}) is given in Appendix \ref{a:1-2}.
Equation (\ref{e:1-2}) states that there is a finite and positive difference between \(\langle \sigma_1\rangle \) and \(\langle \sigma_2\rangle\) for \(T_h \not = T_c\), which is the hidden entropy production between \ref{m:1} and \ref{e:ud}.
This means that the dissipation is underestimated if we assume \ref{e:ud} as the description of the FS ratchet.

Next, we evaluate the right hand side of Eq.~(\ref{e:1-2}) at the overdamped limit.
Taking the ensemble average of \(\sigma_2\) with respect to the asymptotic form of \(P(\theta, \Pi)\) in the overdamped limit, we obtain
\begin{align}
 \langle\sigma_{2}\rangle = \langle\sigma_{3}\rangle + \left\langle \frac{T_\mathrm{eff}(\theta)}{2\mathcal{G}(\theta)} \left(\frac{T_\mathrm{eff}'(\theta)}{T_\mathrm{eff}(\theta)}\right)^2 \right\rangle
 .
 \label{e:2-3}
\end{align}
The derivation of Eq.~(\ref{e:2-3}) is given in Appendix \ref{a:2-3}.
The hidden entropy production between Models-2 and 3, \( \langle\sigma_{2}\rangle -\langle\sigma_{3}\rangle \), is 
positive unless \(T_\mathrm{eff}(\theta)\) is a constant value.
The positivity of these hidden entropy productions are consistent with the general condition discussed in \cite{kawaguchi_fluctuation_2013}.

Since $\Pi^2/m$ relaxes to \(T_\mathrm{eff}(\theta)\) in the overdamped limit, we finally obtain
\begin{align}
 \langle\sigma_{1}\rangle =&
 \langle\sigma_{3}\rangle 
 + \left\langle \frac{T_\mathrm{eff}(\theta)}{2\mathcal{G}(\theta)} \left(\frac{T_\mathrm{eff}'(\theta)}{T_\mathrm{eff}(\theta)}\right)^2 \right\rangle
 \nonumber \\
 &+ \left\langle \frac{\Gamma (\mathcal{G}(\theta) - \Gamma)}{m \mathcal{G}(\theta)} \left(\frac{1}{T_c} - \frac{1}{T_h}\right) (T_h - T_c) \right\rangle
   .
   \label{e:1-2-3}
\end{align}
We note that Eq.~(\ref{e:1-2},\ref{e:2-3},\ref{e:1-2-3}) hold even in the non-steady states by taking into consideration the entropy increment in the system.

Equation~(\ref{e:1-2-3}) indicates that the true entropy production rate \(\langle \sigma_1 \rangle\) is positive in the regime where the FS ratchet operates as a heat engine (\(T_h \not = T_c\) and \(\phi(\theta) \not =  {const.}\)), even when \(\langle \sigma_3\rangle = 0\) holds.
This is consistent with the previous studies of the B\"uttiker-Landauer motor system \cite{derenyi_efficiency_1999,hondou_unattainability_2000,benjamin_inertial_2008,celani_anomalous_2012}.
Indeed, the second term of the right hand side of Eq.~(\ref{e:2-3}) may be considered as a generalization of the results to the case where the frictional coefficient is state-dependent. 
Our results show that the FS ratchet carries another hidden dissipation expressed as the last term of the right hand side of Eq.~(\ref{e:1-2-3}).

The last term of Eq.~(\ref{e:1-2-3}) also has a significant impact on \(\langle \sigma_1\rangle\).
Evaluating the order of each term in Eq.~(\ref{e:1-2-3}), we obtain
\begin{align}
 & \langle \sigma_3 \rangle = \int_0^L d\theta \frac{-U'_\mathrm{eff}(\theta) + f}{T_\mathrm{eff}(\theta)} \left\langle \frac{\dot \theta}{L} \right\rangle
 = \Delta \left\langle \frac{\dot \theta}{L}\right\rangle \sim \tau^{-1}
 ,
 \label{e:apparent}
 \\
 &\left\langle \frac{T_\mathrm{eff}(\theta)}{2\mathcal{G}(\theta)} \left(\frac{T_\mathrm{eff}'(\theta)}{T_\mathrm{eff}(\theta)}\right)^2 \right\rangle
 \sim \left\langle \frac{T_\mathrm{eff}(\theta)}{\mathcal{G}(\theta)} \right\rangle \frac{1}{L^2} \sim \tau^{-1}
 ,
 \label{e:non-dominant}
 \\
 &\left\langle \frac{\Gamma (\mathcal{G}(\theta) - \Gamma)}{m \mathcal{G}(\theta)} \left(\frac{1}{T_c} - \frac{1}{T_h}\right) (T_h - T_c) \right\rangle
 \sim \frac{\Gamma}{m} = \tau_\Pi^{-1}
 ,
 \label{e:dominant}
\end{align}
where we use the fact that \(T_\mathrm{eff}(\theta) / \mathcal{G}(\theta)\) is the effective diffusion coefficient in \ref{e:od}.
Since \(\tau_\Pi / \tau = \varepsilon\), the ratio of Eq.~(\ref{e:dominant}) to Eqs.~(\ref{e:apparent}, \ref{e:non-dominant}) diverges in the limit of \(\varepsilon\to0\).
Therefore, \(\langle \sigma_1\rangle\) is dominated by the hidden entropy production [Eq.~(\ref{e:dominant})] between \ref{m:1} and \ref{e:ud}.

The effect of these hidden entropy productions on the thermodynamic efficiency is numerically investigated in the next section.

\subsection{Hidden Entropy Production in Coarse-graining to \ref{e:another}} \label{ss:4-5}
We discuss the entropy production rate in the route of coarse-graining to obtain \ref{e:another}.
We first have
\begin{align}
\langle \sigma_1 \rangle = \langle \sigma_4 \rangle
 ,
 \label{e:1-4}
\end{align}
in the overdamped limit, since the elimination of the momentum variable from isothermal dynamics does not involve hidden entropy production.
In the tightly confined limit \(\varepsilon' := \tau_x/\tau\rightarrow 0\),
\begin{align}
 \langle \sigma_4 \rangle = \left\langle \frac{\phi'(\theta)^2}{\Gamma T_c + \gamma \phi'(\theta)^2 T_h}\left(\frac{1}{T_c} - \frac{1}{T_h}\right) (T_h - T_c)\left(\rd{U_I(x - \phi(\theta))}{x}\right)^2 \right\rangle
 ,
 \label{e:4-5}
\end{align}
which is derived in Appendix \ref{a:4-5}.
The leading order of \(\langle \sigma_4 \rangle\) is estimated as
\begin{align}
 \langle \sigma_4 \rangle &\sim \left\langle \frac{\Gamma T_c + \gamma \phi'(\theta)^2 T_h}{\mathcal{G}(\theta)^2} \left(\frac{\mathcal{G}(\theta)}{\Gamma T_c + \gamma \phi'(\theta)^2 T_h} \rd{U_I}{x}\right)^2 \right\rangle
 \nonumber \\
 &\lesssim \left\langle \frac{T_h}{\Gamma} \left(\frac{1}{T_c} \rd{U_I}{x}\right)^2 \right\rangle \sim \frac{T_h}{\Gamma} \frac{1}{L_x^2} = \tau_x^{-1}
 .
\end{align}
Since \(\langle \sigma_5 \rangle = O(\tau^{-1})\), the leading order of \(\langle \sigma_4\rangle\) does not include \(\langle \sigma_5\rangle\).
Thus, \(\langle \sigma_1\rangle\) in the limit of \ref{e:another} is dominated by the hidden entropy production between \ref{e:od-od} and 5. 

Let us compare \(\langle \sigma_1\rangle\) in the two coarse-graining limits.
As we saw in the previous subsection, we have
\begin{align}
  \langle \sigma_1 \rangle  \sim \left\langle \frac{\Gamma} {m}\frac{\mathcal{G}(\theta) - \Gamma}{ \mathcal{G}(\theta)} \left(\frac{1}{T_c} - \frac{1}{T_h}\right) (T_h - T_c) \right\rangle
 ,
 \label{e:dominant3}
\end{align}
as the leading order in the limit of obtaining \ref{e:od}. 
Choosing \(U_I(x - \phi(\theta)) = \lambda(x - \phi(\theta))^2/2\), the coarse-graining toward \ref{e:another} leads to
\begin{align}
 \langle \sigma_1 \rangle \sim \left\langle \frac{\lambda}{\gamma} \frac{\mathcal{G}(\theta) - \Gamma}{\mathcal{G}(\theta)} \left(\frac{1}{T_c} - \frac{1}{T_h}\right) (T_h - T_c)  \right\rangle
 ,
 \label{e:dominant5}
\end{align}
since \(x - \phi(\theta)\) follows the canonical distribution characterized by \(T_s(\theta) = [\Gamma T_c + \gamma\phi'(\theta)^2 T_h] / [\Gamma + \gamma\phi'(\theta)^2]\) at the leading order of \(\varepsilon'\) (see Appendix \ref{ss:SPT-another}).
The only difference between Eq.~(\ref{e:dominant3}) and Eq.~(\ref{e:dominant5}) is which time scale is rate-limiting
(\(\tau_\Pi^{-1} = \Gamma/m\) or \(\tau_x^{-1} = \lambda / \gamma\) ).

\subsection{Relation with Feynman's Argument} \label{ss:feynman}
We set \(T_h > T_c\) and \(f < 0\).
In Feynman's argument, the forward (backward) stepping rotation, which produces positive (negative) work, 
is initiated by the absorption of heat from the hotter (colder) bath, 
and the excess energy is released to the colder (hotter) bath as the dissipated heat. 
According to these phenomenological assumptions, 
he estimated the heat absorbed from the hotter and colder baths per forward step as \(\mathcal{Q}^h = \Delta U - f L\) and \(\mathcal{Q}^c = \Delta U\), respectively, where the work per step is \(-fL\) and the energy required to lift the pawl is \(\Delta U\).
Then, the rate of forward and backward steps were considered as
\begin{align}
 R^F_f &= \tau_s^{-1} \exp (-\mathcal{Q}^h/T_h) \\
 R^B_f &= \tau_s^{-1} \exp (-\mathcal{Q}^c/T_c)
 , \label{e:R_FB}
\end{align}
where \(\tau_s\) is the characteristic time scale of the steps.
Taking into account the backward step,
the thermodynamic efficiency  is written as
\begin{align}
 \eta_f = \frac{-fL}{\mathcal{Q}^h} = 1 - \frac{\mathcal{Q}^c}{\mathcal{Q}^h} \leq \eta_{C} 
 .
 \label{e:efficiency in Feynman's argument}
\end{align}
where \(\eta_{\rm C} := 1 - T_c / T_h\) is the Carnot efficiency. The equality is satisfied at the stalled condition \( R^F = R^B \).

A similar discrete-stepping model may be obtained from Models-3 and 5 in the limit of \(\Delta U / T_\mathrm{eff}(\theta) \to \infty\) \cite{millonas_self-consistent_1995}, where \(\Delta U = \max_\theta U_\mathrm{eff}(\theta) - \min_\theta U_\mathrm{eff}(\theta)\) is the effective energy barrier.
From Kramers theory, the forward and backward transition rates of \ref{e:od} in this limit are obtained as
\begin{align}
 R^{F}_{(3)} &= \tau_s^{-1} \exp\left[-\int_{\theta_\mathrm{min}}^{\theta_\mathrm{max}^+} \frac{1}{T_\mathrm{eff}(\theta)} \left(\rd{U_\mathrm{eff}(\theta)}{\theta} - f\right) \mathrm d \theta\right]
 \label{e:kramers_F}
 \\
 R^{B}_{(3)} &= \tau_s^{-1} \exp\left[-\int_{\theta_\mathrm{min}}^{\theta_\mathrm{max}^-} \frac{1}{T_\mathrm{eff}(\theta)} \left(\rd{U_\mathrm{eff}(\theta)}{\theta} - f\right) \mathrm d \theta\right]
 .
 \label{e:kramers_B}
\end{align}
Here, \(\theta_\mathrm{min}\) is a local minimum of \(U_\mathrm{eff}(\theta)\), and \(\theta_\mathrm{max}^\pm\) are the local maxima of \(U_\mathrm{eff}(\theta)\) that are closest to  \(\theta_\mathrm{min}\) ( \(\theta_\mathrm{max}^+ > \theta_\mathrm{min}\), \(\theta_\mathrm{max}^- < \theta_\mathrm{min}\), and \(\theta_\mathrm{max}^- + L = \theta_\mathrm{max}^+ \) ).
The derivation of Eqs.~(\ref{e:kramers_F},\ref{e:kramers_B}) is given in Appendix \ref{a:kramers}.
Assuming a sawtooth shape for the gear:
\begin{align}
 \phi'(\theta_\mathrm{max}^- < \theta \leq \theta_\mathrm{min}) =& \alpha_c >0 \\
 \phi'(\theta_\mathrm{min} < \theta \leq \theta_\mathrm{max}^+) =& \alpha_h <0,
\end{align}
the rates reduce to
\begin{align}
 R^F_{(3)} &= \tau_s^{-1} \exp (-\mathcal{Q}^h_3/T_{h(3)}) \\
 R^B_{(3)} &= \tau_s^{-1} \exp (-\mathcal{Q}^c_3/T_{c(3)})
 , \label{e:R_FB3}
\end{align}
where we can interpret that the heat
\begin{align}
 \mathcal{Q}^h_3 =& \Delta U - f (\theta_\mathrm{max}^+ - \theta_\mathrm{min}) \\
 \mathcal{Q}^c_3 =& \Delta U - f (\theta_\mathrm{max}^- - \theta_\mathrm{min})
 \label{e:Qs}
\end{align}
are exchanged from the baths with effective temperatures
\begin{align}
 T_{h(3)} =& \frac{\Gamma T_h + \gamma \alpha_h^2 T_c}{\Gamma + \gamma \alpha_h^2} \leq T_h \\
 T_{c(3)} =& \frac{\Gamma T_h + \gamma \alpha_c^2 T_c}{\Gamma + \gamma \alpha_c^2} \geq T_c,
 \label{e:Ts}
\end{align}
respectively. The efficiency then satisfies
\begin{align}
 \eta_3 := \frac{-fL}{\mathcal{Q}^h_3} \leq 1 - \frac{T_{c(3)}}{T_{h(3)}} \leq \eta_C. \label{e:stepeff}
\end{align}

The first equality in Eq.~(\ref{e:stepeff}) is met at the stalled state.
For the second equality, however, we must take the limit $\alpha_c \to \infty, \alpha_h \to 0$, which corresponds to an asymmetric sawtooth with $\theta_\mathrm{max}^- = \theta_\mathrm{min}$ and $\theta_\mathrm{max}^+ - \theta_\mathrm{min} \to \infty$.
We also note that even under this asymmetric limit, where Feynman's rates and efficiency are reproduced, the real thermodynamic efficiency is much lower (effectively zero) since there is a large hidden entropy production between Model-1 and 3.
The same situation holds for Model-5.


\subsection{Relation with Parrondo and Espa\~nol's Model}
Choosing \(U(\theta, x) = \lambda (x - \theta)^2 / 2\) and \(\Gamma = \gamma\), Model-0 is written as
\begin{align}
  \dot \theta &= \frac{\Pi}{m},&
  \dot \Pi &= - \frac{\Gamma}{m} \Pi + \lambda (x - \theta) + \sqrt{2\Gamma T_h} \xi,
  \nonumber \\
  \dot x &= \frac{p}{m_x},&
  \dot p &= - \frac{\Gamma}{m_x} p - \lambda (x - \theta) + \sqrt{2\Gamma T_c} \zeta.
  \label{e:linearlized}
\end{align}
Note that there is no net rotation in this model due to the symmetry of the potential.
Parrondo and Espa\~nol identified the continuous heat flow in a case of \(m = m_x\) as \cite{parrondo_criticism_1996}
\begin{align}
 J_ {PE} = \frac{\Gamma}{2m} \frac{\lambda m / \Gamma^2}{(1 + \lambda m / \Gamma^2)} (T_h - T_c)
 .
 \label{e:P_E}
\end{align}
By taking the tightly confined limit, we obtain the coarse-grained description (\ref{e:ud}) for this model as
\begin{align}
 \begin{split}
 \dot \theta =& \frac{\Pi}{m},
 \\
 \dot \Pi =& - \frac{2\Gamma}{m} \Pi + \sqrt{2\Gamma(T_h + T_c)} \Xi
 .
 \label{e:PE dynamics}
 \end{split}
\end{align}
Correspondingly, the hidden entropy production rate becomes
\begin{align}
 \langle \sigma_1\rangle - \langle \sigma_2 \rangle 
 =& \frac{\Gamma}{2m} \left(\frac{1}{T_c} - \frac{1}{T_h}\right) (T_h - T_c)
 ,
 \label{e:1-2_PE}
\end{align}
which means that the heat flow converges to \(\Gamma (T_h - T_c) / (2m)\).
As pointed out by Parrondo and Espa\~nol \cite{parrondo_criticism_1996}, this non-vanishing heat flow prevents the FS ratchet to acquire Carnot efficiency.
In their argument, the deviation from Gibbs-Boltzmann distribution played a crucial role. This deviation, however, seems to disappear when taking the tightly confined limit [Eqs.~(\ref{e:PE dynamics},\ref{e:LG})]. In fact, our result suggests that even an infinitesimal deviation from the Gibbs-Boltzmann distribution can contribute to a finite entropy production. 

\section{Numerical Simulation} \label{s:numerical}
\begin{figure}
 \includegraphics[width=.75\hsize]{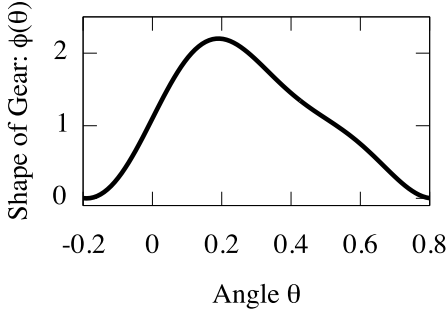}
 \caption{The functional forms of \(\phi(\theta) = \sin(2\pi\theta) + 0.25 \sin(4\pi\theta) + 1.1\) we used for the numerical simulation.}
 \label{f:phi}
\end{figure}

\begin{figure}
 \begin{overpic}[width=.75\hsize]{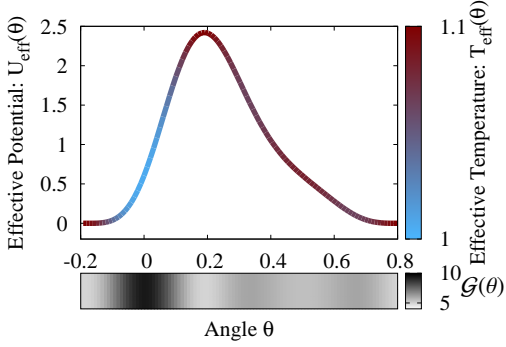}
  \put(93, 10){{\(\mathcal{G}(\theta)\)}}
 \end{overpic}
 \caption{The functional forms of \(U_\mathrm{eff}(\theta), \mathcal{G}(\theta)\) and \(T_\mathrm{eff}(\theta)\) [Eqs.~(\ref{e:U_eff}-\ref{e:T_eff})] which follow from the setup in the numerical simulation, \(U(\theta, x) = kx^2/2 + \lambda[x - \phi(\theta)]^2/2, \phi(\theta) = \sin(2\pi\theta) + 0.25 \sin(4\pi\theta) + 1.1\).
 The temperature difference between the positions of positive and negative potential slopes causes a net flow.}
 \label{f:functional forms}
\end{figure}

We performed numerical simulations of the FS ratchet 
with
\(U(\theta, x) = kx^2 / 2 + \lambda [x - \phi(\theta)]^2 / 2, \phi(\theta) = \sin(2\pi\theta) + 0.25 \sin(4\pi\theta) + 1.1\),
\(\Gamma = 5.0, \gamma = 0.05, k = 1.0, T_h = 1.1, T_c = 0.9\).
In this setting, \(L_x = \sqrt{T_c/\lambda}\) and the shortest time scale included in \(\tau\) is \(\Gamma / T_h = 4.5\).
To obtain the limit of \ref{e:od}, we introduced \(\lambda_0\) as \(\lambda = \lambda_0 / m\) and changed \(m\) and \(\lambda_0\) as parameters to control the separation of time scales.
The limit of \(\lambda_0\to\infty\) corresponds to the tight confinement of the pawl to the ratchet, \(\varepsilon \simeq \gamma / \lambda \tau_\Pi \to 0\). The limit of \(m\to0\) realizes the overdamped limit, \(\tau_\Pi / \tau = m T_h / \Gamma^2\to 0\), while keeping the ratio of \(\gamma/\lambda\) to \(\tau_\Pi\) proportional to \(\lambda_0\).
The functional form of \(\phi(\theta)\) is shown in \figurename~\ref{f:phi}.
By choosing \(\phi(\theta)\) to be asymmetric, \(U_\mathrm{eff}(\theta)\) and \(T_\mathrm{eff}(\theta)\) obtained from Eqs.~(\ref{e:U_eff},\ref{e:T_eff}) become out of phase as shown in \figurename~\ref{f:functional forms}.
The other details of the numerical simulations are illustrated in Appendix \ref{a:numerical}.

\subsection{Entropy Production Rates}
\begin{figure}
 \begin{overpic}[width=\hsize, trim=29 0 90 0, clip]{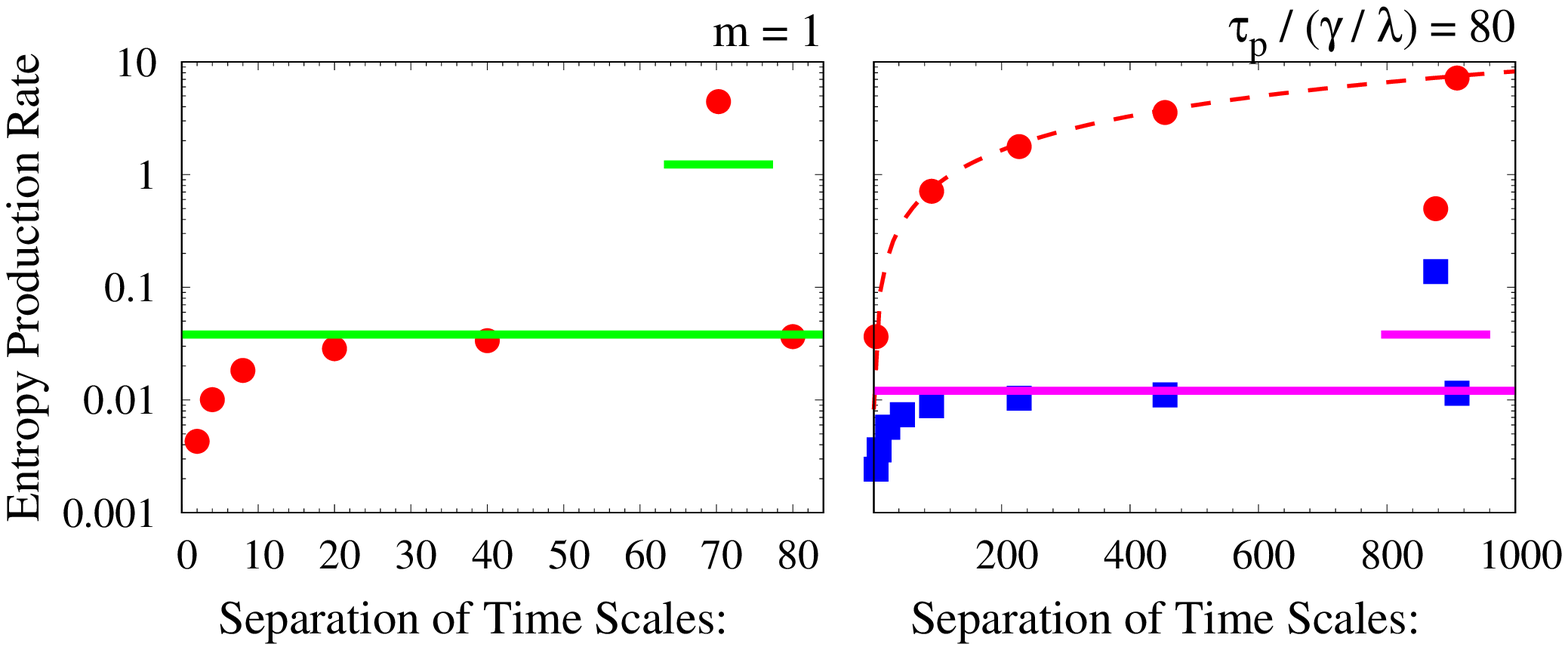}
  \put(24, 35.4){\(\langle \sigma_{1} \rangle\)}
  \put(15, 31.1){R.h.s of Eq.~(\ref{e:1-2})}
  \put(46.5, 2){\(\frac{\tau_\Pi}{\gamma / \lambda} \)}

  \put(72, 28.5){\(\langle \sigma_{1} \rangle\)}
  \put(72, 24.0){\(\langle \sigma_{2} \rangle\)}
  \put(62, 20.0){R.h.s. of Eq.~(\ref{e:2-3})}
  \put(90.5, 1.8){\footnotesize \(\epsilon^{-1} = \frac{\Gamma / T_h }{\tau_\Pi} \)}

  \put(83, 34){\scriptsize \(\sim \epsilon^{-1}\)}
 \end{overpic}
 \caption{Steady state entropy production rates. Filled circles and squares are the numerical results. Solid lines are obtained through Eqs.~(\ref{e:1-2},\ref{e:2-3}), respectively.
 The dashed line \(\sim \epsilon^{-1}\) shows the asymptotic dependence of \(\langle \sigma_1\rangle\) on \(\epsilon\).}
 \label{f:entropy production}
\end{figure}

Numerical results of the steady-state entropy production rates are plotted in \figurename~\ref{f:entropy production}.
In the tightly confined regime \(\varepsilon^{-1} \simeq \tau_\Pi / (\gamma/\lambda) \gtrsim 40\),
\(\langle \sigma_1\rangle\) converges to the right hand side of Eq.~(\ref{e:1-2}).

Next, fixing the parameter at the tightly confined regime, \(\tau_\Pi / (\gamma/\lambda) = 80\), we see the convergence of \(\langle \sigma_2\rangle\) to the right hand side of Eq.~(\ref{e:2-3}), in the overdamped limit \(\epsilon := \tau_\Pi / \tau \to 0\).
Furthermore, we see that \(\langle\sigma_{1}\rangle\) diverges with \(\epsilon^{-1}\), consistent with Eq.~(\ref{e:1-2-3}).

\subsection{Thermodynamic Efficiencies}
To demonstrate the impact of hidden entropy production, we calculated the thermodynamic efficiencies defined at Models-1, 2 and 3.
In \ref{m:1}, the thermodynamic efficiency is \(\eta_1 := 1 - \langle Q^c \rangle / \langle Q^h_1 \rangle\).
For Models-2 and 3, however, the definition of the efficiency is not trivial, since there is only a single heat bath with non-uniform continuous temperature.
We here adopt a generalized definition of efficiency \cite{bo_entropic_2013}.
The average heat flux conditional on the effective temperature $T$ is introduced by
\begin{align}
 \langle Q_{2,3}(T)\rangle := \langle Q_{2,3} \delta(T_\mathrm{eff}(\theta) - T)\rangle.
\end{align}
The averaged heat release and absorption rates are then defined as
\begin{align}
 \begin{split}
 \left\langle Q_{2,3}^ {rel}\right\rangle &= \int   dT \langle Q_{2,3}(T)\rangle \Theta(\langle Q_{2,3}(T)\rangle)
 ,
 \\
 \left\langle Q_{2,3}^ {abs}\right\rangle &= -\int   dT \langle Q_{2,3}(T)\rangle \Theta(-\langle Q_{2,3}(T)\rangle)
 ,\label{e:heatbo}
 \end{split}
\end{align}
where \(\Theta\) represents the Heaviside step function.
The generalized efficiencies \(\eta_2\) and \(\eta_3\) are
\begin{align}
 \eta_{2,3} = 1 - \frac{\left\langle Q_{2,3}^ {rel}\right\rangle}{\left\langle Q_{2,3}^ {abs}\right\rangle}
 .\label{e:effbo}
\end{align}
For Model-3, this definition agrees with the efficiency introduced as Eq.~(\ref{e:stepeff}) for the special case of sawtooth potential in the discrete stepping limit.

\begin{figure}
 \begin{overpic}[width=\hsize]{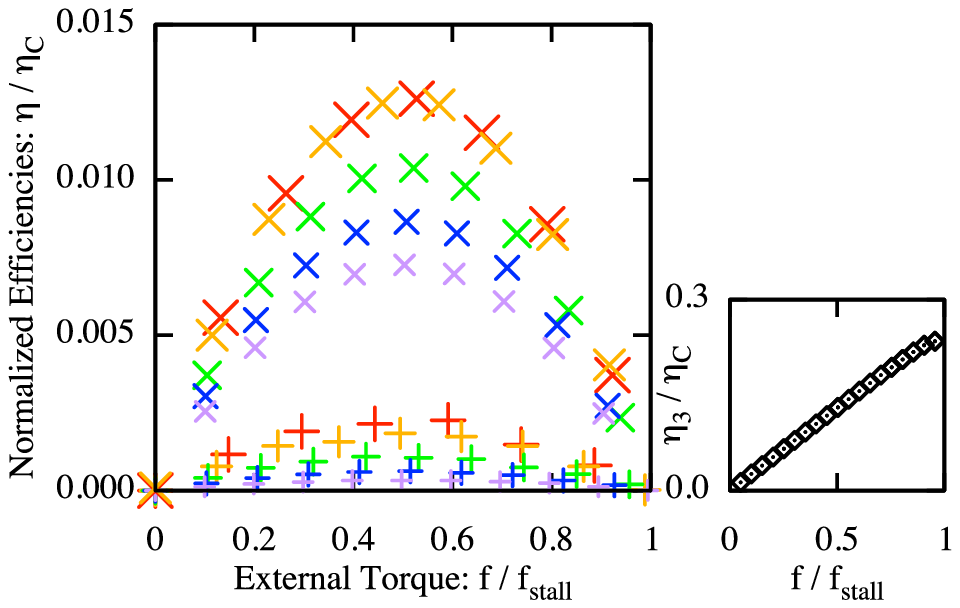}
  \put(84,40){\includegraphics[scale=0.75, trim=230 100 0 20, clip]{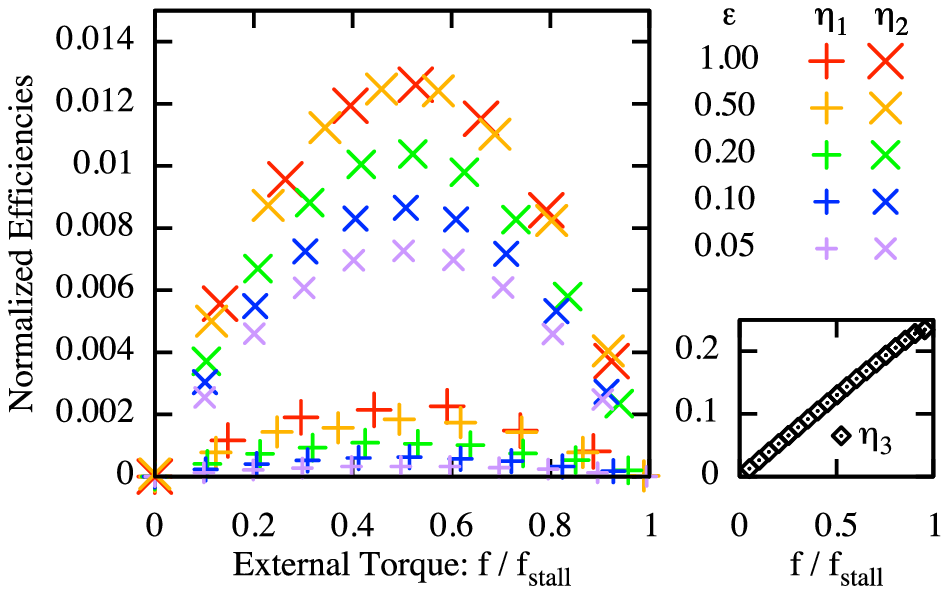}}
  \put(79,65.5){\(\epsilon\)}
  \put(77,61){\(0.20\)}
  \put(77,56.8){\(0.10\)}
  \put(77,52.6){\(0.04\)}
  \put(77,48.4){\(0.02\)}
  \put(77,44.2){\(0.01\)}
  \put(87.5,65.5){\(\eta_1\)}
  \put(92.5,65.5){\(\eta_2\)}
 \end{overpic}
 \caption{Thermodynamic efficiencies as functions of the external torque \(f\). 
 In the left figure, \(\eta_1\) and \(\eta_2\) are parametrized by the separation of time scales \(\epsilon\).
 Smaller symbols correspond to cases of smaller epsilon.
 In the right figure, \(\eta_3\) is plotted.
 The efficiencies are normalized by Carnot efficiency \(\eta_C = 1 - T_c / T_h\).}
 \label{f:efficiency}
\end{figure}

\begin{figure}[tb]
 \begin{overpic}[width=.75\hsize, trim= 0 0 634.5 430, clip]{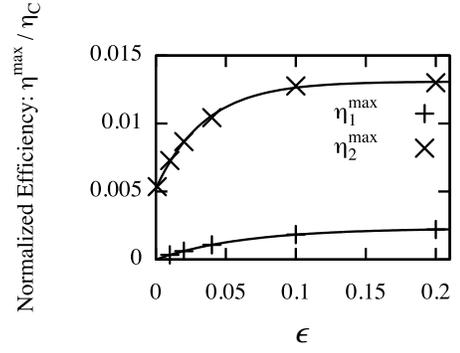}
  \put(60,6){\large \(\epsilon\)}
 \end{overpic}
 \caption{Maximal efficiency, \(\eta^ {max}_i = \max_f \eta_i\). Maximal value of \(\eta_1\) and \(\eta_2\) obtained from the fitting of \figurename~\ref{f:efficiency} by parabolic functions are plotted against \(\epsilon\). Solid lines \(A \exp(-\epsilon/\epsilon_0) + C\) are also plotted as guide for eyes, where \(A, \epsilon_0\) and \(C\) are fitting parameters. The efficiencies are normalized by Carnot efficiency of \ref{m:origin}, \(\eta_C = 1 - T_c/T_h\).}\label{f:maximal}
\end{figure}

The dependence of the efficiencies on \(f\) and \(\epsilon\) are shown in \figurename~\ref{f:efficiency}.
The behavior of \(\eta_1\) and \(\eta_2\) are different from that of \(\eta_3\) at \(\epsilon\to0\).
First, \(\eta_1\) approaches to \(0\) regardless of \(f\), since it follows from Eq.~(\ref{e:vel}) that the steady-state power, \(\dot W = - f \langle \dot \theta \rangle\), is of \(O(\tau^{-1})\), while the heat flows at the rate with \(O(\tau_\Pi^{-1})\).
Second, \(\eta_2\) does not converge to zero (\figurename~\ref{f:efficiency}). 
The finite efficiency means that \(\langle Q_2^{abs}\rangle\) is of the same order as \(\dot W = O(\tau^{-1})\).
Here, \(\langle Q_2^{abs}\rangle\) fails to capture the heat flow of \(O(\tau_\Pi^{-1})\) which contributes to the hidden entropy production \(\langle \sigma_1 \rangle - \langle \sigma_2 \rangle\).
Nevertheless, \(\eta_2\) vanishes at the stalled state, since \(\langle Q_2^{abs}\rangle\) is kept finite while \(\dot W\to0\).
Third, \(\eta_3\) monotonically increases and reaches the maximal value at the stalled state. This corresponds to the seemingly reversible situation, \(\langle \sigma_3 \rangle = 0\). 
However, \(\eta_3\) does not reach Carnot efficiency \(\eta_C\), because \(1.0 < T_\mathrm{eff}(\theta) < 1.1\) as is shown in \figurename~\ref{f:functional forms}, which implies \(\eta_3 < 1 - 1.0/1.1 < \eta_C\).
In \figurename~\ref{f:maximal}, we show the \(\epsilon\)-dependence of maximal efficiency, \(\eta^ {max}_i := \max_f \eta_i\), obtained from the fitting of torque-efficiency curves.
This result indicates that, in the limit of \(\epsilon\to0\), \(\eta_1\) vanishes irrespective of \(f\), and \(\eta_2\) converges to a certain torque-dependent curve.

These results highlight the effects of coarse-graining on the qualitative behaviors of thermodynamic efficiency; one may assume a significantly higher efficiency of an engine by neglecting the dissipative contributions of the fast variables.

\section{Recovery of Entropy Production based on Decomposition of Coarse-grained Langevin Dynamics} \label{s:decomposition}
The exact expression of the entropy production rate in the tightly confined limit [Eq.~(\ref{e:1-2})] inspires us to
consider if it is possible to reconstruct the thermodynamic irreversibility defined at the fine-grained description from the observation at the coarse-grained scale.
In a system where the time scales of variables are well-separated, it is
challenging to probe the dynamics of the fast variable, meaning that the hidden entropy production and the real thermodynamic efficiency are almost impossible to measure~\cite{wang_entropy_2016}.
Although there is no general workaround to the problem of inaccessible fast variables, 
we here describe a way to evaluate \(\langle\sigma_1\rangle\) from \ref{e:ud} of the FS ratchet. 

This is achieved by considering the dynamics as a mixture of two Langevin dynamics with different temperatures and frictions corresponding to the two heat baths (\figurename~\ref{f:schematic}d),
instead of a single set of effective temperature and friction [Eqs.~(\ref{e:G_eff},\ref{e:T_eff})].
The dynamics we consider consists of two Langevin equations,
\begin{align}
 \begin{split}
  \dot \theta =& \frac{\Pi}{m}
  ,
  \\
  \dot \Pi =& - \frac{\Gamma_b(\theta)}{m} \Pi - \rd{U_\mathrm{eff}(\theta)}{\theta} + f + \sqrt{2\Gamma_b(\theta) T_b} \xi
  ,
 \end{split}
 \label{e:constituent-1}
\end{align}
and stochastic switching of an auxiliary variable \(b = h, c\), which controls which heat bath [$(\Gamma_h,T_h)$ or $(\Gamma_c,T_c)$] the Langevin dynamics should be governed by.
The stochastic process of \(\theta, \Pi\) and \(b\) is described by the master equation:
\begin{align}
 &\rd{P(\theta, \Pi, b)}{t} =
 -\rd{}{\theta}\left(\frac{\Pi}{m} P(\theta, \Pi, b)\right) 
 \nonumber \\
 &- \rd{}{\Pi} \left[\left(-\frac{\Gamma_b(\theta)}{m} \Pi - \rd{U_\mathrm{eff}(\theta)}{\theta} + f - \Gamma_b(\theta) T_b \rd{}{\Pi}\right)P(\theta, \Pi, b)\right]
 \nonumber \\
 &- \Lambda P(\theta, \Pi, b) + \Lambda P(\theta, \Pi, b')
 ,
 \tag{Model-6}
 \label{e:decomp}
\end{align}
where, \(b' = c, h\) for \(b = h, c\), 
\(P(\theta, \Pi, b)\) is the joint probability density of \(\theta, \Pi\) and \(b\),
and \(\Lambda\) is the rate of stochastic switching of the heat baths.
According to the singular perturbation theory, in the limit where \(\Lambda^{-1}\) is separated from \(\tau_\Pi\) and \(\tau\), \ref{e:decomp} will give effective dynamics that follows
\begin{align}
 \begin{split}
 \dot \theta =& \frac{\Pi}{m}
 \\
 \dot \Pi =& - \frac{\Gamma_h(\theta) + \Gamma_c(\theta)}{2m} \Pi - \rd{U_\mathrm{eff}(\theta)}{\theta} + f 
 \\
 & + \sqrt{[\Gamma_h(\theta) T_h + \Gamma_c(\theta) T_c]} \Xi
 .
 \end{split}
 \label{e:rep}
\end{align}
Therefore, 
by setting \(\Gamma_h(\theta) = 2\Gamma\) and \(\Gamma_c(\theta) = 2\gamma\phi'(\theta)^2\), Eq.~(\ref{e:rep}) will reproduce the dynamics of \ref{e:ud}.

The entropy production of \ref{e:decomp} is
\begin{align}
 \sigma_6 = -\frac{1}{T_b} \left(\dot \Pi + \rd{U_\mathrm{eff}(\theta)}{\theta} - f\right) \circ \frac{\Pi}{m} + \frac{1}{t' - t}\ln \frac{\Lambda_{b_t\to b_{t'}}(\theta)}{\Lambda_{b_{t'}\to b_t}(\theta)}
 .
 \label{e:decompEP}
\end{align}
In the limit of fast switching, \(\langle \sigma_6\rangle\) converges to the weighted average of the contributions from the two dynamics,
\begin{align}
 \langle \sigma_6 \rangle \xrightarrow{\{\Lambda \tau_\Pi, \Lambda \tau\} \to \infty}& \frac{1}{2} \left\langle -\frac{1}{T_h} \left( \dot \Pi + \rd{U_\mathrm{eff}(\theta)}{\theta} - f\right) \circ \frac{\Pi}{m}\right\rangle_h
 \nonumber \\
 & + \frac{1}{2} \left\langle -\frac{1}{T_c} \left( \dot \Pi + \rd{U_\mathrm{eff}(\theta)}{\theta} - f\right) \circ \frac{\Pi}{m}\right\rangle_c
 ,
 \label{e:decomp of EP}
\end{align}
where the subscripts \(h, c\) indicate which Langevin dynamics are used to calculate the ensemble average.
By comparing Eq.~(\ref{e:1-2}) [or Eq.~(\ref{e:1-2-A})] with Eq.~(\ref{e:decomp of EP}),
we obtain
\begin{align}
 \lim_{\{\Lambda\tau_\Pi, \Lambda \tau\}\to\infty}\langle \sigma_{6}\rangle
 = 
 \lim_{\varepsilon\to0} \langle \sigma_{1}\rangle
 .
 \label{e:result of decomposition}
\end{align}
The details of the derivation of Eq.~(\ref{e:rep}) and Eq.~(\ref{e:result of decomposition}) are given in Appendix \ref{a:decomp}. 

Equation (\ref{e:result of decomposition}) is useful when we know the original temperatures of the heat baths but can only observe the dynamics at the coarse-grained scale.
Since \(\mathcal{G}(\theta)\) and \(T_{\rm eff}(\theta)\) can be measured at the coarse-grained scale, we may solve Eqs.~(\ref{e:G_eff}, \ref{e:T_eff}) using \(T_h\) and \(T_c\) to obtain \(\Gamma_b(\theta)\) in such a situation, which allows the evaluation of \(\langle \sigma_6\rangle\). 
We note that the decomposition of \ref{e:ud} into dynamics involving \(T_h\) and \(T_c\) is not unique if we are allowed to use general \(\theta\)-dependent switching rates.
Nevertheless, we may show that Eq.~(\ref{e:result of decomposition}) always holds as far as \ref{e:ud} is obtained in the fast switching limit (see Appendix \ref{a:decomp}).
The formulation of entropy production based on the decomposition of the stochastic transition  
has been previously discussed \cite{parrondo_criticism_1996,sekimoto_stochastic_2010,esposito_stochastic_2012}.
The approach here is a natural extension of these strategies to the case of a heat engine described by continuous variables.

\section{Discussion and Conclusion} \label{s:conclusion}
We derived the coarse-grained descriptions of the FS ratchet starting from \ref{m:origin} along the routes shown in \figurename~\ref{f:complete}.
We obtained the exact expressions for the entropy production in each model
and clarified the existence of the hidden entropy productions,
which correspond to the differences in the entropy production between the different descriptions.
The impact of the hidden entropy production on the thermodynamic efficiency was investigated numerically, to track how the efficiency of Models-2 and 3 significantly overestimate the true thermodynamic efficiency of \ref{m:origin}.
Additionally, we proposed a way to reconstruct the entropy production for \ref{m:origin} from the coarse-grained scale by introducing pseudo-dynamics described by \ref{e:decomp}.

In this paper, some of the coarse-grained descriptions obtained in this work have been studied previously, e.g. BL motor (Models-2 and 3), two-variable overdamped model (\ref{e:od-od}), single-variable model (Model-3), and the discrete stepping model (Sec. \ref{ss:feynman}). This means that the previous works \cite{feynman_feynman_2010,magnasco_feynmans_1998,sekimoto_kinetic_1997,hondou_unattainability_2000,asfaw_exploring_2007,derenyi_efficiency_1999,velasco_feynmans_2001,salas_unified_2003,tu_efficiency_2008,lin_performance_2009,chen_optimum_2011,luo_optimum_2013,apertet_revisiting_2014,shi-qi_coefficient_2014} based on phenomenological arguments were correct when taking appropriate timescale separation limits.
However, the thermodynamic efficiency of the FS ratchet based on these models have been controversial and often misleading.
For instance, the analysis regarding the FS ratchet as the BL motor suffers (Models-2 and 3) from the hidden entropy production.
This means that the thermodynamic efficiency of the BL motor is always an overestimation.
Similarly, the efficiencies of single-variable models such as Models-3 or 5 \cite{derenyi_efficiency_1999,velasco_feynmans_2001,salas_unified_2003,tu_efficiency_2008,lin_performance_2009,chen_optimum_2011,luo_optimum_2013,apertet_revisiting_2014,shi-qi_coefficient_2014} are also overestimations.
Although a part of studies \cite{derenyi_efficiency_1999,velasco_feynmans_2001,salas_unified_2003,tu_efficiency_2008,lin_performance_2009,chen_optimum_2011} take into consideration the dissipation corresponding the hidden entropy production, it has been overlooked that hidden dissipations can exist for every coarse-graining step.
A two-variable overdamped model used in \cite{sekimoto_kinetic_1997} corresponds \ref{e:od-od}.
Since the coarse-graining from \ref{m:origin} to \ref{e:od-od} does not accompany the hidden entropy production, the efficiency obtained in \ref{e:od-od} appropriately reflects the efficiency of the FS ratchet.
In \cite{sekimoto_kinetic_1997}, it is concluded that the efficiency is lower than Carnot efficiency.
In summary, previous works stating that the FS ratchet can attain Carnot efficiency have all used a coarse-grained version of the model and neglected the hidden dissipation. When appropriate models and dissipations are taken to account, Carnot efficiency cannot be obtained.

The problem of hidden entropy production is inevitable when analysing the thermodynamic aspect of nonequilibrium system, since any model is considered to be constructed phenomenologically.
In this sense, \ref{e:decomp} points at a promising solution for the hidden entropy production.
It enables us to evaluate the true entropy production without knowing the true fine-grained description (\ref{m:origin}).
Therefore, it is important to develop such a framework, which may extract thermodynamic properties from the coarse-grained descriptions.


\begin{acknowledgments}
We thank E. Muneyuki, S. Wang, Y. Murashita and M. Itami for fruitful comments.
This work is supported by the Grant-in-Aids for JSPS Fellow No.~24-3258 (YN), No.~24-8031 and No.~28-908 (KK), the Grant-in-Aid For Young Scientists (B) No.~17K14373 (YN), and JSPS KAKENHI Grants No.~15K05196 and No.~25103002 (NN).
\end{acknowledgments}

\appendix

\newcommand{\mm}{\varpi}

\section{Derivation of Coarse-grained Dynamics} \label{s:SPT}

\subsection{Derivation of \ref{e:ud}} \label{ss:SPT-1}
In this subsection, we describe the details of the derivation of \ref{e:ud}.
Our goal here is to obtain the time evolution equation of the joint probability density \(P(\theta, \Pi)\) from Eq.~(\ref{e:KFP}).
We evaluate 
the last term of the right hand side of Eq.~(\ref{e:formal})
 in the limit of \(\varepsilon := \tau_x / \tau_\Pi \to 0\).
The heart of the singular perturbation theory is to decompose the time-dependence of \(P(\theta, \Pi, x)\) into the explicit part and the implicit part through \(P(\theta, \Pi)\).
In the limit of \(\varepsilon\to0\), the explicit part decays quickly and the right hand side of Eq.~(\ref{e:formal}) essentially turns into a functional of \(P(\theta, \Pi)\).

A singular perturbation problem is mapped to an ordinary perturbation theory by introducing
\(M\) which describes \(P(\theta, \Pi, x)\)
and \(\Omega\) which describes the dynamics of \(P(\theta, \Pi)\).
For this purpose,
we first switch the variables from \((t,x)\) to the dimensionless time and distance
\begin{align}
 \mathcal{T}&:= \frac{t}{\tau_x}, & s &:= \frac{x - \phi(\theta)}{L_x},
 \label{e:ts}
\end{align}
and rewrite Eq.~(\ref{e:KFP}) as
\begin{align}
 &\tau_x^{-1}\rd{P(\theta, \Pi, s)}{\mathcal{T}} = (\mathcal{L}^ {\theta} + \mathcal{L}^ {\Pi}) P(\theta, \Pi, s)
 \nonumber \\
 &+ \tau_x^{-1/2} \left[\sqrt{\frac{\gamma}{T_c}} \phi'(\theta) \left(\rd{}{s} \frac{\Pi}{m}
 -\rd{}{\Pi} \rd{\tilde U_I(s)}{s}\right) P(\theta, \Pi, s) \right]
 \nonumber \\
 &- \tau_x^{-1}\rd{}{s} \left[-\frac{1}{T_c} \rd{\tilde U_I(s)}{s} P(\theta, \Pi, s)
 - \rd{P(\theta, \Pi, s)}{s}\right]
 \nonumber \\
 &- \tau_x^{-1/2}\rd{}{s}\left[
 - \sqrt{\frac{T_c}{\gamma}} \left(\frac{U_0'(\phi(\theta))}{T_c}
 + O\left(\frac{|U_0''| L_x}{|U_0'|}\right)\right) P(\theta, \Pi, s)\right]
 ,
 \label{e:scaled}
\end{align}
where
\begin{align}
 \mathcal{L}^ {\theta} :=& - \rd{}{\theta} \frac{\Pi}{m},
 &
 \mathcal{L}^ {\Pi} :=& -\rd{}{\Pi} \left[\left(-\frac{\Gamma}{m} \Pi + f\right) - \Gamma T_h \rd{}{\Pi}\right]
 .
\end{align}
In terms of \(\tau_x, \tau_\Pi\) and \(\tau\), 
we may estimate the order of the terms in the right hand side of Eq.~(\ref{e:scaled}) as \(O(\tau^{-1}) + O(\tau_\Pi^{-1}), O(\tau_x^{-1/2}\tau_\Pi^{-1/2}), O(\tau_x^{-1})\) and \(O(\tau_x^{-1/2}\tau^{-1/2})\), respectively.
In addition, \(O(|U''_0| L_x / |U'_0|) = O[(\tau_x / \tau)^{1/2}]\).
Based on this order estimation, we may consider \(\varepsilon\) as the small parameter which controls the perturbative analysis.

The explicit and implicit dependence of \(P(\theta, \Pi, s)\) on \(\mathcal{T}\) is implemented by describing \(P(\theta, \Pi, s)\) as output of a \(\mathcal{T}\)-dependent operator, \(M\), that acts on \(P(\theta, \Pi)\): 
\begin{align}
 P(\theta, \Pi, s) = M[P(\theta', \Pi'); \mathcal{T}](\theta, \Pi, s),
 \label{e:M}
\end{align}
where \(\theta'\) and \(\Pi'\) are dummy variables placed only to indicate that \(M\) depends on the joint probability density of \(\theta\) and \(\Pi\).
Furthermore, we represent the time-evolution of \(P(\theta, \Pi)\) by a \(\mathcal{T}\)-dependent operator \(\Omega\) that acts on \(P(\theta, \Pi)\):
\begin{align}
 &\rd{P(\theta, \Pi)}{\mathcal{T}} = \Omega[P(\theta', \Pi'); \mathcal{T}](\theta, \Pi) 
 := \tau_x (\mathcal{L}^ {\theta} + \mathcal{L}^ {\Pi}) P(\theta, \Pi)
 \nonumber \\
 &-\rd{}{\Pi} \left[\tau_x^{1/2}\sqrt{\frac{\gamma}{T_c}} \phi'(\theta)\int   ds \rd{\tilde U_I(s)}{s} M[P(\theta', \Pi'); \mathcal{T}](\theta, \Pi, s)\right]
 ,
 \label{e:Omega}
\end{align}
which is obtained by integrating Eq.~(\ref{e:scaled}) with respect to \(s\).
Since \(M\) depends on \(\mathcal{T}\) explicitly and implicitly [through \(P(\theta, \Pi)\)], the substitution of \(M\) into the left hand side of Eq.~(\ref{e:scaled}) gives
\begin{align}
 &[\mbox{l.h.s. of Eq.~(\ref{e:scaled})}] = 
 \rd{M[P(\theta', \Pi'); \mathcal{T}](\theta, \Pi, s)}{\mathcal{T}}
 \nonumber \\
 &\hspace{10ex}
 + \int   d\theta''   d\Pi'' \rd{P(\theta'', \Pi'')}{\mathcal{T}} \frac{\delta M[P(\theta'', \Pi''); \mathcal{T}](\theta, \Pi, s)}{\delta P(\theta'', \Pi'')} 
 \nonumber \\
 &= \rd{M[P(\theta', \Pi'); \mathcal{T}](\theta, \Pi, s)}{\mathcal{T}}
 \nonumber \\
 &+ \int   d\theta''   d\Pi'' \Omega[P(\theta', \Pi'); \mathcal{T}](\theta'', \Pi'') \frac{\delta M[P(\theta'', \Pi''); \mathcal{T}](\theta, \Pi, s)}{\delta P(\theta'', \Pi'')} ,
\end{align}
according to the chain rule.
Applying Eq.~(\ref{e:M}) also in the right hand side of Eq.~(\ref{e:scaled}), we obtain
\begin{align}
 &\rd{M[P(\theta', \Pi'); \mathcal{T}](\theta, \Pi, s)}{\mathcal{T}}
 \nonumber \\
 &+ \int   d\theta''   d\Pi'' \Omega[P(\theta', \Pi'); \mathcal{T}](\theta'', \Pi'') \frac{\delta M[P(\theta'', \Pi''); \mathcal{T}](\theta, \Pi, s)}{\delta P(\theta'', \Pi'')} 
 \nonumber \\
 &= \tau_x (\mathcal{L}^ {\theta} + \mathcal{L}^ {\Pi}) M[P(\theta', \Pi'); \mathcal{T}](\theta, \Pi, s)
 \nonumber \\
 &+ \tau_x^{1/2} \sqrt{\frac{\gamma}{T_c}} \phi'(\theta) \left(\rd{}{s} \frac{\Pi}{m} 
 - \rd{}{\Pi} \rd{\tilde U_I(s)}{s}\right) M[P(\theta', \Pi'); \mathcal{T}](\theta, \Pi, s)
 \nonumber \\
 &- \rd{}{s} \left[-\frac{1}{T_c} \rd{\tilde U_I(s)}{s} - \rd{}{s}\right]M[P(\theta', \Pi'); \mathcal{T}](\theta, \Pi, s)
 \nonumber \\
 &- \rd{}{s} \left[-\tau_x^{1/2} \sqrt{\frac{T_c}{\gamma}}\left(\frac{U_0'(\phi(\theta))}{T_c} 
 + O\left(\varepsilon^{1/2}\right)\right) M[P(\theta', \Pi'); \mathcal{T}](\theta, \Pi, s)\right]
 .
 \label{e:non-singular}
\end{align}

The remaining task is to apply the standard procedure of perturbation theory.
We expand \(M\) and \(\Omega\) into series of \(\varepsilon^{1/2}\):
\begin{align}
 M[P(\theta',\Pi');\mathcal{T}](\theta, \Pi, s) &= \sum_{n=0} \varepsilon^{n/2} M^{(n)}[P(\theta',\Pi');\mathcal{T}](\theta, \Pi, s),
 \\
 \Omega[P(\theta',\Pi');\mathcal{T}](\theta, \Pi) &= \sum_{n=0} \varepsilon^{(n+1)/2} \Omega^{(n)}[P(\theta',\Pi');\mathcal{T}](\theta, \Pi).
\end{align}
Here, the difference in the lowest order for \(M\) and \(\Omega\) is due to Eqs.~(\ref{e:Omega}).
The leading order of Eq.~(\ref{e:non-singular}) gives
\begin{align}
 &\rd{M^{(0)}[P(\theta', \Pi'); \mathcal{T}](\theta, \Pi, s)}{\mathcal{T}}
 \nonumber \\
 &=- \rd{}{s} \left[-\frac{1}{T_c} \rd{\tilde U_I(s)}{s}
 - \rd{}{s}\right]M^{(0)}[P(\theta', \Pi'); \mathcal{T}](\theta, \Pi, s)
 ,
\end{align}
from which we obtain 
\begin{align}
 &{M^{(0)}[P(\theta', \Pi'); \mathcal{T}](\theta, \Pi, s)} = P(\theta, \Pi) \frac{\exp\left(-{\tilde U_I(s)}/{T_c}\right)}{Z} + ...
 ,
 \label{e:LG}
\end{align}
where \(Z = \int   ds \exp\left(-\tilde U_I(s)/T_c\right)\). The additional terms ... depend on \(\mathcal{T}\) explicitly, and can be neglected since they decay exponentially with the time scale of \(O(\tau_x)\).
Under this assumption of the time scale, \(\Omega^{(0)}[P(\theta', \Pi'); \mathcal{T}](\theta, \Pi)\) vanishes, since the last term in the right hand side of Eq.~(\ref{e:Omega}) is zero in the leading order.
The sub-leading order of Eq.~(\ref{e:non-singular}) is
\begin{align}
 &\rd{M^{(1)}[P(\theta', \Pi'); \mathcal{T}](\theta, \Pi, s)}{\mathcal{T}}
 \nonumber \\
 &= \left(\frac{\tau_x}{\varepsilon}\right)^{1/2} \sqrt{\frac{\gamma}{T_c}} \phi'(\theta) \left(\rd{}{s} \frac{\Pi}{m} - \rd{}{\Pi} \rd{\tilde U_I(s)}{s}\right) M^{(0)}[P(\theta', \Pi'); \mathcal{T}](\theta, \Pi, s)
 \nonumber \\
 &- \rd{}{s} \left[\left(-\frac{1}{T_c} \rd{\tilde U_I(s)}{s}
 - \rd{}{s}\right)M^{(1)}[P(\theta', \Pi'); \mathcal{T}](\theta, \Pi, s) \right.
 \nonumber \\
 &\phantom{-\frac{1}{\gamma}\rd{}{2} \Bigg(}
 \left.- \left(\frac{\tau_x}{\varepsilon}\right)^{1/2} \sqrt{\frac{T_c}{\gamma}} \frac{U_0'(\phi(\theta))}{T_c} M^{(0)}[P(\theta', \Pi'); \mathcal{T}](\theta, \Pi, s)\right]
 ,
\end{align}
which has a particular solution
\begin{align}
 &{M^{(1)}[P(\theta', \Pi'); \mathcal{T}](\theta, \Pi, s)}
 \nonumber \\
 &\propto
 s\frac{\exp\left(-{\tilde U_I(s)}/{T_c}\right)}{Z} 
 \left[-\gamma \phi'(\theta) \left(\frac{\Pi}{m} + T_c\rd{}{\Pi}\right) - {U_0'(\phi(\theta))} \right]P(\theta, \Pi)
 \nonumber \\
 &+ [\mbox{exponentially decaying terms}]
 .
 \label{e:constituent of Kramers}
\end{align}
By substituting Eq.~(\ref{e:constituent of Kramers}) into Eq.~(\ref{e:Omega}),
\begin{align}
 &\Omega^{(1)}[P(\theta', \Pi'); \mathcal{T}]
 \nonumber \\
 &= -\frac{\tau_x}{\varepsilon} \left\{\rd{}{\theta}\left[\frac{\Pi}{m} P(\theta, \Pi) \right]
 + \rd{}{\Pi} \left[\left(-\frac{\Gamma}{m} \Pi + f\right) - \Gamma T_h \rd{}{\Pi}\right] P(\theta, \Pi) \right\}
 \nonumber \\
 &- \rd{}{\Pi} \left[\left(\frac{\tau_x}{\varepsilon}\right)^{1/2} \sqrt{\frac{\gamma}{T_c}} \phi'(\theta)\int   ds \rd{\tilde U_I(s)}{s} M^{(1)}[P(\theta', \Pi'); \mathcal{T}](\theta, \Pi, s)\right]
 \nonumber \\
 =& - \frac{\tau_x}{\varepsilon} \Bigg\{\rd{}{\theta}\left[\frac{\Pi}{m} P(\theta, \Pi) \right]
 + \rd{}{\Pi} \left[\left(-\frac{\Gamma}{m} \Pi + f\right) - \Gamma T_h \rd{}{\Pi}\right] P(\theta, \Pi)
 \nonumber \\
 & + \rd{}{\Pi} \left[-\gamma\phi'(\theta)^2\left( \frac{\Pi}{m} + T_c \rd{}{\Pi}\right) - \phi'(\theta){U'_0(\phi(\theta))} \right] P(\theta, \Pi) \Bigg\}
 .
\end{align}
The Kramers equation [Eq.~(\ref{e:Kramers})] immediately follows from the relation, \(\partial P(\theta, \Pi) / \partial \mathcal{T} = \Omega[P(\theta', \Pi'); \mathcal{T}]\)
with Eqs.~(\ref{e:U_eff}, \ref{e:G_eff}, \ref{e:T_eff}).

\subsection{Quick Derivation of \ref{e:ud}} \label{ss:TCG}
In the procedure of temporal coarse-graining, we first 
formally solve the equation of motion of eliminated variable, \(x\), [Eq.~(\ref{em:x})] as
\begin{align}
 x_t =& \frac{1}{\gamma}\int_{-\infty}^t   dt' e^{-\frac{k + \lambda}{\gamma}(t - t')}
  \left[ \lambda \phi(\theta_{t'}) + \sqrt{2\gamma T_c} \tilde\zeta_{t'}\right]
 \nonumber \\
 =& \frac{\lambda}{k + \lambda} \phi(\theta) - \frac{\gamma \lambda}{(k + \lambda)^2} \phi'(\theta_{t}) \frac{\Pi_{t}}{m} \left[1 + o\left(\varepsilon\right)\right]
 \nonumber \\
  &+ \frac{ \sqrt{2\gamma T_c} }{\gamma}\int_{-\infty}^t   dt' e^{-\frac{k + \lambda}{\gamma}(t - t')} \tilde\zeta_{t'} 
  .
  \label{e:formal solution}
\end{align}
Here, we performed integration by part twice.
The substitution of Eq.~(\ref{e:formal solution}) into the equation of motion of \(\Pi\) gives Eq~(\ref{em:p formal}).
As done in \cite{sekimoto_temporal_1999}, the underdamped Langevin equation is obtained by integrating Eq.~(\ref{em:p formal}) over \(t \in [t_0, t_0+\Delta t]\), using the identity
\begin{align}
 \int_{t_0}^{{t_0}+\Delta t}   dt \int_{-\infty}^t   dt' =& \int_{-\infty}^{t_0}   dt' \int_{t_0}^{{t_0}+\Delta t}   dt + \int_{t_0}^{{t_0}+\Delta t}   dt' \int_{t'}^{{t_0}+\Delta t}   dt
 .
 \label{e:equality}
\end{align}
By neglecting the \(O\left( \sqrt{2\gamma T_c} \frac{\lambda}{k + \lambda} \int_{-\infty}^{t_0}   dt' e^{-\frac{k + \lambda}{\gamma} ({t_0} - t')} \tilde\zeta_{t'}\right)\) term,
we obtain 
\begin{align}
 \frac{\Pi_{{t_0}+\Delta t} - \Pi_{t_0}}{\Delta t}
 =& -\frac{\Gamma + \frac{\gamma \lambda^2}{(k + \lambda)^2} \phi'(\theta_t)^2}{m} \Pi + f - \frac{k\lambda}{k + \lambda} \phi'(\theta) \phi(\theta)
 \nonumber \\
 &+ \sqrt{2 \Gamma T_h} \xi_t + \sqrt{2\gamma \phi'(\theta_t)^2 T_c} \frac{\lambda}{k + \lambda}\frac{1}{\Delta t} \int_{t_0}^{{t_0}+\Delta t}   dt' \tilde\zeta_{t'}
 .
 \label{e:lambda-dependent}
\end{align}
Since \(\sqrt{m/k}\) should be included in the set of slow time scales, \(k/\lambda = O(\tau_x / \tau) = O(\varepsilon)\).
Therefore, Eq.~(\ref{e:lambda-dependent}) results in \ref{e:ud} in the limit of \(\varepsilon\to0\).

\subsection{Derivation of \ref{e:od}} \label{ss:SPT-OD}
The coarse-graining from \ref{e:ud} to \ref{e:od} can also be formulated through the framework of Appendix \ref{ss:SPT-1}.
By introducing the dimensionless time and momentum
\begin{align}
 \tilde{\mathcal{T}} &= \frac{t}{\tau_\Pi},& \mm &= \frac{\Pi}{\sqrt{mT_0}},
\end{align}
where \(T_0\) is the reference point of temperature,
the Kramers equation [Eq.~(\ref{e:Kramers})] corresponding to \ref{e:ud}
may be rewritten as
\begin{align}
 \frac{\Gamma}{m}\rd{P(\theta, \mm)}{\tilde{\mathcal{T}}} =& -\rd{}{\theta} \left(\sqrt{\frac{T_0}{m}} \mm P(\theta, \mm)\right)
 \nonumber \\
 &- \rd{}{\mm}\left[ \left(- \frac{1}{\sqrt{mT_0}}\rd{U_\mathrm{eff}(\theta)}{\theta} + \frac{1}{\sqrt{mT_0}}f\right) P(\theta, \mm)\right.
 \nonumber \\
 &\left.-\frac{\Gamma}{m} \frac{\mathcal{G}(\theta)}{\Gamma} \left(\mm P(\theta, \mm) - \frac{T_\mathrm{eff}(\theta)}{T_0}\rd{P(\theta, \mm)}{\mm}\right)\right].
 \label{e:scaled2}
\end{align}
The first, second and third lines are 
\(O(\tau^{-1}), O(\tau^{-1/2} \tau_\Pi^{-1/2})\) and \(O(\tau_\Pi^{-1})\), respectively.
Following the procedure in Appendix \ref{ss:SPT-1}, we define 
\begin{align}
 \tilde M[P(\theta'); \tilde{\mathcal{T}}](\theta, \mm) &:= P(\theta, \mm)
 \\
 \tilde \Omega[P(\theta'); \tilde{\mathcal{T}}](\theta) &:
 = - \tau_\Pi \int   d\mm\rd{}{\theta} \left(\sqrt{\frac{T_0}{m}}\mm \tilde M[P(\theta'); \tilde{\mathcal{T}}](\theta, \mm)\right)
 \label{e:Omega2}
 \\
 &= \int   d\mm \rd{P(\theta, \mm)}{\tilde{\mathcal{T}}} = \rd{P(\theta)}{\tilde{\mathcal{T}}} 
 ,
\end{align}
where \(P(\theta) = \int   d\mm P(\theta, \mm)\).
In the standard perturbation theory of Eq.~(\ref{e:scaled2}) expressed in terms of \(\tilde M\) and \(\tilde \Omega\) with a small parameter \(\epsilon := \tau_\Pi / \tau\),
the leading order gives, 
\begin{align}
 &\rd{\tilde M^{(0)}[P(\theta'); \tilde{\mathcal{T}}](\theta, \mm)}{\tilde{\mathcal{T}}}
 \nonumber \\
 &= -\rd{}{\mm}\left[ -\frac{\mathcal{G}(\theta)}{\Gamma}\mm  - \frac{\mathcal{G}(\theta)}{\Gamma} \frac{T_\mathrm{eff}(\theta)}{T_0}\rd{}{\mm}\right]{\tilde M^{(0)}[P(\theta'); \tilde{\mathcal{T}}](\theta, \mm)}
 ,
\end{align}
which has a solution
\begin{align}
 \tilde M^{(0)}[P(\theta'); \tilde{\mathcal{T}}](\theta, \mm)
 =& P(\theta) \frac{\exp\left(-\frac{T_0 \mm^2}{2T(\theta)}\right)}{\sqrt{2\pi T(\theta)}}
 \nonumber \\
 &+ [\mbox{exponentially decaying terms}]
 .
\end{align}
Since \(\tilde \Omega^{(0)}[P(\theta'); \tilde{\mathcal{T}}](\theta)\) vanishes again, we proceed to the sub-leading order of Eq.~(\ref{e:scaled2}), 
\begin{align}
 &\rd{\tilde M^{(1)}[P(\theta'); \tilde{\mathcal{T}}](\theta, \mm)}{\tilde{\mathcal{T}}} = 
 - \tau\rd{}{\theta} \left(\sqrt{\frac{T_0}{m}}{\mm} \tilde M^{(0)}[P(\theta'); \tilde{\mathcal{T}}](\theta, \mm)\right)
 \nonumber \\
 &- \tau\rd{}{\mm}\left[ \frac{1}{\sqrt{mT_0}}\left(-\rd{U_\mathrm{eff}(\theta)}{\theta} + f\right)\tilde M^{(0)}[P(\theta'); \tilde{\mathcal{T}}](\theta, \mm)\right]
 \nonumber \\
 &-\rd{}{\mm}\left[ -\frac{\mathcal{G}(\theta)}{\Gamma}\mm - \frac{\mathcal{G}(\theta)}{\Gamma}\frac{T_\mathrm{eff}(\theta)}{T_0} \rd{}{\mm}\right]{\tilde M^{(1)}[P(\theta'); \tilde{\mathcal{T}}](\theta, \mm)},
\end{align}
which has a particular solution
\begin{align}
 &\tilde M^{(1)}[P(\theta'); \tilde{\mathcal{T}}](\theta, \mm) = \left\{ -\rd{P(\theta)}{\theta} \right.
 \nonumber \\
 &\left.- \left[\left(\frac{T_0\mm^2}{6 T_\mathrm{eff}(\theta)} + \frac{1}{2}\right) \frac{T_\mathrm{eff}'(\theta)}{T_\mathrm{eff}(\theta)} + \frac{1}{T_\mathrm{eff}(\theta)} \left(\rd{U_\mathrm{eff}(\theta)}{\theta} - f\right)\right]P(\theta)\right\}
 \nonumber \\
 & \cdot \tau \sqrt{\frac{T_0}{m}} \frac{\Gamma}{\mathcal{G}(\theta)} \mm \frac{\exp\left(-{T_0\mm^2}/{2T(\theta)}\right)}{\sqrt{2\pi T(\theta)}}
 \nonumber \\
 &+ [\mbox{exponentially decaying terms}]
 .
 \label{e:sol}
\end{align}
By substituting Eq.~(\ref{e:sol}) into Eq.~(\ref{e:Omega2}),
\begin{align}
 &\tilde \Omega^{(1)}[P(\theta'); \tilde{\mathcal{T}}](\theta) := - \tau\int   d\mm\rd{}{\theta} \left(\sqrt{\frac{T_0}{m}}\mm \tilde M^{(1)}[P(\theta'); \tilde{\mathcal{T}}](\theta, \mm)\right)
 \nonumber \\
 =& -\tau\rd{}{\theta}\left\{\frac{1}{\mathcal{G}(\theta)}\left[\left(-\rd{U_\mathrm{eff}(\theta)}{\theta} + f \right) P(\theta) - \rd{}{\theta}[T_\textrm{eff}(\theta){P(\theta)}]\right]\right\}
 .
\end{align}
We finally reach
\begin{align}
 &\rd{P(\theta)}{t} 
 \label{e:Fokker-Planck}
 \\
 =&-\rd{}{\theta}\left\{\frac{1}{\mathcal{G}(\theta)}\left[\left(-\rd{U_\mathrm{eff}(\theta)}{\theta} + f\right) P(\theta) - \rd{}{\theta}[{T_\textrm{eff}(\theta)P(\theta)}]\right]\right\}
 \nonumber 
 .
\end{align}

In order to obtain the overdamped Langevin equation corresponding to Eq.~(\ref{e:Fokker-Planck}), we first recall that \cite{van_kampen_stochastic_1992}
\begin{align}
 \dot X = A(X) + C(X) \circ \tilde\Xi,
\end{align}
can be mapped to an additive Langevin equation
\begin{align}
 \dot {\bar X} = \bar A(\bar X) + \tilde\Xi,
 \label{e:linear}
\end{align}
where \(\tilde\Xi\) is a white Gaussian noise with zero mean and unit variance, and \(\bar X = \int^X   dX / C(X), \bar A(\bar X) = A(X) / C(X)\).
Since Eq.~(\ref{e:linear}) has a corrsponding Fokker-Planck equation
\begin{align}
 \rd{P(\bar X)}{t} = -\rd{}{\bar X} \left( \bar A(\bar X) P(\bar X) - \frac{1}{2} \rd{P(\bar X)}{\bar X}\right),
\end{align}
we obtain the Fokker-Planck equation for \(P(X)\) through variable transformation [note that \(P(\bar X) = P(X) C(X)\)]:
\begin{align}
 \rd{P(X)}{t} = -\rd{}{X} \left( A(X) P(X) - \frac{C(X)}{2} \rd{}{X}[C(X)P(X)]\right)
 .
 \label{e:st form}
\end{align}
Therefore, by rewriting Eq.~(\ref{e:Fokker-Planck}) in the form of Eq.~(\ref{e:st form}),
the Langevin equation corresponding to Eq.~(\ref{e:Fokker-Planck}) is obtained as
\begin{align}
 \dot \theta =& \frac{1}{\mathcal{G}(\theta)} \left(-\rd{U_\mathrm{eff}(\theta)}{\theta} + f\right) - \frac{1}{2 \mathcal{G}(\theta)^2} \rd{}{\theta} \left[{\mathcal{G}(\theta)}{T_\mathrm{eff}(\theta)}\right]
 \nonumber \\
 &+ \sqrt{2\frac{T_\mathrm{eff}(\theta)}{\mathcal{G}(\theta)}} \circ \tilde\Xi
 .
 \label{e:od-st}
\end{align}
By changing the product, we have
\begin{align}
 \dot \theta = \frac{1}{\mathcal{G}(\theta)}\left(-\rd{U_\mathrm{eff}(\theta)}{\theta} + f\right) + \sqrt{\frac{2}{\mathcal{G}(\theta)}} \odot \sqrt{{T_\mathrm{eff}(\theta)}} \cdot \tilde\Xi
 .
\end{align}
Multiplying \(\mathcal{G}(\theta)\) in the sense of anti-It\^o to both sides
\begin{align}
 {\mathcal{G}(\theta)} \odot \dot \theta = \left(-\rd{U_\mathrm{eff}(\theta)}{\theta} + f\right) + \sqrt{2{\mathcal{G}(\theta)}} \odot \sqrt{{T_\mathrm{eff}(\theta)}} \cdot \tilde\Xi
 .
\end{align}

\subsection{Derivation of \ref{e:another}} \label{ss:SPT-another}
We start by rescaling the variables in Eq.~(\ref{e:od-od FP}) by  Eq.~(\ref{e:ts}).
\begin{align}
 &\tau_x^{-1}\rd{P(\theta, s)}{\mathcal{T}} = -\rd{}{\theta}\left[\frac{f}{\Gamma} P(\theta, s) - \frac{T_h}{\Gamma}\rd{}{\theta}{P(\theta, s)}\right]
 \nonumber \\
 & + \tau_x^{-1/2} \sqrt{\frac{\gamma}{T_c}}\rd{}{s}\left[\phi'(\theta) \left(\frac{f}{\Gamma} P(\theta, s) - \frac{T_h}{\Gamma}\rd{}{\theta}{P(\theta, s)}\right) \right]
 \nonumber \\
 & -\tau_x^{-1/2} \sqrt{\frac{\gamma}{T_c}} \rd{}{\theta}\left[\frac{\phi'(\theta)}{\Gamma}  \rd{U_I(s)}{s} P(\theta, s) + \frac{\phi'(\theta)}{\Gamma} T_h \rd{}{s}{P(\theta, s)}\right]
 \nonumber \\
 & + \tau_x^{-1/2} \sqrt{\frac{\gamma}{T_c}}\rd{}{s}\left\{\frac{1}{\gamma} \left[\rd{U_0(\phi(\theta))}{\phi(\theta)} + O\left(\frac{\tau_x}{\tau}\right) \right]P(\theta, s)\right\}
 \nonumber \\
 & - \tau_x^{-1} \frac{1}{\Gamma T_c} \rd{}{s}\left[-\left( {\Gamma + \gamma \phi'(\theta)^2} \right) \rd{U_I(s)}{s} P(\theta, s) \right.
 \nonumber \\
 & \phantom{- \tau_x^{-1} \frac{1}{\Gamma T_c} \rd{}{s}\Bigg[} \left. - \left({\Gamma T_c + \gamma\phi'(\theta)^2 T_h}\right) \rd{}{s}{P(\theta, s)}\right]
 .
 \label{e:sorted}
\end{align}
The first line of Eq.~(\ref{e:sorted}) is \(O(\tau^{-1})\), the last two lines are \(O(\tau_x^{-1})\), and the remaining terms are \(O(\tau^{-1/2} \tau_x^{-1/2})\), respectively.
Again, following the procedure in Appendix \ref{ss:SPT-1},
we define
\begin{align}
 &\hat M[P(\theta'); \mathcal{T}](\theta, s) := P(\theta, s)
 \\
 &\hat \Omega[P(\theta'); \mathcal{T}](\theta) := - \tau_x\rd{}{\theta}\left[\frac{f}{\Gamma} P(\theta) - \frac{T_h}{\Gamma}\rd{}{\theta}{P(\theta)}\right]
 \nonumber \\
 &-\tau_x^{1/2} \sqrt{\frac{\gamma}{T_c}} \rd{}{\theta}\left[\int   ds\frac{\phi'(\theta)}{\Gamma} \rd{U_I(s)}{s} \hat M[P(\theta');\mathcal{T}](\theta, s) \right]
 .
 \label{e:Omega3}
\end{align}

Now, we apply the standard perturbation theory to Eq.~(\ref{e:sorted}) expressed in terms of \(\hat M\) and \(\hat \Omega\), with a small parameter \(\varepsilon' = \tau_x / \tau\).
The leading order of Eq.~(\ref{e:sorted}) gives
\begin{align}
 \rd{\hat M^{(0)}}{\mathcal{T}} = 
 & - \frac{1}{\Gamma T_c}\rd{}{s}\left[-\mathcal{G}(\theta) \rd{U_I(s)}{s} \hat M^{(0)} - \mathcal{G}(\theta) T_s(\theta) \rd{}{s}{\hat M^{(0)}} \right]
 ,
\end{align}
where
\begin{align}
 T_s(\theta) = \frac{T_h \gamma \phi'(\theta)^2 + T_c \Gamma}{\Gamma + \gamma \phi'(\theta)^2}
 .
\end{align}
We obtain
\begin{align}
 \hat M^{(0)} = P(\theta) \frac{\exp{\left(-\frac{U_I(s)}{T_s(\theta)}\right)}}{z(\theta)}
 + [\mbox{exponentially decaying terms}]
 ,
\end{align}
with
\begin{align}
 z(\theta) = \int   ds \, {\exp{\left(-\frac{U_I(s)}{T_s(\theta)}\right)}}
 .
\end{align}
In the time scale of \( \tau \), 
the \(O(\varepsilon'^{1/2})\) term of \(\hat \Omega\) vanishes.
The sub-leading order of Eq.~(\ref{e:sorted}) becomes
\begin{align}
 \rd{\hat M^{(1)}}{\mathcal{T}} =
  \tau^{-1/2} \sqrt{\frac{\gamma}{T_c}}\rd{}{s}\left[\phi'(\theta) \left(\frac{f}{\Gamma} \hat M^{(0)} - \frac{T_h}{\Gamma}\rd{}{\theta}{\hat M^{(0)}}\right)\right]
 \nonumber \\
 - \tau^{-1/2} \sqrt{\frac{\gamma}{T_c}} \rd{}{\theta}\left[\frac{\phi'(\theta)}{\Gamma} \rd{U_I(s)}{s} \hat M^{(0)} + \frac{\phi'(\theta)}{\Gamma} T_h \rd{}{s}{\hat M^{(0)}}\right]
 \nonumber \\
 + \tau^{-1/2} \sqrt{\frac{\gamma}{T_c}}\rd{}{s}\left[ \frac{1}{\gamma} \rd{U_0(\phi(\theta))}{\phi(\theta)}{\hat M^{(0)}}\right]
  \nonumber \\
  - \frac{1}{\Gamma T_c}\rd{}{s}\left[-\mathcal{G}(\theta) \rd{U_I(s)}{s} \hat M^{(1)} - \mathcal{G}(\theta) T_s(\theta) \rd{}{s}{\hat M^{(1)}} \right]
 ,
\end{align}
which gives
\begin{align}
 \hat M^{(1)} \propto &
 \frac{- s \hat M^{(0)}}{\mathcal{G}(\theta) T_s(\theta)}
 \left\{-\rd{}{\theta}\left[\frac{\phi'(\theta)}{\Gamma}(T_h - T_s(\theta)) \right] + \frac{f}{\Gamma} \phi'(\theta) \right.
 \nonumber \\
 & \left.+ \frac{1}{\gamma} \rd{U_0(\phi(\theta))}{\phi(\theta)}\right\}
 -\frac{\phi'(\theta)}{\Gamma}(2T_h - T_s(\theta)) \frac{1}{\mathcal{G}(\theta) T_s(\theta)}
 \nonumber \\
 &\left[{\rd{}{\theta}(s \hat M^{(0)})} - \frac{1}{T_s(\theta)^2}\rd{T_s(\theta)}{\theta} \mathcal{I}(s) \hat M^{(0)}\right]
 \nonumber \\
 &+ [\mbox{exponentially decaying terms}]
 ,
 \label{e:constituent of Fokker-Planck}
\end{align}
where
\begin{align}
 \mathcal{I}(s) = s U_I(s) - \int^s   ds' U_I(s')
 .
\end{align}
Substitution of Eq.~(\ref{e:constituent of Fokker-Planck}) into Eq.~(\ref{e:Omega3}) results in
\begin{align}
 &\hat \Omega^{(1)}[P(\theta); \mathcal{T}] := - \tau\rd{}{\theta}\left[\frac{f}{\Gamma} P(\theta) - \frac{T_h}{\Gamma}\rd{}{\theta}{P(\theta)}\right]
 \nonumber \\
 &- \tau^{1/2}\sqrt{\frac{\gamma}{T_c}}\rd{}{\theta}\left[\int   ds\frac{\phi'(\theta)}{\Gamma} \rd{U_I(s)}{s} \hat M^{(1)}[P(\theta');\mathcal{T}](\theta, s) \right]
 \nonumber \\
 =&
 - \tau\rd{}{\theta}\left[ \frac{f}{\mathcal{G}(\theta)} P(\theta) + \rd{}{\theta}\left( \frac{\gamma \phi'(\theta)^2}{2\mathcal{G}(\theta)^2}\right) (T_h - T_c) P(\theta)\right.
 \nonumber \\
 &\left.- \frac{1}{\mathcal{G}(\theta)} \rd{U_\mathrm{eff}(\theta)}{\theta} P(\theta)
 - \frac{T_\mathrm{eff}(\theta)}{\mathcal{G}(\theta)}\rd{}{\theta}{P(\theta)} \right]
 ,
\end{align}
which gives the Fokker-Planck equation,
\begin{align}
 \rd{P(\theta)}{t}
 =& - \rd{}{\theta}\left[ -\frac{1}{\mathcal{G}(\theta)} \rd{\ln \mathcal{G}(\theta)}{\theta} (T_h - T_c) P(\theta)\right.
 \nonumber \\
 &\left.+ \frac{1}{\mathcal{G}(\theta)} \left(f - \rd{U_\mathrm{eff}(\theta)}{\theta}\right)P(\theta) - \frac{1}{\mathcal{G}(\theta)}\rd{}{\theta}{\left({T_\mathrm{eff}(\theta)}P(\theta)\right)} \right]
 ,
\end{align}
corresponding to \ref{e:another}.

\section{Asymptotic Behavior of Entropy Production Rates} \label{s:hep}
\begin{widetext}
\subsection{Transition Probabilities in Eqs.~(\ref{e:W-first}-\ref{e:W-last})} \label{ss:Ws}
\begin{align}
 W_1(\theta_{t'}, \Pi_{t'}, x_{t'}| \theta_t, \Pi_t, x_t) =& \delta\left(\theta_{t'} - \theta_t - \frac{\Pi_t + \Pi_{t'}}{2m} (t' - t)\right)
 W_ {UD}\left(\Pi_{t'}\bigg| \Pi_t; -\rd{U(\bar\theta, \bar x)}{\bar \theta} + f, \Gamma, T_h\right)
 \nonumber \\
 &\frac{1}{\sqrt{4\pi (t' - t) T_c / \gamma}}
 \exp\left(-\frac{[\gamma(x_{t'} - x_t) + \partial{U(\bar \theta, \bar x)}/\partial{\bar x}(t' - t)]^2}{4\gamma T_c (t' - t)} + \rd[2]{U(\bar \theta, \bar x)}{\bar x} \frac{(t' - t)}{2\gamma}\right)
 ,
 \label{e:TP of Model-1}
 \\
 W_2(\theta_{t'}, \Pi_{t'}| \theta_t, \Pi_t) =& \delta\left(\theta_{t'} - \theta_t - \frac{\Pi_t + \Pi_{t'}}{2m} (t' - t)\right)
 W_ {UD}\left(\Pi_{t'}\bigg| \Pi_t; -\rd{U_\mathrm{eff}(\bar\theta)}{\bar\theta} + f, \mathcal{G}(\bar\theta), T_\mathrm{eff}(\bar\theta)\right)
 ,
 \\
 W_3(\theta_{t'}| \theta_t) =&
 \frac{1}{\sqrt{4\pi (t' - t) T_\mathrm{eff}(\bar\theta)/\mathcal{G}(\bar\theta)}}
 \exp\left(-\frac{\{\mathcal{G}(\bar\theta)(\theta_{t'} - \theta_t) + [\partial{(U_\mathrm{eff}(\bar\theta) + T_\mathrm{eff}(\bar\theta))}/\partial{\bar\theta} - f] (t' - t)\}^2}{4 \mathcal{G}(\bar\theta) T_\mathrm{eff}(\bar\theta) (t' - t)}\right.
 \nonumber \\
 &\left. + \frac{1}{2}\rd{}{\bar\theta}\left[\frac{1}{\mathcal{G}(\bar\theta)}\left(\rd{U_\mathrm{eff}(\bar\theta)}{\bar\theta} - f\right)
 + \frac{T_\mathrm{eff}(\bar\theta)}{\mathcal{G}(\bar\theta)^2} \rd{\mathcal{G}(\bar\theta)}{\bar\theta} + \frac{1}{2}\rd{}{\bar\theta}\frac{T_\mathrm{eff}(\bar\theta)}{\mathcal{G}(\bar\theta)}\right] (t' - t)\right)
 ,
 \\
 W_4(\theta_{t'}, x_{t'}| \theta_t, x_t) =& 
 \frac{1}{\sqrt{4\pi (t' - t) T_h / \Gamma}}
 \exp\left(-\frac{[\Gamma(\theta_{t'} - \theta_t) + (\partial{U(\bar \theta, \bar x)}/\partial{\bar \theta} - f) (t' - t)]^2}{4\Gamma T_h (t' - t)} + \rd[2]{U(\bar \theta, \bar x)}{\bar \theta} \frac{(t' - t)}{2\Gamma}\right)
 \nonumber \\
 &\frac{1}{\sqrt{4\pi (t' - t) T_c / \gamma}}
 \exp\left(-\frac{[\gamma(x_{t'} - x_t) + \partial{U(\bar \theta, \bar x)}/\partial{\bar x}(t' - t)]^2}{4\gamma T_c (t' - t)} + \rd[2]{U(\bar \theta, \bar x)}{\bar x} \frac{(t' - t)}{2\gamma}\right)
 ,
 \\
 W_5(\theta_{t'}| \theta_t) =&
 \frac{1}{\sqrt{4\pi (t' - t) T_\mathrm{eff}(\bar\theta)/\mathcal{G}(\bar\theta)}}
 \exp\left(-\frac{\{\mathcal{G}(\bar\theta)(\theta_{t'} - \theta_t) + [\partial{(U_\mathrm{eff}(\bar\theta) + T_\mathrm{eff}(\bar\theta)) + {\ln \mathcal{G}(\bar \theta)} (T_h - T_c)}/\partial{\bar\theta} - f] (t' - t)\}^2}{4 \mathcal{G}(\bar\theta) T_\mathrm{eff}(\bar\theta) (t' - t)} \right.
 \nonumber \\
 &\left. + \frac{1}{2}\rd{}{\bar\theta}\left[\frac{1}{\mathcal{G}(\bar\theta)}\left(\rd{U_\mathrm{eff}(\bar\theta)}{\bar\theta} + \rd{\ln\mathcal{G}(\bar\theta)}{\bar\theta} (T_h - T_c) - f\right)
 + \frac{T_\mathrm{eff}(\bar\theta)}{\mathcal{G}(\bar\theta)^2} \rd{\mathcal{G}(\bar\theta)}{\bar\theta} + \frac{1}{2}\rd{}{\bar\theta}\frac{T_\mathrm{eff}(\bar\theta)}{\mathcal{G}(\bar\theta)}\right] (t' - t)\right)
 ,
\end{align}
where 
\(\bar \theta = (\theta_t + \theta_{t'}) / 2, \bar x = (x_t + x_{t'}) / 2\), and 
\begin{align}
 W_ {UD}(\Pi_{t'}|\Pi_t; F, g, T) = \frac{1}{\sqrt{4\pi (t' - t) T / g }} \exp\left(-\frac{\left[\Pi_{t'} - \Pi_t + \left(\frac{g}{m}\frac{\Pi_t + \Pi_{t'}}{2} - F\right) (t' - t)\right]^2}{4 g T (t' - t)}\right),
\end{align}
is the transition probability of momentum degree of freedom following the underdamped Langevin equation.
\end{widetext}
\subsection{Derivation of Eq.~(\ref{e:1-2})} \label{a:1-2}
Based on the results of Appendix \ref{s:SPT}, we evaluate the ensemble average of the entropy production rate \(\langle \sigma_1\rangle\) in the limit of \(\varepsilon\to0\). 
Since \(\langle Q_1^h\rangle\) may be rewritten as
\begin{align}
 \langle Q_1^h\rangle =& \left\langle -\left(\dot \Pi + \rd{U(\theta, x)}{\theta} - f\right) \circ \frac{\Pi}{m}\right\rangle = \left\langle \frac{\Gamma}{m} \left(\frac{\Pi^2}{m} - T_h\right)\right\rangle
 ,
\end{align}
we consider the ensemble average of \(Q_1^c\)
with respect to \(M[P(\theta',\Pi');\mathcal{T}](\theta, \Pi, s)\),
by replacing \(x\) in \(Q_1^c\) by \(\phi(\theta) + L_x s\):
\begin{align}
 Q_1^c =& - \rd{U(\theta, x)}{x} \circ \dot x
 \nonumber \\
 =&- \left(\frac{1}{L_x}\rd{U_I(s)}{s} + \rd{U_0(\phi(\theta))}{\phi(\theta)} + O(L_x)\right)\circ (\phi'(\theta) \dot \theta + L_x \dot s)
 \nonumber \\
 =& \left(- \frac{1}{L_x} \rd{U_I(s)}{s} - \rd{U_0(\phi(\theta))}{\phi(\theta)}\right) \phi'(\theta) \frac{\Pi}{m} - \rd{U_I(s)}{s} \circ \dot s + O(L_x)
 .
 \label{e:qc}
\end{align}
The ensemble average of the first term of Eq.~(\ref{e:qc}) is [up to \(O(L_x^0)\)], 
\begin{align}
 &\left\langle -\frac{\phi'(\theta)}{L_x} \rd{U_I(s)}{s} \frac{\Pi}{m}\right\rangle
 \nonumber \\
 &\simeq \int  d\theta  d\Pi  ds \left[-\frac{\phi'(\theta)}{L_x} \frac{\Pi}{m} \rd{U_I(s)}{s} M^{(0)}[P(\theta', \Pi'); \mathcal{T}](\theta, \Pi, s) \right.
 \nonumber \\
 &\left. - \phi'(\theta) \frac{\Pi}{m} \rd{U_I(s)}{s} \sqrt{\frac{\gamma}{T_c}\frac{\varepsilon}{\tau_x}}M^{(1)}[P(\theta', \Pi'); \mathcal{T}](\theta, \Pi, s) \right]
 \nonumber \\
 &= -\int   d\theta   d\Pi   ds \phi'(\theta) \frac{\Pi}{m} \frac{1}{T_c}\rd{U_I(s)}{s} s \frac{\exp(-U_I(s)/T_c)}{Z}
 \nonumber \\
 &\quad \left[-\gamma \phi'(\theta) \left(\frac{\Pi}{m} + T_c \rd{}{\Pi}\right) - U_0'(\phi(\theta))\right] P(\theta, \Pi)
 \nonumber \\
 &= \left\langle\frac{\gamma \phi'(\theta)^2}{m} \left(\frac{\Pi^2}{m} - T_c \right) + \rd{U_0(\phi(\theta))}{\theta}\frac{\Pi}{m}\right\rangle 
 .
\end{align}
Since \(U_I(s)\) and \(T_c\) are fixed, 
\(\left\langle- U_I'(s) \circ \dot s \right\rangle = \frac{ d}{ dt}\langle -U_I(s) \rangle = 0\).
Putting these altogether, we obtain
\begin{align}
 \left\langle \sigma_1 \right\rangle =& \left\langle \frac{1}{T_h} \frac{\Gamma}{m} \left(\frac{\Pi^2}{m} - T_h\right) \right\rangle + \left\langle \frac{1}{T_c} \frac{\gamma\phi'(\theta)^2}{m} \left(\frac{\Pi^2}{m} - T_c\right) \right\rangle
 .
 \label{e:1-2-A}
\end{align}
By comparing this with
\begin{align}
 \langle \sigma_2 \rangle = \left\langle \frac{1}{T_\mathrm{eff}(\theta)} \frac{\mathcal{G}(\theta)}{m} \left(\frac{\Pi^2}{m} - T_\mathrm{eff}(\theta)\right)\right\rangle
 ,
\end{align}
we obtain Eq.~(\ref{e:1-2}).

\subsection{Derivation of Eq.~(\ref{e:2-3})} \label{a:2-3}
Next, we evaluate the ensemble average of the entropy production rate, \(\langle \sigma_2\rangle\), in the limit of \(\epsilon\to0\).
The entropy production rate may be rewritten as
\begin{align}
 \sigma_2 =& \frac{1}{T_\mathrm{eff}(\theta)} \left(\dot \Pi + \rd{U_0(\phi(\theta))}{\theta} - f\right)\circ \frac{\Pi}{m}
 \nonumber \\
 =& - \frac{1}{T_\mathrm{eff}(\theta)} \left[\frac{  d}{  dt} \frac{\Pi^2}{2m} + \left(\rd{U_0(\phi(\theta))}{\theta} - f\right)\frac{\Pi}{m}\right]
 \nonumber \\
 =& -\frac{  d}{  dt}  \left(\frac{1}{T_\mathrm{eff}(\theta)} \frac{\Pi^2}{2m}\right) - \frac{\Pi^2}{2m} \frac{1}{T_\mathrm{eff}(\theta)^2} \frac{\Pi}{m} \rd{}{\theta}T_\mathrm{eff}(\theta)
 \nonumber \\
 &- \frac{1}{T_\mathrm{eff}(\theta)}\left(\rd{U_0(\phi(\theta))}{\theta} - f\right)\frac{\Pi}{m}
 ,
 \label{e:cbea}
\end{align}
where we use that \(T_\mathrm{eff}(\theta)\) does not depend on time explicitly.
Since it immediately follows from the oddness of Eq.~(\ref{e:cbea}) as the function of \(\Pi\) that the ensemble average with respect to \(\tilde M^{(0)}[P(\theta'); \tilde{\mathcal{T}}](\theta, \mm)\) vanishes, we obtain a finite contribution from that with respect to \(\tilde M^{(1)}[P(\theta'); \tilde{\mathcal{T}}](\theta, \mm)\) as
\begin{widetext}
\begin{align}
 \langle\sigma_2\rangle =&\int   d\theta \frac{1}{\mathcal{G}(\theta)T_\mathrm{eff}(\theta)}\left\{\left[ - T_\mathrm{eff}(\theta) \rd{P(\theta)}{\theta} + \left(- \rd{U_0(\phi(\theta))}{\theta} - {T_\mathrm{eff}'(\theta)} + f\right) P(\theta) \right] 
 \left(-\rd{U_0(\phi(\theta))}{\theta} - \frac{3}{2} T_\mathrm{eff}'(\theta) + f\right)
  + \frac{1}{2} {T_\mathrm{eff}'(\theta)}^2 P(\theta) \right\}
 \nonumber \\
 =& \left\langle \frac{1}{T_\mathrm{eff}(\theta)}\left( - \rd{U_0(\phi(\theta))}{\theta} - \frac{3}{2} T_\mathrm{eff}'(\theta) + f\right) \circ \dot \theta + \frac{T_\mathrm{eff}(\theta)}{2\mathcal{G}(\theta)} \left(\frac{T_\mathrm{eff}'(\theta)}{T_\mathrm{eff}(\theta)}\right)^2\right\rangle \nonumber \\
 =&
 \langle \sigma_3 \rangle + 
 \left\langle\frac{T_\mathrm{eff}(\theta)}{2\mathcal{G}(\theta)} \left(\frac{T_\mathrm{eff}'(\theta)}{T_\mathrm{eff}(\theta)}\right)^2\right\rangle
 .
\end{align}
\end{widetext}
The third line is obtained by using the overdamped Langevin equation of \ref{e:od}.

\subsection{Derivation of Eq.~(\ref{e:1-2}) based on Coarse-graining in \ref{ss:quick}}
Here, we present a different coarse-graining method based on temporal coarse-graining, which does not involve the ensemble average.
We first substitute Eq.~(\ref{e:formal solution}) into the expression of \(\sigma_1\),
\begin{align}
 \sigma_1 =& -\frac{1}{T_h} \left(\dot \Pi - \lambda \phi'(\theta)(x - \phi(\theta))\right) \circ \frac{\Pi}{m}
 \nonumber \\
 &- \frac{1}{T_c} (\lambda + k) \left(x - \frac{\lambda}{\lambda + k} \phi(\theta)\right) \circ \dot x
 \nonumber \\
 =& -\frac{1}{T_h} \left(\dot \Pi - k \frac{\lambda}{k+\lambda} \phi'(\theta) \phi(\theta) \right) \circ \frac{\Pi}{m}
 \nonumber \\
 & +\frac{1}{T_h} \lambda \phi'(\theta) \left( - \frac{\gamma \lambda}{(k + \lambda)^2} \phi'(\theta) \frac{\Pi}{m} + \frac{\sqrt{2\gamma T_c}}{k + \lambda} \tilde\zeta^c\right) \circ \frac{\Pi}{m}
 \nonumber \\
 & -\frac{1}{T_c} (\lambda + k) \left( - \frac{\gamma \lambda}{(k + \lambda)^2} \phi'(\theta) \frac{\Pi}{m} + \frac{\sqrt{2\gamma T_c}}{k+\lambda} \tilde\zeta^c\right)
 \nonumber \\
 & \circ \left[\frac{\lambda}{k + \lambda} \phi'(\theta) \frac{\Pi}{m} + \frac{\sqrt{2\gamma T_c}}{\gamma} \left(\tilde\zeta - \tilde\zeta^c\right)\right]
 ,
\end{align}
where \(\tilde\zeta^c = \int^t_{-\infty} e^{-\frac{k+\lambda}{\gamma} (t - t')} \tilde\zeta_{t'}\).
Integrating this over \(t \in [t_0, t_0+\Delta t]\) and neglecting the higher order terms in the limit of \(\varepsilon\to0\) as done in Appendix \ref{ss:TCG}, we obtain
\begin{align}
 &\int_{t_0}^{{t_0}+\Delta t}   dt\sigma_1 = \int_{t_0}^{{t_0}+\Delta t}   dt\left\{
 -\frac{1}{T_c} \phi'(\theta)\left( - \gamma \phi'(\theta) \frac{\Pi}{m} + \sqrt{2\gamma T_c} \Xi\right)
 \right.
 \nonumber \\
 & \left.
 -\frac{1}{T_h} \left[\dot \Pi - k \phi'(\theta) \phi(\theta) + \phi'(\theta) \left( \gamma \phi'(\theta) \frac{\Pi}{m} - {\sqrt{2\gamma T_c}} \Xi\right)\right]
 \right\}
 \circ \frac{\Pi}{m}
 .
 \label{e:temporal asymptotic}
\end{align}
Here, we use \(\int_{t_0}^{{t_0}+\Delta t}   dt\tilde\zeta^c (\tilde\zeta - \tilde\zeta^c)\to 0\) in the sense of the convergence in mean square.
The ensemble average of Eq.~(\ref{e:temporal asymptotic}) is the same as the right hand side of Eq.~(\ref{e:1-2}).

\subsection{Derivation of Eq.~(\ref{e:4-5})} \label{a:4-5}
We here show that \(\langle \sigma_4\rangle\) is dominated by the \(O(\tau_x^{-1})\) terms in the tightly confined limit.
We rewrite \(\langle\sigma_4\rangle\) in terms of \(s\),
\begin{align}
 \langle\sigma_4\rangle =& \frac{1}{T_h} \left\langle\left( f - \rd{U(\theta, x)}{\theta} \right) \circ \dot \theta\right\rangle - \frac{1}{T_c} \left\langle\rd{U(\theta, x)}{x} \circ \dot x\right\rangle
 \nonumber \\
 =& \frac{1}{\Gamma T_h} \left\langle -\left( f - \rd{U(\theta, x)}{\theta} \right) \rd{U(\theta, x)}{\theta} - T_h \rd[2]{U(\theta, x)}{\theta}\right\rangle
 \nonumber \\
 & - \frac{1}{\gamma T_c} \left\langle -\left(\rd{U(\theta, x)}{x}\right)^2 + T_c \rd[2]{U(\theta, x)}{x}\right\rangle
 \nonumber \\
 =& \left\langle \frac{\mathcal{G}(\theta)T_\mathrm{eff}(\theta)}{\Gamma \gamma T_h T_c}\left(\frac{1}{L} \rd{U_I(s)}{s}\right)^2 - \frac{\mathcal{G}(\theta)}{\Gamma\gamma} \frac{1}{L^2} \rd[2]{U_I(s)}{s}\right\rangle
 \nonumber \\
 =& \left\langle \frac{\mathcal{G}(\theta)}{\Gamma \gamma}\left(\frac{T_\mathrm{eff}(\theta)}{T_h T_c} - \frac{1}{T_s(\theta)}\right)\left(\frac{1}{L} \rd{U_I(s)}{s}\right)^2\right\rangle
 \nonumber \\
 & + \left\langle \frac{\mathcal{G}(\theta)}{\Gamma\gamma} \left[\frac{1}{T_s(\theta)}\left(\frac{1}{L} \rd{U_I(s)}{s}\right)^2 - \frac{1}{L^2} \rd[2]{U_I(s)}{s}\right]\right\rangle + O(L_x^{-1})
 .
\end{align}
The ensemble average of the second term in the last line [with respect to \(\hat M^{(0)} + \varepsilon' \hat M^{(1)}\)]  is smaller than \(O(L_x)\).
Therefore, we obtain Eq.~(\ref{e:4-5}) as the leading term.

\section{Derivation of Eqs.~(\ref{e:kramers_F},\ref{e:kramers_B})} \label{a:kramers}
We calculate the forward and backward transition rates of Models-3 the limit of \(\Delta U_\mathrm{eff} / T_\mathrm{eff}(\theta) \to\infty\).
In the case of \ref{e:od}, we may obtain an additive Langevin equation from Eq.~(\ref{e:od-st}) by tranforming the variable from \(\theta\) to \(q\) as
\begin{align}
 \dot q =& -\rd{\psi(q)}{q}
 + \frac{1}{2} \rd{}{q} \ln \left(\frac{T_\mathrm{eff}(q)}{\mathcal{G}(q)}\right) + \sqrt{2} \tilde \Xi
 ,
 \label{e:od-additive}
\end{align}
where 
\begin{align}
 q :=& \int^\theta \sqrt{\frac{\mathcal{G}(\theta')}{T_\mathrm{eff}(\theta')}} \mathrm d\theta',
\end{align}
and \(\psi\) [defined in Eq.~(\ref{e:psi})], \(T_\mathrm{eff}\) and \(\mathcal{G}\) are regarded as functions of \(q\).
By applying Kramers theory \cite{kramers_brownian_1940} to Eq.~(\ref{e:od-additive}), the forward and backward transition rates are given as
\begin{align}
 R^{F,B}_{(3)} 
 =& \frac{1}{2\pi}\sqrt{-\left.\rd[2]{\tilde\psi(q)}{q}\right|_{q=q_\mathrm{min}}\left.\rd[2]{\tilde\psi(q)}{q}\right|_{q=q_\mathrm{max}}}
 \nonumber \\
 &\exp\left[- \tilde\psi(q_\mathrm{max}) + \tilde\psi(q_\mathrm{min}) \right]
 \nonumber \\
 =& \frac{1}{2\pi}\sqrt{-\left.\rd[2]{\tilde\psi(q)}{q}\right|_{q=q_\mathrm{min}}\left.\rd[2]{\tilde\psi(q)}{q}\right|_{q=q_\mathrm{max}}}
 \nonumber \\
 &\sqrt{\frac{\mathcal{G}(q_\mathrm{min}) T_\mathrm{eff}(q_\mathrm{max})}{ T_\mathrm{eff}(q_\mathrm{min}) \mathcal{G}(q_\mathrm{max})}}
 \exp\left[- \psi(q_\mathrm{max}) + \psi(q_\mathrm{min}) \right]
 \nonumber \\
 &\propto 
 \exp\left[- \psi(q_\mathrm{max}) + \psi(q_\mathrm{min}) \right]
 ,
\end{align}
where
\(\tilde\psi(q) = \psi(q) + \frac{1}{2} [\ln \mathcal{G}(q) - \ln T_\mathrm{eff}(q)]\),
\(q_\mathrm{min}\) is a local minimum of \(\tilde\psi(q)\) and \(q_\mathrm{max}\) is the nearest local maximum of \(\tilde\psi(q)\) so that \(q_\mathrm{max} > q_\mathrm{min}\) for \(R^F_{(3)}\) and \(q_\mathrm{max} < q_\mathrm{min}\) for \(R^B_{(3)}\).
Since, in the limit of \(\Delta U_\mathrm{eff} / T_\mathrm{eff}(\theta) \to\infty\), the local maxima and minima of \(\tilde\psi(q)\) agree with those of \(U_\mathrm{eff}(\theta(q))\), \(\theta_\mathrm{max}\) and \(\theta_\mathrm{min}\),
\begin{align}
 R^{F,B}_{(3)} &\propto \exp\left[-\int_{\theta_\mathrm{min}}^{\theta_\mathrm{max}} \frac{1}{T_\mathrm{eff}(\theta)} \left(\rd{U_\mathrm{eff}(\theta)}{\theta} + \rd{T_\mathrm{eff}(\theta)}{\theta} - f\right) \mathrm d \theta\right]
 \nonumber \\
 &= \frac{T_\mathrm{eff}(\theta_\mathrm{min})}{T_\mathrm{eff}(\theta_\mathrm{max})} \exp\left[-\int_{\theta_\mathrm{min}}^{\theta_\mathrm{max}} \frac{1}{T_\mathrm{eff}(\theta)} \left(\rd{U_\mathrm{eff}(\theta)}{\theta} - f\right) \mathrm d \theta\right]
 .
\end{align}
By defining the common prefactor as \(\tau_s^{-1}\), we obtain
\begin{align}
 R^{F,B}_{(3)} &= \tau_s^{-1}  \exp\left[-\int_{\theta_\mathrm{min}}^{\theta_\mathrm{max}} \frac{1}{T_\mathrm{eff}(\theta)} \left(\rd{U_\mathrm{eff}(\theta)}{\theta} - f\right) \mathrm d \theta\right]
 .
\end{align}
We may estimate the transition rates of \ref{e:another} in the same mannar.

\section{Details of Numerical Simulation} \label{a:numerical}
The numerical simulations are mainly carried out based on the Langevin equation of Model-1. In the numerical integration of Langevin equation, we employ the velocity Verlet method for the underdamped part and the Euler method for the overdamped part. The time step is set to \(2 \times 10^{-3}\) and the total length of simulations is set to \(2^{12}\).
The ensemble averages of the entropy production are calculated from \(2^{12}\)-independent runs, and the average entropy production rates are obtained from linear fitting.

In the numerical investigation of efficiency (\figurename~\ref{f:efficiency}, \ref{f:maximal}), we use the numerical integration of the Kramers equation of Model-2 together with the Langevin equation of Model-1. 
The phase space with a cut-off of momentum at \(\Pi = \pm8\) is discretized into \(2^8 \times (2^7+1)\) elements along the position and momentum axes, respectively.
The derivatives with respect to \(\theta\) or \(\Pi\) are approximated by the central difference. 
The time step is set to \(0.056 \times 10^{-5}\) and the total length of simulations is set to \(2^{3}\).

\section{Definition of Heat and its Effect on Thermodynamic Efficiency} \label{a:heat}
In Sec.~\ref{s:hidden entropy production}, we introduced heat flux by respecting the energy balance, and used them to discuss the thermodynamic efficiencies of the FS ratchet at the coarse-grained scales.
We here note on how these results will be affected when adopting a different definition for heat flux $\tilde Q_{3,5}$ which satisfies
\begin{align}
 \sigma_{3}(\theta_{t'}|\theta_{t}) =& \frac{1}{T_\mathrm{eff}(\theta)} \circ \tilde Q_3
 ,
 \label{e:OD1 heat another}
 \\
 \sigma_{5}(\theta_{t'}|\theta_{t}) =& \frac{1}{T_\mathrm{eff}(\theta)} \circ \tilde Q_5
 .
 \label{e:OD2 heat another}
\end{align}
while keeping the definitions of $\sigma_{3,5}$.

The difference between \(Q_3\) [Eq.~(\ref{e:OD1 heat})] and \(\tilde Q_3\) is
\begin{align}
 Q_3 - \tilde Q_3 = \rd{T_\mathrm{eff}(\theta)}{\theta} \circ \dot \theta
 .
\end{align}
By multiplying \(\delta(T_\mathrm{eff}(\theta) - T)\) in the sense of Stratonovich and taking the ensemble average, we obtain
\begin{align}
 \langle Q_3(T) \rangle - \langle \tilde Q_3(T) \rangle &= \left\langle \rd{T_\mathrm{eff}(\theta)}{\theta} \delta(T_\mathrm{eff}(\theta) - T) \circ \dot \theta \right\rangle
 \nonumber \\
 &= \int \rd{T_\mathrm{eff}(\theta)}{\theta} \delta(T_\mathrm{eff}(\theta) - T) J   d\theta
 \nonumber \\
 &= \sum_j \frac{T'_\mathrm{eff}(\theta_j)}{\left|T'_\mathrm{eff}(\theta_j)\right|}  J = 0
 , \label{e:heatdiff}
\end{align}
where 
\(\langle \tilde Q_3(T) \rangle := \langle \delta(T_\mathrm{eff}(\theta) - T) \circ \tilde Q_3 \rangle\),
\(J\) is the probability current at the steady state, 
and \(\theta_j\) are the angles satisfying \(T_\mathrm{eff}(\theta_j) = T\).
(The stochastic product for \(\langle Q_3(T)\rangle\) was not specified in \cite{bo_entropic_2013}.)
Equation (\ref{e:heatdiff}) suggests that the average heat flux under the condition of \(T_\mathrm{eff}(\theta) = T\) is the same between the two definitions of heat flux, which means that the thermodynamic efficiency \(\eta_3\) [Eq.~(\ref{e:effbo})] is uneffected by the change from $Q_3$ to $\tilde Q_3$.
The generalization to multi-dimensional cases is straightforward.

By the same argument, \(\eta_5\) is independent on which heat flux ($Q_5$ or $\tilde Q_5$) is used.

\section{Derivation of Eqs.~(\ref{e:rep}, \ref{e:result of decomposition})} \label{a:decomp}
We derive Eqs.~(\ref{e:rep}, \ref{e:result of decomposition}) based on the singular perturbation theory starting from a generalized version of \ref{e:decomp}:
\begin{align}
 &\rd{P(\theta, \Pi, b)}{t} =
 -\rd{}{\theta}\left(\frac{\Pi}{m} P(\theta, \Pi, b)\right) 
 \nonumber \\
 &- \rd{}{\Pi} \left[\left(-\frac{\Gamma_b(\theta)}{m} \Pi - \rd{U_\mathrm{eff}(\theta)}{\theta} + f - \Gamma_b(\theta) T_b \rd{}{\Pi}\right)P(\theta, \Pi, b)\right]
 \nonumber \\
 &- \Lambda_{b\to b'}(\theta) P(\theta, \Pi, b) + \Lambda_{b'\to b}(\theta) P(\theta, \Pi, b')
 .
 \label{e:decompG}
\end{align}
The limit of fast switching is when \(\tau_\Lambda = \max_\theta [\Lambda_{h\to c}(\theta) + \Lambda_{c\to h}(\theta)]^{-1}\) is separated from \(\tau_\Pi\) and \(\tau\) while the ratio \(\tau_\Pi / \tau\) is kept fixed.
Under this condition, the first and second lines of the right hand side of Eq.~(\ref{e:decompG}) are \(O(\tau^{-1}) + O(\tau_\Pi^{-1})\), and the third line is \(O(\tau_\Lambda^{-1})\).
By introducing
\begin{align}
 &\check M[P(\theta', \Pi'); \mathscr{T}](\theta, \Pi, b) := P(\theta, \Pi, b)
 \nonumber \\
 &\check \Omega[P(\theta', \Pi'); \mathscr{T}](\theta, \Pi) := 
 -\rd{}{\theta}\left(\frac{\Pi}{m} P(\theta, \Pi)\right) 
 \nonumber \\
 &- \rd{}{\Pi} \left[\left(- \rd{U_\mathrm{eff}(\theta)}{\theta} + f\right)P(\theta, \Pi)\right]
 \nonumber \\
 &
 - \rd{}{\Pi} \left[\sum_b\left(-\frac{\Gamma_b(\theta)}{m} \Pi - \Gamma_b(\theta) T_b \rd{}{\Pi}\right)\check M[P(\theta', \Pi'); \mathscr{T}](\theta, \Pi, b)\right]
 ,
\end{align}
with \(\mathscr{T} := t / \tau_\Lambda\), and expanding \(\check M\) and \(\check \Omega\) into series of \(\delta = \tau_\Lambda / \tau_\Pi \sim \tau_\Lambda / \tau\), 
we obtain the leading order of Eq.~(\ref{e:decompG}) expressed in terms of \(\check M\) and \(\check \Omega\),
\begin{align}
 &\rd{\check M^{(0)}[P(\theta', \Pi'); \mathscr{T}](\theta, \Pi, b)}{\mathscr{T}}
 \nonumber \\
 &= - \Lambda_{b\to b'}(\theta) {\check M^{(0)}[P(\theta', \Pi'); \mathscr{T}](\theta, \Pi, b)}
 \nonumber \\
 &\phantom{=} + \Lambda_{b'\to b}(\theta) {\check M^{(0)}[P(\theta', \Pi'); \mathscr{T}](\theta, \Pi, b')}
 ,
\end{align}
and a solution
\begin{align}
 {\check M^{(0)}[P(\theta', \Pi'); \mathscr{T}](\theta, \Pi, b)} =& P(\theta, \Pi) \frac{\Lambda_{b'\to b}(\theta)}{\Lambda_{h\to c}(\theta) + \Lambda_{c\to h}(\theta)}
 \label{e:}
 \\
 & + [\mbox{exponentially decaying terms}]
 \nonumber ,
\end{align}
which gives 
\begin{align}
 &\check \Omega^{(0)}[P(\theta', \Pi'); \mathscr{T}](\theta, \Pi) := 
 -\rd{}{\theta}\left(\frac{\Pi}{m} P(\theta, \Pi)\right) 
 \nonumber \\
 &- \rd{}{\Pi} \left[\left(- \rd{U_\mathrm{eff}(\theta)}{\theta} + f\right)P(\theta, \Pi)\right]
 \nonumber \\
 &
 - \rd{}{\Pi} \left[\sum_b\left(-\frac{\Gamma_b(\theta)}{m} \Pi - \Gamma_b(\theta) T_b \rd{}{\Pi}\right)\frac{\Lambda_{b'\to b}(\theta)}{\Lambda_{h\to c}(\theta) + \Lambda_{c\to h}(\theta)} P(\theta, \Pi)\right]
 .
\end{align}
Therefore, Eq.~(\ref{e:decompG}) will be equivalent to \ref{e:ud} if \(\Gamma_b(\theta)\) and the transition rates satisfy
\begin{align}
 \sum_b \Gamma_b(\theta) \frac{\Lambda_{b'\to b}(\theta)}{\Lambda_{h\to c}(\theta) + \Lambda_{c\to h}(\theta)} &= \mathcal{G}(\theta)
 ,
 \label{e:Gammas}
 \\
 \sum_b \Gamma_b(\theta) T_b \frac{\Lambda_{b'\to b}(\theta)}{\Lambda_{h\to c}(\theta) + \Lambda_{c\to h}(\theta)} &= \mathcal{G}(\theta) T_\mathrm{eff}(\theta)
 .
 \label{e:GTs}
\end{align}
\ref{e:decomp} satisfies this condition since \(\Gamma_h(\theta) = 2\Gamma, \Gamma_c(\theta) = 2\gamma\phi'(\theta)^2\), and \(\Lambda_{h\to c}(\theta) = \Lambda_{c\to h}(\theta) = \Lambda\).

The entropy production rate defined through the transition probability of Eq.~(\ref{e:decompG}) is Eq.~(\ref{e:decompEP}).
The ensemble average of the last term of Eq.~(\ref{e:decompEP}) with respect to \({\check M^{(0)}}\) vanishes, and the average with respect to \(\check M^{(1)}\) does not contribute to \(\langle \sigma_6\rangle\) at the steady state. Therefore, in the limit of fast switching, we obtain
\begin{align}
 \langle\sigma_6\rangle =& \left\langle-\frac{1}{T_b} \left(\dot \Pi + \rd{U_\mathrm{eff}(\theta)}{\theta} - f\right) \circ \frac{\Pi}{m}\right\rangle
 = \left\langle\frac{\Gamma_b(\theta)}{m} \left(\frac{\Pi^2}{mT_b} - 1\right)\right\rangle
 \nonumber \\
 &= \left\langle\sum_b \frac{\Lambda_{b'\to b}(\theta)}{\Lambda_{h\to c}(\theta) + \Lambda_{c\to h}(\theta)}\frac{\Gamma_b(\theta)}{m} \left(\frac{\Pi^2}{mT_b} - 1\right)\right\rangle
 \nonumber \\
 &= \left\langle \frac{\Gamma}{m} \left(\frac{\Pi^2}{mT_h} - 1\right)
 + \frac{\gamma\phi'(\theta)^2}{m} \left(\frac{\Pi^2}{mT_c} - 1\right)\right\rangle \\
 &= \langle \sigma_1 \rangle
 .
\end{align}
To obtain the third line, we used the solutions of Eqs.~(\ref{e:Gammas},\ref{e:GTs}).


\begin{thebibliography}{57}%
\makeatletter
\providecommand \@ifxundefined [1]{%
 \@ifx{#1\undefined}
}%
\providecommand \@ifnum [1]{%
 \ifnum #1\expandafter \@firstoftwo
 \else \expandafter \@secondoftwo
 \fi
}%
\providecommand \@ifx [1]{%
 \ifx #1\expandafter \@firstoftwo
 \else \expandafter \@secondoftwo
 \fi
}%
\providecommand \natexlab [1]{#1}%
\providecommand \enquote  [1]{``#1''}%
\providecommand \bibnamefont  [1]{#1}%
\providecommand \bibfnamefont [1]{#1}%
\providecommand \citenamefont [1]{#1}%
\providecommand \href@noop [0]{\@secondoftwo}%
\providecommand \href [0]{\begingroup \@sanitize@url \@href}%
\providecommand \@href[1]{\@@startlink{#1}\@@href}%
\providecommand \@@href[1]{\endgroup#1\@@endlink}%
\providecommand \@sanitize@url [0]{\catcode `\\12\catcode `\$12\catcode
  `\&12\catcode `\#12\catcode `\^12\catcode `\_12\catcode `\%12\relax}%
\providecommand \@@startlink[1]{}%
\providecommand \@@endlink[0]{}%
\providecommand \url  [0]{\begingroup\@sanitize@url \@url }%
\providecommand \@url [1]{\endgroup\@href {#1}{\urlprefix }}%
\providecommand \urlprefix  [0]{URL }%
\providecommand \Eprint [0]{\href }%
\providecommand \doibase [0]{http://dx.doi.org/}%
\providecommand \selectlanguage [0]{\@gobble}%
\providecommand \bibinfo  [0]{\@secondoftwo}%
\providecommand \bibfield  [0]{\@secondoftwo}%
\providecommand \translation [1]{[#1]}%
\providecommand \BibitemOpen [0]{}%
\providecommand \bibitemStop [0]{}%
\providecommand \bibitemNoStop [0]{.\EOS\space}%
\providecommand \EOS [0]{\spacefactor3000\relax}%
\providecommand \BibitemShut  [1]{\csname bibitem#1\endcsname}%
\let\auto@bib@innerbib\@empty
\bibitem [{\citenamefont {Toyabe}\ \emph {et~al.}(2010)\citenamefont {Toyabe},
  \citenamefont {Sagawa}, \citenamefont {Ueda}, \citenamefont {Muneyuki},\ and\
  \citenamefont {Sano}}]{toyabe_experimental_2010}%
  \BibitemOpen
  \bibfield  {author} {\bibinfo {author} {\bibfnamefont {Shoichi}\ \bibnamefont
  {Toyabe}}, \bibinfo {author} {\bibfnamefont {Takahiro}\ \bibnamefont
  {Sagawa}}, \bibinfo {author} {\bibfnamefont {Masahito}\ \bibnamefont {Ueda}},
  \bibinfo {author} {\bibfnamefont {Eiro}\ \bibnamefont {Muneyuki}}, \ and\
  \bibinfo {author} {\bibfnamefont {Masaki}\ \bibnamefont {Sano}},\ }\bibfield
  {title} {\enquote {\bibinfo {title} {Experimental demonstration of
  information-to-energy conversion and validation of the generalized
  {Jarzynski} equality},}\ }\href {\doibase 10.1038/nphys1821} {\bibfield
  {journal} {\bibinfo  {journal} {Nature Phys.}\ }\textbf {\bibinfo {volume}
  {6}},\ \bibinfo {pages} {988--992} (\bibinfo {year} {2010})}\BibitemShut
  {NoStop}%
\bibitem [{\citenamefont {Evans}\ \emph {et~al.}(1993)\citenamefont {Evans},
  \citenamefont {Cohen},\ and\ \citenamefont
  {Morriss}}]{evans_probability_1993}%
  \BibitemOpen
  \bibfield  {author} {\bibinfo {author} {\bibfnamefont {Denis}\ \bibnamefont
  {Evans}}, \bibinfo {author} {\bibfnamefont {E.}~\bibnamefont {Cohen}}, \ and\
  \bibinfo {author} {\bibfnamefont {G.}~\bibnamefont {Morriss}},\ }\bibfield
  {title} {\enquote {\bibinfo {title} {Probability of second law violations in
  shearing steady states},}\ }\href {\doibase 10.1103/PhysRevLett.71.2401}
  {\bibfield  {journal} {\bibinfo  {journal} {Phys. Rev. Lett.}\
  }\textbf {\bibinfo {volume} {71}},\ \bibinfo {pages} {2401--2404} (\bibinfo
  {year} {1993})}\BibitemShut {NoStop}%
\bibitem [{\citenamefont {Jarzynski}(1997)}]{jarzynski_nonequilibrium_1997}%
  \BibitemOpen
  \bibfield  {author} {\bibinfo {author} {\bibfnamefont {C.}~\bibnamefont
  {Jarzynski}},\ }\bibfield  {title} {\enquote {\bibinfo {title}
  {Nonequilibrium {Equality} for {Free} {Energy} {Differences}},}\ }\href
  {\doibase 10.1103/PhysRevLett.78.2690} {\bibfield  {journal} {\bibinfo
  {journal} {Phys. Rev. Lett.}\ }\textbf {\bibinfo {volume} {78}},\
  \bibinfo {pages} {2690--2693} (\bibinfo {year} {1997})}\BibitemShut {NoStop}%
\bibitem [{\citenamefont {Wang}\ \emph {et~al.}(2002)\citenamefont {Wang},
  \citenamefont {Sevick}, \citenamefont {Mittag}, \citenamefont {Searles},\
  and\ \citenamefont {Evans}}]{wang_experimental_2002}%
  \BibitemOpen
  \bibfield  {author} {\bibinfo {author} {\bibfnamefont {G.~M.}\ \bibnamefont
  {Wang}}, \bibinfo {author} {\bibfnamefont {E.~M.}\ \bibnamefont {Sevick}},
  \bibinfo {author} {\bibfnamefont {Emil}\ \bibnamefont {Mittag}}, \bibinfo
  {author} {\bibfnamefont {Debra~J.}\ \bibnamefont {Searles}}, \ and\ \bibinfo
  {author} {\bibfnamefont {Denis~J.}\ \bibnamefont {Evans}},\ }\bibfield
  {title} {\enquote {\bibinfo {title} {Experimental {Demonstration} of
  {Violations} of the {Second} {Law} of {Thermodynamics} for {Small} {Systems}
  and {Short} {Time} {Scales}},}\ }\href {\doibase
  10.1103/PhysRevLett.89.050601} {\bibfield  {journal} {\bibinfo  {journal}
  {Phys. Rev. Lett.}\ }\textbf {\bibinfo {volume} {89}},\ \bibinfo
  {pages} {050601} (\bibinfo {year} {2002})}\BibitemShut {NoStop}%
\bibitem [{\citenamefont {Sagawa}\ and\ \citenamefont
  {Ueda}(2008)}]{sagawa_second_2008}%
  \BibitemOpen
  \bibfield  {author} {\bibinfo {author} {\bibfnamefont {Takahiro}\
  \bibnamefont {Sagawa}}\ and\ \bibinfo {author} {\bibfnamefont {Masahito}\
  \bibnamefont {Ueda}},\ }\bibfield  {title} {\enquote {\bibinfo {title}
  {Second {Law} of {Thermodynamics} with {Discrete} {Quantum} {Feedback}
  {Control}},}\ }\href {\doibase 10.1103/PhysRevLett.100.080403} {\bibfield
  {journal} {\bibinfo  {journal} {Phys. Rev. Lett.}\ }\textbf {\bibinfo
  {volume} {100}},\ \bibinfo {pages} {080403} (\bibinfo {year}
  {2008})}\BibitemShut {NoStop}%
\bibitem [{\citenamefont {Esposito}\ \emph {et~al.}(2009)\citenamefont
  {Esposito}, \citenamefont {Lindenberg},\ and\ \citenamefont {Van~den
  Broeck}}]{esposito_universality_2009}%
  \BibitemOpen
  \bibfield  {author} {\bibinfo {author} {\bibfnamefont {Massimiliano}\
  \bibnamefont {Esposito}}, \bibinfo {author} {\bibfnamefont {Katja}\
  \bibnamefont {Lindenberg}}, \ and\ \bibinfo {author} {\bibfnamefont
  {Christian}\ \bibnamefont {Van~den Broeck}},\ }\bibfield  {title} {\enquote
  {\bibinfo {title} {Universality of {Efficiency} at {Maximum} {Power}},}\
  }\href {\doibase 10.1103/PhysRevLett.102.130602} {\bibfield  {journal}
  {\bibinfo  {journal} {Phys. Rev. Lett.}\ }\textbf {\bibinfo {volume}
  {102}},\ \bibinfo {pages} {130602} (\bibinfo {year} {2009})}\BibitemShut
  {NoStop}%
\bibitem [{\citenamefont {Shiraishi}\ \emph {et~al.}(2016)\citenamefont
  {Shiraishi}, \citenamefont {Saito},\ and\ \citenamefont
  {Tasaki}}]{shiraishi_universal_2016}%
  \BibitemOpen
  \bibfield  {author} {\bibinfo {author} {\bibfnamefont {Naoto}\ \bibnamefont
  {Shiraishi}}, \bibinfo {author} {\bibfnamefont {Keiji}\ \bibnamefont
  {Saito}}, \ and\ \bibinfo {author} {\bibfnamefont {Hal}\ \bibnamefont
  {Tasaki}},\ }\bibfield  {title} {\enquote {\bibinfo {title} {Universal
  trade-off relation between power and efficiency for heat engines},}\ }\href
  {http://arxiv.org/abs/1605.00356} {\bibfield  {journal} {\bibinfo  {journal}
  {arXiv: {1605.00356}}\ } (\bibinfo {year} {2016})}
  \BibitemShut {NoStop}%
\bibitem [{\citenamefont {Hatano}\ and\ \citenamefont
  {Sasa}(2001)}]{hatano_steady-state_2001}%
  \BibitemOpen
  \bibfield  {author} {\bibinfo {author} {\bibfnamefont {Takahiro}\
  \bibnamefont {Hatano}}\ and\ \bibinfo {author} {\bibfnamefont {Shin-ichi}\
  \bibnamefont {Sasa}},\ }\bibfield  {title} {\enquote {\bibinfo {title}
  {Steady-{State} {Thermodynamics} of {Langevin} {Systems}},}\ }\href {\doibase
  10.1103/PhysRevLett.86.3463} {\bibfield  {journal} {\bibinfo  {journal}
  {Phys. Rev. Lett.}\ }\textbf {\bibinfo {volume} {86}},\ \bibinfo
  {pages} {3463--3466} (\bibinfo {year} {2001})}\BibitemShut {NoStop}%
\bibitem [{\citenamefont {Komatsu}\ \emph {et~al.}(2008)\citenamefont
  {Komatsu}, \citenamefont {Nakagawa}, \citenamefont {Sasa},\ and\
  \citenamefont {Tasaki}}]{komatsu_steady-state_2008}%
  \BibitemOpen
  \bibfield  {author} {\bibinfo {author} {\bibfnamefont {Teruhisa~S.}\
  \bibnamefont {Komatsu}}, \bibinfo {author} {\bibfnamefont {Naoko}\
  \bibnamefont {Nakagawa}}, \bibinfo {author} {\bibfnamefont {Shin-ichi}\
  \bibnamefont {Sasa}}, \ and\ \bibinfo {author} {\bibfnamefont {Hal}\
  \bibnamefont {Tasaki}},\ }\bibfield  {title} {\enquote {\bibinfo {title}
  {Steady-{State} {Thermodynamics} for {Heat} {Conduction}: {Microscopic}
  {Derivation}},}\ }\href {\doibase 10.1103/PhysRevLett.100.230602} {\bibfield
  {journal} {\bibinfo  {journal} {Phys. Rev. Lett.}\ }\textbf {\bibinfo
  {volume} {100}},\ \bibinfo {pages} {230602} (\bibinfo {year}
  {2008})}\BibitemShut {NoStop}%
\bibitem [{\citenamefont {Maes}\ and\ \citenamefont
  {Netočný}(2014)}]{maes_nonequilibrium_2014}%
  \BibitemOpen
  \bibfield  {author} {\bibinfo {author} {\bibfnamefont {Christian}\
  \bibnamefont {Maes}}\ and\ \bibinfo {author} {\bibfnamefont {Karel}\
  \bibnamefont {Netočný}},\ }\bibfield  {title} {{
  \enquote {\bibinfo {title} {A {Nonequilibrium} {Extension} of the
  {Clausius} {Heat} {Theorem}},}\ }}\href {\doibase 10.1007/s10955-013-0822-9}
  {\bibfield  {journal} {\bibinfo  {journal} {J. Stat. Phys.}\
  }\textbf {\bibinfo {volume} {154}},\ \bibinfo {pages} {188--203} (\bibinfo
  {year} {2014})}\BibitemShut {NoStop}%
\bibitem [{\citenamefont {Seifert}(2012)}]{seifert_stochastic_2012}%
  \BibitemOpen
  \bibfield  {author} {\bibinfo {author} {\bibfnamefont {Udo}\ \bibnamefont
  {Seifert}},\ }\bibfield  {title} {{\enquote {\bibinfo
  {title} {Stochastic thermodynamics, fluctuation theorems and molecular
  machines},}\ }}\href {\doibase 10.1088/0034-4885/75/12/126001} {\bibfield
  {journal} {\bibinfo  {journal} {Rep. Prog. Phys.}\ }\textbf
  {\bibinfo {volume} {75}},\ \bibinfo {pages} {126001} (\bibinfo {year}
  {2012})}\BibitemShut {NoStop}%
\bibitem [{\citenamefont {Crooks}(1999)}]{crooks_entropy_1999}%
  \BibitemOpen
  \bibfield  {author} {\bibinfo {author} {\bibfnamefont {Gavin~E.}\
  \bibnamefont {Crooks}},\ }\bibfield  {title} {\enquote {\bibinfo {title}
  {Entropy production fluctuation theorem and the nonequilibrium work relation
  for free energy differences},}\ }\href {\doibase 10.1103/PhysRevE.60.2721}
  {\bibfield  {journal} {\bibinfo  {journal} {Phys. Rev. E}\ }\textbf
  {\bibinfo {volume} {60}},\ \bibinfo {pages} {2721--2726} (\bibinfo {year}
  {1999})}\BibitemShut {NoStop}%
\bibitem [{\citenamefont {Gomez-Marin}\ \emph {et~al.}(2008)\citenamefont
  {Gomez-Marin}, \citenamefont {Parrondo},\ and\ \citenamefont {Van~den
  Broeck}}]{gomez-marin_lower_2008}%
  \BibitemOpen
  \bibfield  {author} {\bibinfo {author} {\bibfnamefont {A.}~\bibnamefont
  {Gomez-Marin}}, \bibinfo {author} {\bibfnamefont {J.~M.~R.}\ \bibnamefont
  {Parrondo}}, \ and\ \bibinfo {author} {\bibfnamefont {C.}~\bibnamefont
  {Van~den Broeck}},\ }\bibfield  {title} {\enquote {\bibinfo {title} {Lower
  bounds on dissipation upon coarse graining},}\ }\href {\doibase
  10.1103/PhysRevE.78.011107} {\bibfield  {journal} {\bibinfo  {journal}
  {Phys. Rev. E}\ }\textbf {\bibinfo {volume} {78}},\ \bibinfo {pages}
  {011107} (\bibinfo {year} {2008})}\BibitemShut {NoStop}%
\bibitem [{\citenamefont {Puglisi}\ \emph {et~al.}(2010)\citenamefont
  {Puglisi}, \citenamefont {Pigolotti}, \citenamefont {Rondoni},\ and\
  \citenamefont {Vulpiani}}]{puglisi_entropy_2010}%
  \BibitemOpen
  \bibfield  {author} {\bibinfo {author} {\bibfnamefont {A.}~\bibnamefont
  {Puglisi}}, \bibinfo {author} {\bibfnamefont {S.}~\bibnamefont {Pigolotti}},
  \bibinfo {author} {\bibfnamefont {L.}~\bibnamefont {Rondoni}}, \ and\
  \bibinfo {author} {\bibfnamefont {A.}~\bibnamefont {Vulpiani}},\ }\bibfield
  {title} {{\enquote {\bibinfo {title} {Entropy production
  and coarse graining in {Markov} processes},}\ }}\href {\doibase
  10.1088/1742-5468/2010/05/P05015} {\bibfield  {journal} {\bibinfo  {journal}
  {J. Stat. Mech. Theory E.}\ }\textbf {\bibinfo
  {volume} {2010}},\ \bibinfo {pages} {P05015} (\bibinfo {year}
  {2010})}\BibitemShut {NoStop}%
\bibitem [{\citenamefont {Roldán}\ and\ \citenamefont
  {Parrondo}(2010)}]{roldan_estimating_2010}%
  \BibitemOpen
  \bibfield  {author} {\bibinfo {author} {\bibfnamefont {Édgar}\ \bibnamefont
  {Roldán}}\ and\ \bibinfo {author} {\bibfnamefont {Juan M.~R.}\ \bibnamefont
  {Parrondo}},\ }\bibfield  {title} {\enquote {\bibinfo {title} {Estimating
  {Dissipation} from {Single} {Stationary} {Trajectories}},}\ }\href {\doibase
  10.1103/PhysRevLett.105.150607} {\bibfield  {journal} {\bibinfo  {journal}
  {Phys. Rev. Lett.}\ }\textbf {\bibinfo {volume} {105}},\ \bibinfo
  {pages} {150607} (\bibinfo {year} {2010})}\BibitemShut {NoStop}%
\bibitem [{\citenamefont {Celani}\ \emph {et~al.}(2012)\citenamefont {Celani},
  \citenamefont {Bo}, \citenamefont {Eichhorn},\ and\ \citenamefont
  {Aurell}}]{celani_anomalous_2012}%
  \BibitemOpen
  \bibfield  {author} {\bibinfo {author} {\bibfnamefont {Antonio}\ \bibnamefont
  {Celani}}, \bibinfo {author} {\bibfnamefont {Stefano}\ \bibnamefont {Bo}},
  \bibinfo {author} {\bibfnamefont {Ralf}\ \bibnamefont {Eichhorn}}, \ and\
  \bibinfo {author} {\bibfnamefont {Erik}\ \bibnamefont {Aurell}},\ }\bibfield
  {title} {\enquote {\bibinfo {title} {Anomalous {Thermodynamics} at the
  {Microscale}},}\ }\href {\doibase 10.1103/PhysRevLett.109.260603} {\bibfield
  {journal} {\bibinfo  {journal} {Phys. Rev. Lett.}\ }\textbf {\bibinfo
  {volume} {109}},\ \bibinfo {pages} {260603} (\bibinfo {year}
  {2012})}\BibitemShut {NoStop}%
\bibitem [{\citenamefont {Crisanti}\ \emph {et~al.}(2012)\citenamefont
  {Crisanti}, \citenamefont {Puglisi},\ and\ \citenamefont
  {Villamaina}}]{crisanti_nonequilibrium_2012}%
  \BibitemOpen
  \bibfield  {author} {\bibinfo {author} {\bibfnamefont {A.}~\bibnamefont
  {Crisanti}}, \bibinfo {author} {\bibfnamefont {A.}~\bibnamefont {Puglisi}}, \
  and\ \bibinfo {author} {\bibfnamefont {D.}~\bibnamefont {Villamaina}},\
  }\bibfield  {title} {\enquote {\bibinfo {title} {Nonequilibrium and
  information: {The} role of cross correlations},}\ }\href {\doibase
  10.1103/PhysRevE.85.061127} {\bibfield  {journal} {\bibinfo  {journal}
  {Phys. Rev. E}\ }\textbf {\bibinfo {volume} {85}},\ \bibinfo {pages}
  {061127} (\bibinfo {year} {2012})}\BibitemShut {NoStop}%
\bibitem [{\citenamefont {Esposito}(2012)}]{esposito_stochastic_2012}%
  \BibitemOpen
  \bibfield  {author} {\bibinfo {author} {\bibfnamefont {Massimiliano}\
  \bibnamefont {Esposito}},\ }\bibfield  {title} {\enquote {\bibinfo {title}
  {Stochastic thermodynamics under coarse graining},}\ }\href {\doibase
  10.1103/PhysRevE.85.041125} {\bibfield  {journal} {\bibinfo  {journal}
  {Phys. Rev. E}\ }\textbf {\bibinfo {volume} {85}},\ \bibinfo {pages}
  {041125} (\bibinfo {year} {2012})}\BibitemShut {NoStop}%
\bibitem [{\citenamefont {Mehl}\ \emph {et~al.}(2012)\citenamefont {Mehl},
  \citenamefont {Lander}, \citenamefont {Bechinger}, \citenamefont {Blickle},\
  and\ \citenamefont {Seifert}}]{mehl_role_2012}%
  \BibitemOpen
  \bibfield  {author} {\bibinfo {author} {\bibfnamefont {J.}~\bibnamefont
  {Mehl}}, \bibinfo {author} {\bibfnamefont {B.}~\bibnamefont {Lander}},
  \bibinfo {author} {\bibfnamefont {C.}~\bibnamefont {Bechinger}}, \bibinfo
  {author} {\bibfnamefont {V.}~\bibnamefont {Blickle}}, \ and\ \bibinfo
  {author} {\bibfnamefont {U.}~\bibnamefont {Seifert}},\ }\bibfield  {title}
  {\enquote {\bibinfo {title} {Role of {Hidden} {Slow} {Degrees} of {Freedom}
  in the {Fluctuation} {Theorem}},}\ }\href {\doibase
  10.1103/PhysRevLett.108.220601} {\bibfield  {journal} {\bibinfo  {journal}
  {Phys. Rev. Lett.}\ }\textbf {\bibinfo {volume} {108}},\ \bibinfo
  {pages} {220601} (\bibinfo {year} {2012})}\BibitemShut {NoStop}%
\bibitem [{\citenamefont {Kawaguchi}\ and\ \citenamefont
  {Nakayama}(2013)}]{kawaguchi_fluctuation_2013}%
  \BibitemOpen
  \bibfield  {author} {\bibinfo {author} {\bibfnamefont {Kyogo}\ \bibnamefont
  {Kawaguchi}}\ and\ \bibinfo {author} {\bibfnamefont {Yohei}\ \bibnamefont
  {Nakayama}},\ }\bibfield  {title} {\enquote {\bibinfo {title} {Fluctuation
  theorem for hidden entropy production},}\ }\href {\doibase
  10.1103/PhysRevE.88.022147} {\bibfield  {journal} {\bibinfo  {journal}
  {Phys. Rev. E}\ }\textbf {\bibinfo {volume} {88}},\ \bibinfo {pages}
  {022147} (\bibinfo {year} {2013})}\BibitemShut {NoStop}%
\bibitem [{\citenamefont {Strasberg}\ \emph {et~al.}(2013)\citenamefont
  {Strasberg}, \citenamefont {Schaller}, \citenamefont {Brandes},\ and\
  \citenamefont {Esposito}}]{strasberg_thermodynamics_2013}%
  \BibitemOpen
  \bibfield  {author} {\bibinfo {author} {\bibfnamefont {Philipp}\ \bibnamefont
  {Strasberg}}, \bibinfo {author} {\bibfnamefont {Gernot}\ \bibnamefont
  {Schaller}}, \bibinfo {author} {\bibfnamefont {Tobias}\ \bibnamefont
  {Brandes}}, \ and\ \bibinfo {author} {\bibfnamefont {Massimiliano}\
  \bibnamefont {Esposito}},\ }\bibfield  {title} {\enquote {\bibinfo {title}
  {Thermodynamics of a {Physical} {Model} {Implementing} a {Maxwell}
  {Demon}},}\ }\href {\doibase 10.1103/PhysRevLett.110.040601} {\bibfield
  {journal} {\bibinfo  {journal} {Phys. Rev. Lett.}\ }\textbf {\bibinfo
  {volume} {110}},\ \bibinfo {pages} {040601} (\bibinfo {year}
  {2013})}\BibitemShut {NoStop}%
\bibitem [{\citenamefont {Munakata}\ and\ \citenamefont
  {Rosinberg}(2014)}]{munakata_entropy_2014}%
  \BibitemOpen
  \bibfield  {author} {\bibinfo {author} {\bibfnamefont {T.}~\bibnamefont
  {Munakata}}\ and\ \bibinfo {author} {\bibfnamefont {M. L.}\ \bibnamefont
  {Rosinberg}},\ }\bibfield  {title} {\enquote {\bibinfo {title} {Entropy
  {Production} and {Fluctuation} {Theorems} for {Langevin} {Processes} under
  {Continuous} {Non}-{Markovian} {Feedback} {Control}},}\ }\href {\doibase
  10.1103/PhysRevLett.112.180601} {\bibfield  {journal} {\bibinfo  {journal}
  {Phys. Rev. Lett.}\ }\textbf {\bibinfo {volume} {112}},\ \bibinfo
  {pages} {180601} (\bibinfo {year} {2014})}\BibitemShut {NoStop}%
\bibitem [{\citenamefont {Zimmermann}\ and\ \citenamefont
  {Seifert}(2015)}]{zimmermann_effective_2015}%
  \BibitemOpen
  \bibfield  {author} {\bibinfo {author} {\bibfnamefont {Eva}\ \bibnamefont
  {Zimmermann}}\ and\ \bibinfo {author} {\bibfnamefont {Udo}\ \bibnamefont
  {Seifert}},\ }\bibfield  {title} {\enquote {\bibinfo {title} {Effective rates
  from thermodynamically consistent coarse-graining of models for molecular
  motors with probe particles},}\ }\href {\doibase 10.1103/PhysRevE.91.022709}
  {\bibfield  {journal} {\bibinfo  {journal} {Phys. Rev. E}\ }\textbf
  {\bibinfo {volume} {91}},\ \bibinfo {pages} {022709} (\bibinfo {year}
  {2015})}\BibitemShut {NoStop}%
\bibitem [{\citenamefont {Wang}\ \emph {et~al.}(2016)\citenamefont {Wang},
  \citenamefont {Kawaguchi}, \citenamefont {Sasa},\ and\ \citenamefont
  {Tang}}]{wang_entropy_2016}%
  \BibitemOpen
  \bibfield  {author} {\bibinfo {author} {\bibfnamefont {Shou-Wen}\
  \bibnamefont {Wang}}, \bibinfo {author} {\bibfnamefont {Kyogo}\ \bibnamefont
  {Kawaguchi}}, \bibinfo {author} {\bibfnamefont {Shin-ichi}\ \bibnamefont
  {Sasa}}, \ and\ \bibinfo {author} {\bibfnamefont {Lei-Han}\ \bibnamefont
  {Tang}},\ }\bibfield  {title} {\enquote {\bibinfo {title} {Entropy
  {Production} of {Nanosystems} with {Time} {Scale} {Separation}},}\ }\href
  {\doibase 10.1103/PhysRevLett.117.070601} {\bibfield  {journal} {\bibinfo
  {journal} {Phys. Rev. Lett.}\ }\textbf {\bibinfo {volume} {117}},\
  \bibinfo {pages} {070601} (\bibinfo {year} {2016})}\BibitemShut {NoStop}%
\bibitem [{\citenamefont {Bo}\ and\ \citenamefont
  {Celani}(2017)}]{bo_multiple-scale_2017}%
  \BibitemOpen
  \bibfield  {author} {\bibinfo {author} {\bibfnamefont {Stefano}\ \bibnamefont
  {Bo}}\ and\ \bibinfo {author} {\bibfnamefont {Antonio}\ \bibnamefont
  {Celani}},\ }\bibfield  {title} {\enquote {\bibinfo {title} {Multiple-scale
  stochastic processes: {Decimation}, averaging and beyond},}\ }\href {\doibase
  10.1016/j.physrep.2016.12.003} {\bibfield  {journal} {\bibinfo  {journal}
  {Phys. Rep.}\ }\ 
  \textbf {\bibinfo {volume} {670}},\
  \bibinfo {pages} {1--59} (\bibinfo {year} {2017})}\BibitemShut {NoStop}%
\bibitem [{\citenamefont {Feynman}\ \emph {et~al.}(2010)\citenamefont
  {Feynman}, \citenamefont {Leighton},\ and\ \citenamefont
  {Sands}}]{feynman_feynman_2010}%
  \BibitemOpen
  \bibfield  {author} {\bibinfo {author} {\bibfnamefont {Richard~Phillips}\
  \bibnamefont {Feynman}}, \bibinfo {author} {\bibfnamefont {Robert~B.}\
  \bibnamefont {Leighton}}, \ and\ \bibinfo {author} {\bibfnamefont {Matthew}\
  \bibnamefont {Sands}},\ }\href
  {https://opac.dl.itc.u-tokyo.ac.jp/opac/opac_details.cgi?lang=0&bibid=2003031814&amode=11}
  {\emph {\bibinfo {title} {The {Feynman} lectures on physics}}}\ (\bibinfo
  {publisher} {Basic Books},\ \bibinfo {address} {New York},\ \bibinfo {year}
  {2010})\BibitemShut {NoStop}%
\bibitem [{\citenamefont
  {Smoluchowski}(1912)}]{smoluchowski_experimentell_1912}%
  \BibitemOpen
  \bibfield  {author} {\bibinfo {author} {\bibfnamefont {Marian}\ \bibnamefont
  {Smoluchowski}},\ }\bibfield  {title} {\enquote {\bibinfo {title}
  {Experimentell nachweisbare, der üblichen {Thermodynamik} widersprechende
  {Molekularphänomene}},}\ }\href@noop {} {\bibfield  {journal} {\bibinfo
  {journal} {Phys. Z. XIII}\ ,\ \bibinfo {pages} {1069--1080}}
  (\bibinfo {year} {1912})}\BibitemShut {NoStop}%
\bibitem [{\citenamefont {Parrondo}\ and\ \citenamefont
  {Español}(1996)}]{parrondo_criticism_1996}%
  \BibitemOpen
  \bibfield  {author} {\bibinfo {author} {\bibfnamefont {Juan M.~R.}\
  \bibnamefont {Parrondo}}\ and\ \bibinfo {author} {\bibfnamefont {Pep}\
  \bibnamefont {Español}},\ }\bibfield  {title} {\enquote {\bibinfo {title}
  {Criticism of Feynman’s analysis of the ratchet as an engine},}\ }\href
  {\doibase 10.1119/1.18393} {\bibfield  {journal} {\bibinfo  {journal}
  {Am. J. Phys.}\ }\textbf {\bibinfo {volume} {64}},\ \bibinfo
  {pages} {1125} (\bibinfo {year} {1996})}\BibitemShut {NoStop}%
\bibitem [{\citenamefont {Derényi}\ and\ \citenamefont
  {Astumian}(1999)}]{derenyi_efficiency_1999}%
  \BibitemOpen
  \bibfield  {author} {\bibinfo {author} {\bibfnamefont {Imre}\ \bibnamefont
  {Derényi}}\ and\ \bibinfo {author} {\bibfnamefont {R.~Dean}\ \bibnamefont
  {Astumian}},\ }\bibfield  {title} {\enquote {\bibinfo {title} {Efficiency of
  {Brownian} heat engines},}\ }\href {\doibase 10.1103/PhysRevE.59.R6219}
  {\bibfield  {journal} {\bibinfo  {journal} {Phys. Rev. E}\ }\textbf
  {\bibinfo {volume} {59}},\ \bibinfo {pages} {R6219--R6222} (\bibinfo {year}
  {1999})}\BibitemShut {NoStop}%
\bibitem [{\citenamefont {Hondou}\ and\ \citenamefont
  {Sekimoto}(2000)}]{hondou_unattainability_2000}%
  \BibitemOpen
  \bibfield  {author} {\bibinfo {author} {\bibfnamefont {Tsuyoshi}\
  \bibnamefont {Hondou}}\ and\ \bibinfo {author} {\bibfnamefont {Ken}\
  \bibnamefont {Sekimoto}},\ }\bibfield  {title} {\enquote {\bibinfo {title}
  {Unattainability of {Carnot} efficiency in the {Brownian} heat engine},}\
  }\href {\doibase 10.1103/PhysRevE.62.6021} {\bibfield  {journal} {\bibinfo
  {journal} {Phys. Rev. E}\ }\textbf {\bibinfo {volume} {62}},\ \bibinfo
  {pages} {6021--6025} (\bibinfo {year} {2000})}\BibitemShut {NoStop}%
\bibitem [{\citenamefont {Benjamin}\ and\ \citenamefont
  {Kawai}(2008)}]{benjamin_inertial_2008}%
  \BibitemOpen
  \bibfield  {author} {\bibinfo {author} {\bibfnamefont {Ronald}\ \bibnamefont
  {Benjamin}}\ and\ \bibinfo {author} {\bibfnamefont {Ryoichi}\ \bibnamefont
  {Kawai}},\ }\bibfield  {title} {\enquote {\bibinfo {title} {Inertial effects
  in {Büttiker}-{Landauer} motor and refrigerator at the overdamped limit},}\
  }\href {\doibase 10.1103/PhysRevE.77.051132} {\bibfield  {journal} {\bibinfo
  {journal} {Phys. Rev. E}\ }\textbf {\bibinfo {volume} {77}},\ \bibinfo
  {pages} {051132} (\bibinfo {year} {2008})}\BibitemShut {NoStop}%
\bibitem [{\citenamefont {Shiraishi}(2015)}]{shiraishi_attainability_2015}%
  \BibitemOpen
  \bibfield  {author} {\bibinfo {author} {\bibfnamefont {Naoto}\ \bibnamefont
  {Shiraishi}},\ }\bibfield  {title} {\enquote {\bibinfo {title} {Attainability
  of {Carnot} efficiency with autonomous engines},}\ }\href {\doibase
  10.1103/PhysRevE.92.050101} {\bibfield  {journal} {\bibinfo  {journal}
  {Phys. Rev. E}\ }\textbf {\bibinfo {volume} {92}},\ \bibinfo {pages}
  {050101} (\bibinfo {year} {2015})}\BibitemShut {NoStop}%
\bibitem [{\citenamefont {Açıkkalp}(2016)}]{acikkalp_analysis_2016}%
  \BibitemOpen
  \bibfield  {author} {\bibinfo {author} {\bibfnamefont {Emin}\ \bibnamefont
  {Açıkkalp}},\ }\bibfield  {title} {{\enquote {\bibinfo
  {title} {Analysis of a {Brownian} heat engine with ecological criteria},}\
  }}\href {\doibase 10.1140/epjp/i2016-16426-6} {\bibfield  {journal} {\bibinfo
   {journal} {Eur. Phys. J. Plus}\ }\textbf {\bibinfo {volume}
  {131}},\ \bibinfo {pages} {426} (\bibinfo {year} {2016})}\BibitemShut
  {NoStop}%
\bibitem [{\citenamefont {A. Martínez}\ \emph {et~al.}(2017)\citenamefont
  {A. Martínez}, \citenamefont {Roldán}, \citenamefont {Dinis},\ and\
  \citenamefont {A. Rica}}]{a.martinez_colloidal_2017}%
  \BibitemOpen
  \bibfield  {author} {\bibinfo {author} {\bibfnamefont {Ignacio}\ \bibnamefont
  {A. Martínez}}, \bibinfo {author} {\bibfnamefont {Édgar}\ \bibnamefont
  {Roldán}}, \bibinfo {author} {\bibfnamefont {Luis}\ \bibnamefont {Dinis}}, \
  and\ \bibinfo {author} {\bibfnamefont {Raúl}\ \bibnamefont {A. Rica}},\
  }\bibfield  {title} {{\enquote {\bibinfo {title}
  {Colloidal heat engines: a review},}\ }}\href {\doibase 10.1039/C6SM00923A}
  {\bibfield  {journal} {\bibinfo  {journal} {Soft Matter}\ }\textbf {\bibinfo
  {volume} {13}},\ \bibinfo {pages} {22--36} (\bibinfo {year}
  {2017})}\BibitemShut {NoStop}%
\bibitem [{\citenamefont {Büttiker}(1987)}]{buttiker_transport_1987}%
  \BibitemOpen
  \bibfield  {author} {\bibinfo {author} {\bibfnamefont {M.}~\bibnamefont
  {Büttiker}},\ }\bibfield  {title} {\enquote {\bibinfo {title} {Transport as
  a consequence of state-dependent diffusion},}\ }\href {\doibase
  10.1007/BF01304221} {\bibfield  {journal} {\bibinfo  {journal} {Z.
  Phys. B: Condens. Matter}\ }\textbf {\bibinfo {volume} {68}},\ \bibinfo
  {pages} {161--167} (\bibinfo {year} {1987})}\BibitemShut {NoStop}%
\bibitem [{\citenamefont {Millonas}(1995)}]{millonas_self-consistent_1995}%
  \BibitemOpen
  \bibfield  {author} {\bibinfo {author} {\bibfnamefont {Mark~M.}\ \bibnamefont
  {Millonas}},\ }\bibfield  {title} {\enquote {\bibinfo {title}
  {Self-{Consistent} {Microscopic} {Theory} of {Fluctuation}-{Induced}
  {Transport}},}\ }\href {\doibase 10.1103/PhysRevLett.74.10} {\bibfield
  {journal} {\bibinfo  {journal} {Phys. Rev. Lett.}\ }\textbf {\bibinfo
  {volume} {74}},\ \bibinfo {pages} {10--13} (\bibinfo {year}
  {1995})}\BibitemShut {NoStop}%
\bibitem [{\citenamefont {Sekimoto}(1997)}]{sekimoto_kinetic_1997}%
  \BibitemOpen
  \bibfield  {author} {\bibinfo {author} {\bibfnamefont {Ken}\ \bibnamefont
  {Sekimoto}},\ }\bibfield  {title} {\enquote {\bibinfo {title} {Kinetic
  {Characterization} of {Heat} {Bath} and the {Energetics} of 
  {Thermal} {Ratchet} {Models}},}\ }\href {\doibase 10.1143/JPSJ.66.1234}
  {\bibfield  {journal} {\bibinfo  {journal} {J. Phys. Soc.
  Jpn.}\ }\textbf {\bibinfo {volume} {66}},\ \bibinfo {pages} {1234--1237}
  (\bibinfo {year} {1997})}\BibitemShut {NoStop}%
\bibitem [{\citenamefont {Hondou}\ and\ \citenamefont
  {Takagi}(1998)}]{hondou_irreversible_1998}%
  \BibitemOpen
  \bibfield  {author} {\bibinfo {author} {\bibfnamefont {Tsuyoshi}\
  \bibnamefont {Hondou}}\ and\ \bibinfo {author} {\bibfnamefont {Fumiko}\
  \bibnamefont {Takagi}},\ }\bibfield  {title} {\enquote {\bibinfo {title}
  {Irreversible {Operation} in a {Stalled} {State} of {Feynman}'s {Ratchet}},}\
  }\href {\doibase 10.1143/JPSJ.67.2974} {\bibfield  {journal} {\bibinfo
  {journal} {J. Phys. Soc. Jpn.}\ }\textbf {\bibinfo
  {volume} {67}},\ \bibinfo {pages} {2974--2976} (\bibinfo {year}
  {1998})}\BibitemShut {NoStop}%
\bibitem [{\citenamefont {Jack}\ and\ \citenamefont
  {Tumlin}(2016)}]{jack_intrinsic_2016}%
  \BibitemOpen
  \bibfield  {author} {\bibinfo {author} {\bibfnamefont {M.~W.}\ \bibnamefont
  {Jack}}\ and\ \bibinfo {author} {\bibfnamefont {C.}~\bibnamefont {Tumlin}},\
  }\bibfield  {title} {\enquote {\bibinfo {title} {Intrinsic irreversibility
  limits the efficiency of multidimensional molecular motors},}\ }\href
  {\doibase 10.1103/PhysRevE.93.052109} {\bibfield  {journal} {\bibinfo
  {journal} {Phys. Rev. E}\ }\textbf {\bibinfo {volume} {93}},\ \bibinfo
  {pages} {052109} (\bibinfo {year} {2016})}\BibitemShut {NoStop}%
\bibitem [{\citenamefont {Zwanzig}(1973)}]{zwanzig_nonlinear_1973}%
  \BibitemOpen
  \bibfield  {author} {\bibinfo {author} {\bibfnamefont {Robert}\ \bibnamefont
  {Zwanzig}},\ }\bibfield  {title} {{\enquote {\bibinfo
  {title} {Nonlinear generalized {Langevin} equations},}\ }}\href {\doibase
  10.1007/BF01008729} {\bibfield  {journal} {\bibinfo  {journal} {J.
  Stat. Phys.}\ }\textbf {\bibinfo {volume} {9}},\ \bibinfo {pages}
  {215--220} (\bibinfo {year} {1973})}\BibitemShut {NoStop}%
\bibitem [{\citenamefont {Sekimoto}(1999)}]{sekimoto_temporal_1999}%
  \BibitemOpen
  \bibfield  {author} {\bibinfo {author} {\bibfnamefont {Ken}\ \bibnamefont
  {Sekimoto}},\ }\bibfield  {title} {\enquote {\bibinfo {title} {Temporal
  {Coarse} {Graining} for  {Systems} of {Brownian} {Particles}
   with {Non}-{Constant} {Temperature}},}\ }\href {\doibase
  10.1143/JPSJ.68.1448} {\bibfield  {journal} {\bibinfo  {journal} {J.
  Phys. Soc. Jpn.}\ }\textbf {\bibinfo {volume} {68}},\ \bibinfo
  {pages} {1448--1449} (\bibinfo {year} {1999})}\BibitemShut {NoStop}%
\bibitem [{\citenamefont {Jayannavar}\ and\ \citenamefont
  {Mahato}(1995)}]{jayannavar_macroscopic_1995}%
  \BibitemOpen
  \bibfield  {author} {\bibinfo {author} {\bibfnamefont {A.~M.}\ \bibnamefont
  {Jayannavar}}\ and\ \bibinfo {author} {\bibfnamefont {Mangal~C.}\
  \bibnamefont {Mahato}},\ }\bibfield  {title} {{\enquote
  {\bibinfo {title} {Macroscopic equation of motion in inhomogeneous media: {A}
  microscopic treatment},}\ }}\href {\doibase 10.1007/BF02848625} {\bibfield
  {journal} {\bibinfo  {journal} {Pramana}\ }\textbf {\bibinfo {volume} {45}},\
  \bibinfo {pages} {369--376} (\bibinfo {year} {1995})}\BibitemShut {NoStop}%
\bibitem [{\citenamefont {Sekimoto}(2010)}]{sekimoto_stochastic_2010}%
  \BibitemOpen
  \bibfield  {author} {\bibinfo {author} {\bibfnamefont {Ken}\ \bibnamefont
  {Sekimoto}},\ }\href
  {https://opac.dl.itc.u-tokyo.ac.jp/opac/opac_details.cgi?lang=0&bibid=2002852725&amode=11}
  {\emph {\bibinfo {title} {Stochastic energetics}}},\ \bibinfo {series}
  {Lecture notes in physics}, Vol.\ \bibinfo {volume} {799}\ (\bibinfo
  {publisher} {Springer},\ \bibinfo {address} {Berlin},\ \bibinfo {year}
  {2010})\BibitemShut {NoStop}%
\bibitem [{\citenamefont {Magnasco}\ and\ \citenamefont
  {Stolovitzky}(1998)}]{magnasco_feynmans_1998}%
  \BibitemOpen
  \bibfield  {author} {\bibinfo {author} {\bibfnamefont {Marcelo~O.}\
  \bibnamefont {Magnasco}}\ and\ \bibinfo {author} {\bibfnamefont {Gustavo}\
  \bibnamefont {Stolovitzky}},\ }\bibfield  {title} {{
  \enquote {\bibinfo {title} {Feynman's {Ratchet} and {Pawl}},}\ }}\href
  {\doibase 10.1023/B:JOSS.0000033245.43421.14} {\bibfield  {journal} {\bibinfo
   {journal} {J. Stat. Phys.}\ }\textbf {\bibinfo {volume}
  {93}},\ \bibinfo {pages} {615--632} (\bibinfo {year} {1998})}\BibitemShut
  {NoStop}%
\bibitem [{\citenamefont {Gomez-Marin}\ and\ \citenamefont
  {Sancho}(2006)}]{gomez-marin_ratchet_2006}%
  \BibitemOpen
  \bibfield  {author} {\bibinfo {author} {\bibfnamefont {A.}~\bibnamefont
  {Gomez-Marin}}\ and\ \bibinfo {author} {\bibfnamefont {J.~M.}\ \bibnamefont
  {Sancho}},\ }\bibfield  {title} {\enquote {\bibinfo {title} {Ratchet, pawl
  and spring {Brownian} motor},}\ }\href {\doibase 10.1016/j.physd.2005.12.015}
  {\bibfield  {journal} {\bibinfo  {journal} {Physica D}\
  }
  \textbf {\bibinfo {volume} {216}},\ \bibinfo {pages}
  {214--219} (\bibinfo {year} {2006})}\BibitemShut {NoStop}%
\bibitem [{\citenamefont {Bo}\ and\ \citenamefont
  {Celani}(2013)}]{bo_entropic_2013}%
  \BibitemOpen
  \bibfield  {author} {\bibinfo {author} {\bibfnamefont {Stefano}\ \bibnamefont
  {Bo}}\ and\ \bibinfo {author} {\bibfnamefont {Antonio}\ \bibnamefont
  {Celani}},\ }\bibfield  {title} {\enquote {\bibinfo {title} {Entropic anomaly
  and maximal efficiency of microscopic heat engines},}\ }\href {\doibase
  10.1103/PhysRevE.87.050102} {\bibfield  {journal} {\bibinfo  {journal}
  {Phys. Rev. E}\ }\textbf {\bibinfo {volume} {87}},\ \bibinfo {pages}
  {050102} (\bibinfo {year} {2013})}\BibitemShut {NoStop}%
\bibitem [{\citenamefont {Asfaw}\ and\ \citenamefont
  {Bekele}(2007)}]{asfaw_exploring_2007}%
  \BibitemOpen
  \bibfield  {author} {\bibinfo {author} {\bibfnamefont {Mesfin}\ \bibnamefont
  {Asfaw}}\ and\ \bibinfo {author} {\bibfnamefont {Mulugeta}\ \bibnamefont
  {Bekele}},\ }\bibfield  {title} {\enquote {\bibinfo {title} {Exploring the
  operation of a tiny heat engine},}\ }\href {\doibase
  10.1016/j.physa.2007.05.035} {\bibfield  {journal} {\bibinfo  {journal}
  {Physica A:}\ }\textbf {\bibinfo
  {volume} {384}},\ \bibinfo {pages} {346--358} (\bibinfo {year}
  {2007})}\BibitemShut {NoStop}%
\bibitem [{\citenamefont {Velasco}\ \emph {et~al.}(2001)\citenamefont
  {Velasco}, \citenamefont {Roco}, \citenamefont {Medina},\ and\ \citenamefont
  {Hernández}}]{velasco_feynmans_2001}%
  \BibitemOpen
  \bibfield  {author} {\bibinfo {author} {\bibfnamefont {S}~\bibnamefont
  {Velasco}}, \bibinfo {author} {\bibfnamefont {J~M~M}\ \bibnamefont {Roco}},
  \bibinfo {author} {\bibfnamefont {A}~\bibnamefont {Medina}}, \ and\ \bibinfo
  {author} {\bibfnamefont {A~Calvo}\ \bibnamefont {Hernández}},\ }\bibfield
  {title} {\enquote {\bibinfo {title} {Feynman's ratchet optimization: maximum
  power and maximum efficiency regimes},}\ }\href {\doibase
  10.1088/0022-3727/34/6/323} {\bibfield  {journal} {\bibinfo  {journal}
  {J. Phys. D: Appl. Phys.}\ }\textbf {\bibinfo {volume} {34}},\
  \bibinfo {pages} {1000--1006} (\bibinfo {year} {2001})}\BibitemShut {NoStop}%
\bibitem [{\citenamefont {Salas}\ and\ \citenamefont
  {Hernández}(2003)}]{salas_unified_2003}%
  \BibitemOpen
  \bibfield  {author} {\bibinfo {author} {\bibfnamefont {N.~Sánchez}\
  \bibnamefont {Salas}}\ and\ \bibinfo {author} {\bibfnamefont {A.~Calvo}\
  \bibnamefont {Hernández}},\ }\bibfield  {title} {{
  \enquote {\bibinfo {title} {Unified working regime of nonlinear systems
  rectifying thermal fluctuations},}\ }}\href {\doibase
  10.1209/epl/i2003-00197-2} {\bibfield  {journal} {\bibinfo  {journal} {
  Europhys. Lett.}\ }\textbf {\bibinfo {volume} {61}},\ \bibinfo {pages}
  {287} (\bibinfo {year} {2003})}\BibitemShut {NoStop}%
\bibitem [{\citenamefont {Tu}(2008)}]{tu_efficiency_2008}%
  \BibitemOpen
  \bibfield  {author} {\bibinfo {author} {\bibfnamefont {Z~C}\ \bibnamefont
  {Tu}},\ }\bibfield  {title} {\enquote {\bibinfo {title} {Efficiency at
  maximum power of {Feynman}'s ratchet as a heat engine},}\ }\href {\doibase
  10.1088/1751-8113/41/31/312003} {\bibfield  {journal} {\bibinfo  {journal}
  {J. Phys. A: Math. Theor.}\ }\textbf {\bibinfo
  {volume} {41}},\ \bibinfo {pages} {312003} (\bibinfo {year}
  {2008})}\BibitemShut {NoStop}%
\bibitem [{\citenamefont {Lin}\ and\ \citenamefont
  {Chen}(2009)}]{lin_performance_2009}%
  \BibitemOpen
  \bibfield  {author} {\bibinfo {author} {\bibfnamefont {Bihong}\ \bibnamefont
  {Lin}}\ and\ \bibinfo {author} {\bibfnamefont {Jincan}\ \bibnamefont
  {Chen}},\ }\bibfield  {title} {{\enquote {\bibinfo
  {title} {Performance characteristics and parametric optimum criteria of a
  {Brownian} micro-refrigerator in a spatially periodic temperature field},}\
  }}\href {\doibase 10.1088/1751-8113/42/7/075006} {\bibfield  {journal}
  {\bibinfo  {journal} {J. Phys. A: Math. Theor.}\
  }\textbf {\bibinfo {volume} {42}},\ \bibinfo {pages} {075006} (\bibinfo
  {year} {2009})}\BibitemShut {NoStop}%
\bibitem [{\citenamefont {Chen}\ \emph {et~al.}(2011)\citenamefont {Chen},
  \citenamefont {Ding},\ and\ \citenamefont {Sun}}]{chen_optimum_2011}%
  \BibitemOpen
  \bibfield  {author} {\bibinfo {author} {\bibfnamefont {Lingen}\ \bibnamefont
  {Chen}}, \bibinfo {author} {\bibfnamefont {Zemin}\ \bibnamefont {Ding}}, \
  and\ \bibinfo {author} {\bibfnamefont {Fengrui}\ \bibnamefont {Sun}},\
  }\bibfield  {title} {\enquote {\bibinfo {title} {Optimum performance analysis
  of {Feynman}'s engine as cold and hot ratchets},}\ }\href {\doibase
  10.1515/jnetdy.2011.011} {\bibfield  {journal} {\bibinfo  {journal} {J.
  Non-Equilib. Thermodyn.}\ }\textbf {\bibinfo {volume} {36}},\
  \bibinfo {pages} {155--177} (\bibinfo {year} {2011})}\BibitemShut {NoStop}%
\bibitem [{\citenamefont {Luo}\ \emph {et~al.}(2013)\citenamefont {Luo},
  \citenamefont {Liu},\ and\ \citenamefont {He}}]{luo_optimum_2013}%
  \BibitemOpen
  \bibfield  {author} {\bibinfo {author} {\bibfnamefont {X.~G.}\ \bibnamefont
  {Luo}}, \bibinfo {author} {\bibfnamefont {N.}~\bibnamefont {Liu}}, \ and\
  \bibinfo {author} {\bibfnamefont {J.~Z.}\ \bibnamefont {He}},\ }\bibfield
  {title} {\enquote {\bibinfo {title} {Optimum analysis of a {Brownian}
  refrigerator},}\ }\href {\doibase 10.1103/PhysRevE.87.022139} {\bibfield
  {journal} {\bibinfo  {journal} {Phys. Rev. E}\ }\textbf {\bibinfo
  {volume} {87}},\ \bibinfo {pages} {022139} (\bibinfo {year}
  {2013})}\BibitemShut {NoStop}%
\bibitem [{\citenamefont {Apertet}\ \emph {et~al.}(2014)\citenamefont
  {Apertet}, \citenamefont {Ouerdane}, \citenamefont {Goupil},\ and\
  \citenamefont {Lecoeur}}]{apertet_revisiting_2014}%
  \BibitemOpen
  \bibfield  {author} {\bibinfo {author} {\bibfnamefont {Y.}~\bibnamefont
  {Apertet}}, \bibinfo {author} {\bibfnamefont {H.}~\bibnamefont {Ouerdane}},
  \bibinfo {author} {\bibfnamefont {C.}~\bibnamefont {Goupil}}, \ and\ \bibinfo
  {author} {\bibfnamefont {Ph.}\ \bibnamefont {Lecoeur}},\ }\bibfield  {title}
  {\enquote {\bibinfo {title} {Revisiting {Feynman}'s ratchet with
  thermoelectric transport theory},}\ }\href {\doibase
  10.1103/PhysRevE.90.012113} {\bibfield  {journal} {\bibinfo  {journal}
  {Phys. Rev. E}\ }\textbf {\bibinfo {volume} {90}},\ \bibinfo {pages}
  {012113} (\bibinfo {year} {2014})}\BibitemShut {NoStop}%
\bibitem [{\citenamefont {Shi-Qi}\ \emph {et~al.}(2014)\citenamefont {Shi-Qi},
  \citenamefont {Pan},\ and\ \citenamefont
  {Zhan-Chun}}]{shi-qi_coefficient_2014}%
  \BibitemOpen
  \bibfield  {author} {\bibinfo {author} {\bibfnamefont {Sheng}\ \bibnamefont
  {Shi-Qi}}, \bibinfo {author} {\bibfnamefont {Yang}\ \bibnamefont {Pan}}, \
  and\ \bibinfo {author} {\bibfnamefont {Tu}~\bibnamefont {Zhan-Chun}},\
  }\bibfield  {title} {{\enquote {\bibinfo {title}
  {Coefficient of {Performance} at {Maximum} \(\chi\)-{Criterion} for {Feynman}
  {Ratchet} as a {Refrigerator}},}\ }}\href {\doibase
  10.1088/0253-6102/62/4/16} {\bibfield  {journal} {\bibinfo  {journal}
  {Commun. Theor. Phys.}\ }\textbf {\bibinfo {volume} {62}},\
  \bibinfo {pages} {589} (\bibinfo {year} {2014})}\BibitemShut {NoStop}%
\bibitem [{\citenamefont {van Kampen}(1992)}]{van_kampen_stochastic_1992}%
  \BibitemOpen
  \bibfield  {author} {\bibinfo {author} {\bibfnamefont {N~G}\ \bibnamefont
  {van Kampen}},\ }\href@noop {} {\emph {\bibinfo {title} {Stochastic processes
  in physics and chemistry}}},\ North-{Holland} personal library\ (\bibinfo
  {publisher} {North-Holland},\ \bibinfo {address} {Amsterdam ; Tokyo},\
  \bibinfo {year} {1992})\BibitemShut {NoStop}%
\bibitem [{\citenamefont {Kramers}(1940)}]{kramers_brownian_1940}%
  \BibitemOpen
  \bibfield  {author} {\bibinfo {author} {\bibfnamefont {H.~A.}\ \bibnamefont
  {Kramers}},\ }\bibfield  {title} {\enquote {\bibinfo {title} {Brownian motion
  in a field of force and the diffusion model of chemical reactions},}\ }\href
  {\doibase 10.1016/S0031-8914(40)90098-2} {\bibfield  {journal} {\bibinfo
  {journal} {Physica}\ }\textbf {\bibinfo {volume} {7}},\ \bibinfo {pages}
  {284--304} (\bibinfo {year} {1940})}\BibitemShut {NoStop}%
\end{thebibliography}
%

\end{document}